\documentclass[fleqn,usenatbib]{mnras}

\usepackage{newtxtext,newtxmath}

\usepackage[T1]{fontenc}

\DeclareRobustCommand{\VAN}[3]{#2}
\let\VANthebibliography\thebibliography
\def\thebibliography{\DeclareRobustCommand{\VAN}[3]{##3}\VANthebibliography}

\usepackage{multirow}

\usepackage{amssymb}
\usepackage{caption}
\usepackage{booktabs}
\usepackage{float}
\usepackage{lscape}
\usepackage{tabularray}
\usepackage{longtable}
\usepackage{placeins}

\usepackage{graphicx}	
\usepackage{amsmath}	

\usepackage{url}
\usepackage{hyperref}

\title[Carbon-Chain Inventory of Perseus Cores]{Inventories of Rich Carbon-Chain Chemistry in Prestellar and Starless Cores in the Perseus Molecular Cloud}

\author[A. Pokorny-Yadav et al.]{Anissa Pokorny-Yadav,$^{1,2}$\thanks{E-mail: \href{anissapy@berkeley.edu}{anissapy@berkeley.edu}}\thanks{Summer student at the National Radio Astronomy Observatory.}
Samantha Scibelli,$^{2}$\thanks{Jansky Fellow of the National Radio Astronomy Observatory.}
Judit Ferrer Asensio,$^{3}$
Yancy Shirley,$^{4}$
Andrés Megías,$^{5}$
\newauthor{ and Izaskun Jiménez-Serra$^{5}$}
\\
$^{1}$Department of Astronomy, University of California, Berkeley, University Avenue and Oxford St, Berkeley, CA 94720, USA\\
$^{2}$National Radio Astronomy Observatory, 520 Edgemont Rd, Charlottesville, VA 22903, USA\\
$^{3}$RIKEN Cluster for Pioneering Research, Wako-shi, Saitama, 351-0106, Japan\\
$^{4}$Steward Observatory, University of Arizona, 933 North Cherry Avenue, Tucson, AZ 85721, USA\\
$^{5}$Centro de Astrobiología (CAB), CSIC-INTA, Carretera de Ajalvir, km 4, E-28850 Torrejón de Ardoz, Spain}

\date{Accepted 2026 August 3. Received 2026 July 31; in original form 2026 March 2}

\pubyear{\the\year{}}

\begin{document}
\label{firstpage}
\pagerange{\pageref{firstpage}--\pageref{lastpage}}
\maketitle

\begin{abstract}
Carbon-chain molecules serve as an important reservoir of reactive organic matter that will eventually be incorporated into protoplanetary disks, planets, and cometary material. Prestellar and starless cores are composed of cold ($\sim$ 10 K) and dense ($\sim$ 10$^5$ cm$^{-3}$) clumps of gas and dust within molecular clouds, and are nurseries for low-mass stars and planetary systems. Surveys of starless cores have focused on the study of complex organic molecules, COMs, whereas observations of carbon-chains in starless cores are limited. We analyze the carbon-chain inventories of 15 prestellar and starless cores in the Perseus Molecular Cloud. Using Yebes 40m single-dish observations, we detect CS, CCS, CCCS, HC$_3$N, DC$_3$N, and HC$_5$N in at least 10/15 cores and HC$_7$N in 4/15 cores. Our study also finds related isotopologues, where $^{13}$CS, C$^{34}$S, C$^{13}$CS, CC$^{34}$S, H$^{13}$CCCN, HC$^{13}$CCN, HCC$^{13}$CN, HC$^{13}$CCCCN, HCC$^{13}$CCCN, HCCC$^{13}$CCN, HCCCCC$^{15}$N, and DCCCCCN are detected. We report detection statistics, compare column density ratios with Taurus, Serpens, and protostar sources, examine DC$_3$N/HC$_3$N deuterium fractionation, and investigate the relative abundances and correlations between cyanopolyyne (HC$_n$N) and sulfur-bearing (C$_n$S) carbon-chains. The diverse suite of species detected reveals the richness of carbon-chain chemistry in Perseus and illustrates how local environmental conditions, such as density, temperature, and proximity to protostellar activity, shape each core's molecular inventory and relative evolutionary phase. Our findings provide a glimpse into the carbon-chain reservoir of starless and prestellar cores in Perseus, which may ultimately be inherited by emerging protoplanetary disks and later integrated into planetary systems and biologically relevant material.
\end{abstract} 




\section{Introduction} \label{sec:intro}

Detecting the presence of organic molecules in the interstellar medium (ISM) enhances our understanding of the origins and evolution of organic chemistry, which is fundamental to the basis of life. In the ISM, a complex organic molecule (COM) has at least one carbon atom in a structure of at least six total atoms \citep{Herbst2009, 2023ASPC..534..379C}. Astrochemists and astrobiologists are interested in the prevalence of COMs, especially in low-mass star and planet formation regions, due to their connection to the prebiotic chemistry found on Earth.

Carbon-chain molecules are organic compounds consisting of predominantly linear and occasionally branched arrangements of carbon atoms linked by covalent bonds, often forming the backbone structure of various COMs. Cyclic arrangements of carbon atoms are typically classified separately as closed-ring structures. Understanding carbon-chain chemistry is essential for constraining pathways that build up cyclic ring structures, as these can serve as precursors to aromatic species that play a key role in building molecular complexity in both the ISM and planetary environments. The detection of cyanobenzene (c-C$_6$H$_5$CN) in TMC-1 marked the first unambiguous identification of a simple aromatic ring in space, highlighting the chemical bridge between carbon-chain growth and ring formation \citep{2018Sci...359..202M, McCarthy2021}. Studying the carbon-chain inventories of interstellar regions also allows us to learn about the physical and chemical conditions and evolutionary track of star formation \citep{Taniguchi2024}. Carbon-chain molecules found in molecular clouds act as a reservoir of reactive organic material that will eventually be incorporated into the formation of protoplanetary disks and planets \citep{Law2018}. Some carbon-chain species are thought to be eventually seeded into biologically relevant species, such as amino acids, RNA, and DNA. For instance, cyanoacetylene, HC$_3$N, has been suggested to be a possible precursor for the DNA building blocks cytosine (C$_4$H$_5$N$_3$O), thymine (C$_5$H$_6$N$_2$O$_2$), and uracil (C$_4$H$_4$N$_2$O$_2$; \citealt{Choe2021}).

Clumps of cold ($\sim$ 10 K) and dense ($\sim$ 10$^5$ cm$^{-3}$) gas and dust are used to probe the initial chemical and physical conditions at the earliest stage of star formation \citep{Bergin2007}. Molecular clouds, the nurseries of low- and high-mass stars, are hosts to these embryonic clumps of gas and dust known as starless and prestellar cores. The term starless implies that the cores are composed of dense and cold molecular gas with no evidence of an embedded protostar \citep{2007prpl.conf...17D, 2014prpl.conf...27A}. Prestellar cores constitute a subset of these objects that have overcome turbulence, thermal, and magnetic pressure, and are therefore gravitationally bound and expected to undergo collapse to form protostars \citep{2000prpl.conf...59A,WardThompson2007,2023ASPC..534..233P}.

The conditions in prestellar and starless cores, characterized by low temperatures ($<$ 10 K), ionization rates lower than the diffuse ISM field, and the presence of reactive ions, enable efficient ion-molecule chemistry \citep{1973ApJ...185..505H,Herbst1989, Ohishi1998}. At the same time, barrierless neutral-neutral reactions, particularly those involving radicals, can contribute to the effective formation and maintenance of carbon-chain molecule inventories \citep{2014MNRAS.437..930L, Redaelli2025}. As cold cores evolve, carbon is sequestered into CO as it freezes onto dust grains, limiting the available gaseous carbon. However, cosmic-ray-induced He$^+$ destroys CO and releases ionic carbon C$^+$ back into the gas phase, recycling carbon throughout star formation and allowing for various carbon-chain formation pathways. While chemical models predict that carbon-chain molecules produced in cold, dark clouds are destroyed through reactions with He$^+$, H$^+$, and H$_3$$^+$ \citep{Suzuki1992} or depletion onto dust grains at even faster rates \citep{Hassel2008}, their formation and prevalence continue in lukewarm gas in later protostellar stages via Warm Carbon-Chain Chemistry (WCCC; \citealt{Sakai2013}). Therefore, it is imperative to understand their carbon-chain inventories prior to gravitational collapse to address questions about chemical inheritance and carbon-chain chemistry in the cold gas phase.

COMs and complex carbon-chains have been primarily detected and studied in the Taurus Molecular Cloud, mainly due to the rich history of chemistry in the well-known dark cloud TMC-1 (e.g., \citealt{2025ApJS..281....9X}). It is important to mention that carbon-chains were first detected in TMC-1 in the form of cyanopolyynes (HC$_n$N species), specifically HC$_3$N \citep{1971ApJ...163L..35T, Little1977, Broten1978, Kroto1978}. Cyanopolyynes are a fascinating class of carbon-chain molecules, characterized by a hydrogen atom, a long chain of carbon atoms, and a cyano (CN) group at the end. They have been detected throughout all stages of low-mass star formation, making them an interesting species for studying chemical inheritance into planetary systems. Cyanopolyynes' cyano group, -CN, is known to spatially map to larger ringed molecules, such as c-C$_6$H$_5$CN \citep{2023A&A...674L...4C}, emphasizing their crucial role in the formation of larger, potentially prebiotic species. Sulfur-bearing carbon-chains, CCS and CCCS, were also first identified in the ISM towards TMC-1, which contains the strongest observed lines for these species \citep{1987ApJ...317L.111K, 1987ApJ...317L.115S, 1987ApJ...317L.119Y}. In addition to cyanopolyynes, sulfur-bearing species are of particular interest in the study of prebiotic chemistry because sulfur is a key element in many biological processes on Earth. Studying their chemistry provides crucial constraints on the long-standing ‘missing sulfur’ problem in the astrochemistry community. \cite{Laas2019} predicted that a significant fraction of sulfur may be locked in organo-sulfur species. Therefore, the study of sulfur-bearing molecules, such as CCS and CCCS, is important to constrain how much sulfur is locked into this type of carbon-bearing species.

The Perseus Molecular Cloud is another well-studied active and nearby ($\sim$ 300 pc) star-forming region. Based on infrared observations of embedded young stars and their ammonia, NH$_3$, survey, \cite{Ladd1994} found that Perseus hosts cores with properties that fall between the less massive, more quiescent cores of Taurus and the more massive, highly turbulent cores of Orion A. This intermediate nature makes Perseus a particularly valuable site for investigating the chemical diversity and evolution of low-mass star-forming regions. Perseus is also home to several protostars with CO outflow jets, especially in the well-studied stellar clusters NGC 1333 and IC 348 \citep{1999A&A...343..571G,2013ApJ...774...22P}. Recent studies of Perseus have focused on COM and carbon-chain content in protostellar envelopes \citep{Law2018, Belloche2020}, warm-phase complex chemistry in embedded protostars \citep{Yang2021}, and COM content in starless and prestellar cores \citep{Scibelli2024}.

In \cite{Scibelli2024}, the authors found a prevalence of the COM acetaldehyde, CH$_3$CHO, in 15/35 starless and prestellar cores targeted with the Arizona Radio Observatory (ARO) 12m telescope. These 15 sources were then followed up with Yebes 40m observations, which detected even higher-complexity molecules such as methyl formate, HCOOCH$_3$, and dimethyl ether, CH$_3$OCH$_3$, in at least 20\% of the sample.
The 15 chemically-rich cores in this sample offer an exciting opportunity to compile the chemical inventories and evolutionary histories of carbon-chain molecules in young, evolving prestellar and starless cores.

We begin by detailing the observations and molecular line detection pipeline in Section \ref{sec:2}. In Section \ref{sec:3}, we present the detection statistics and column density calculations. In Section \ref{sec:4}, we investigate fractional abundances, compare column density ratios with other prestellar cores, protostars, and pre-cluster clumps in Perseus, Taurus, and Serpens, analyze isotopic and deuterium fractionation, and discuss the chemistry of sulfur-bearing and cyanopolyyne carbon-chain families. Finally, we summarize and conclude in Section \ref{sec:5}.

\section{Methods} \label{sec:2}
\begin{figure*}
    \centering
    \includegraphics[width=16cm]{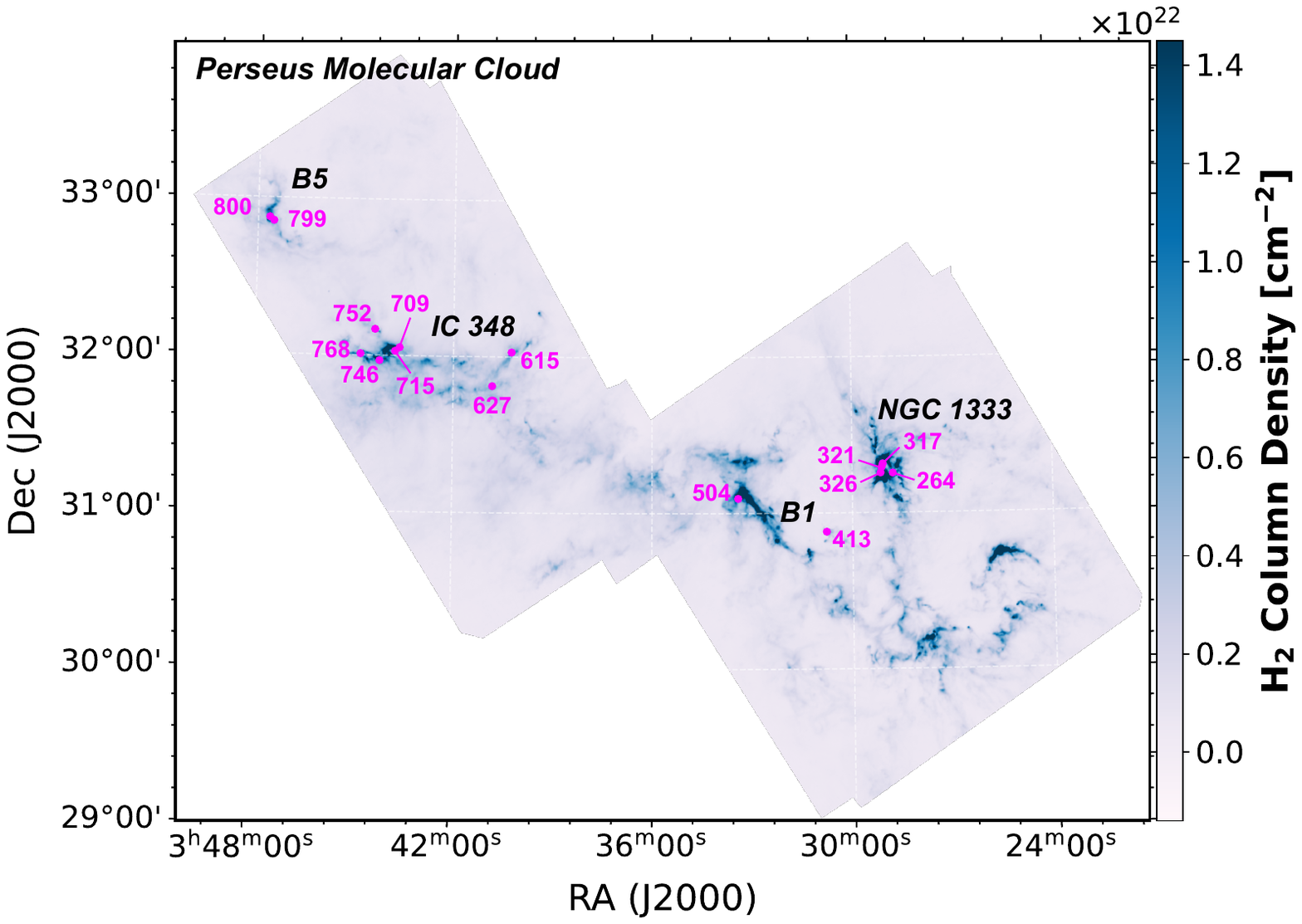}
    \caption{\textit{Herschel} H$_2$ column density map \citep{2012A&A...540A..10S} of Perseus starless and prestellar cores. Labeled in magenta are the locations and names of the 15 starless and prestellar cores analyzed. The star-forming regions within Perseus are denoted in black.}
    \label{cdmap}
\end{figure*}

\subsection{Yebes 40m Observations}

During the spring of 2022 and 2023, observations with the Yebes 40m were performed to investigate 15 starless and prestellar cores in the Perseus Molecular Cloud (projects 22A022 and 23A025; PI: Scibelli). These 15 cores were selected as a subset of a larger 35-core sample, created from cross-referencing ammonia, NH$_3$, observations \citep{2008ApJS..175..509R}, Bolocam 1.1 m continuum maps from the Caltech Submillimeter Observatory \citep{2008ApJ...684.1240E}, and a higher resolution \textit{Herschel} core catalog \citep{2021A&A...645A..55P}. All 35 cores were observed with the ARO 12m, and those `bright' in CH$_3$CHO (detected at $\sigma_{T_{\rm mb}}$ = 6 mK) were then chosen for the Yebes 40m follow-up (see \citealt{Scibelli2024}). Each of the 15 cores was observed with the dual (horizontal and vertical) linear polarization Q-band receiver \citep{2021A&A...645A..37T} using the frequency switching technique with a standard throw of 10.52 MHz. The Q-band receiver allowed for a bandwidth spanning 31.5 -- 50.0 GHz (6 -- 9 mm) with a resolution of 38.0 kHz (0.38 -- 0.23 km s$^{-1}$). For the molecular transitions selected in the study, the beam size ($\theta_\mathrm{beam}$) of the Yebes 40m range from 37$\arcsec$--57$\arcsec$, which is 1.5 -- 2.6 larger than the source sizes of the prestellar/starless cores in this sample ($\theta_\mathrm{src} =$ 22$\arcsec$--25$\arcsec$). The source sizes of these cores were determined by \cite{Scibelli2024}, where the authors first assumed a physical size of 7020 au and calculated a range of angular source sizes based on the specific distances of each core in the Perseus Molecular Cloud (283 -- 325 pc).

The Yebes 40m data were then inspected and reduced using Python-based scripts\footnote{\url{https://github.com/andresmegias/gildas-class-pipeline/}} developed by \cite{2023MNRAS.519.1601M}. To combine spectra, the scripts employ the CLASS program of the GILDAS package \citep{2005sf2a.conf..721P, 2013ascl.soft05010G}. We adopt a 10\% calibration uncertainty for these observations \citep{Cernicharo2020}. See details of \cite{Scibelli2024} for further explanations on the Yebes 40m observations and data reduction process.

Due to the large bandwidth (18.5 GHz) of the Yebes 40m observations, ranging from 31.5 -- 50.0 GHz (6 -- 9 mm), there is a wealth of additional lines that were not looked at by \cite{Scibelli2024}. We therefore conducted a deeper investigation into the carbon-chain chemistry of these 15 Perseus prestellar/starless cores. 

\subsection{Data Pipeline} \label{sec:2.2}

Using Python-based scripts and the Pyspeckit package\footnote{\url{https://pyspeckit.readthedocs.io/en/latest/}} \citep{2011ascl.soft09001G, Ginsburg2022}, we built a molecular detection pipeline to search for bright lines in the 15 starless and prestellar cores for which we had data.
We separated the data into 8 bands with $\sim$2.5 GHz wide intervals and calculated distinct noise (RMS) levels for each band. The pipeline extracts lines with intensities $>$ 3$\sigma$ RMS of the main beam temperature, \textit{$T_{\rm mb}$}, to be considered for potential detection. The RMS ($\sigma$) values range from 2.3 -- 7.1 mK (see Tables \ref{info1}\,and \ref{info2} for specific RMS values). The online tool \textsc{Splatalogue}\footnote{\url{https://splatalogue.online}} was used to compile a list of molecule transitions within the frequency range of 31.5 -- 50.0 GHz. The molecular data originates from the Jet Propulsion Laboratory Millimeter and Submillimeter Spectral Line catalog \citep[JPL;][]{PICKETT1998}. References for laboratory spectroscopic data are listed in Appendix \ref{ap:E}. The selection criteria restricted transitions to be energetically favorable for cold starless cores, where transitions have upper energies of \textit{E$_u$} $<$ 30 K. 

\begin{figure*} 
    \centering
    \includegraphics[width=\textwidth]{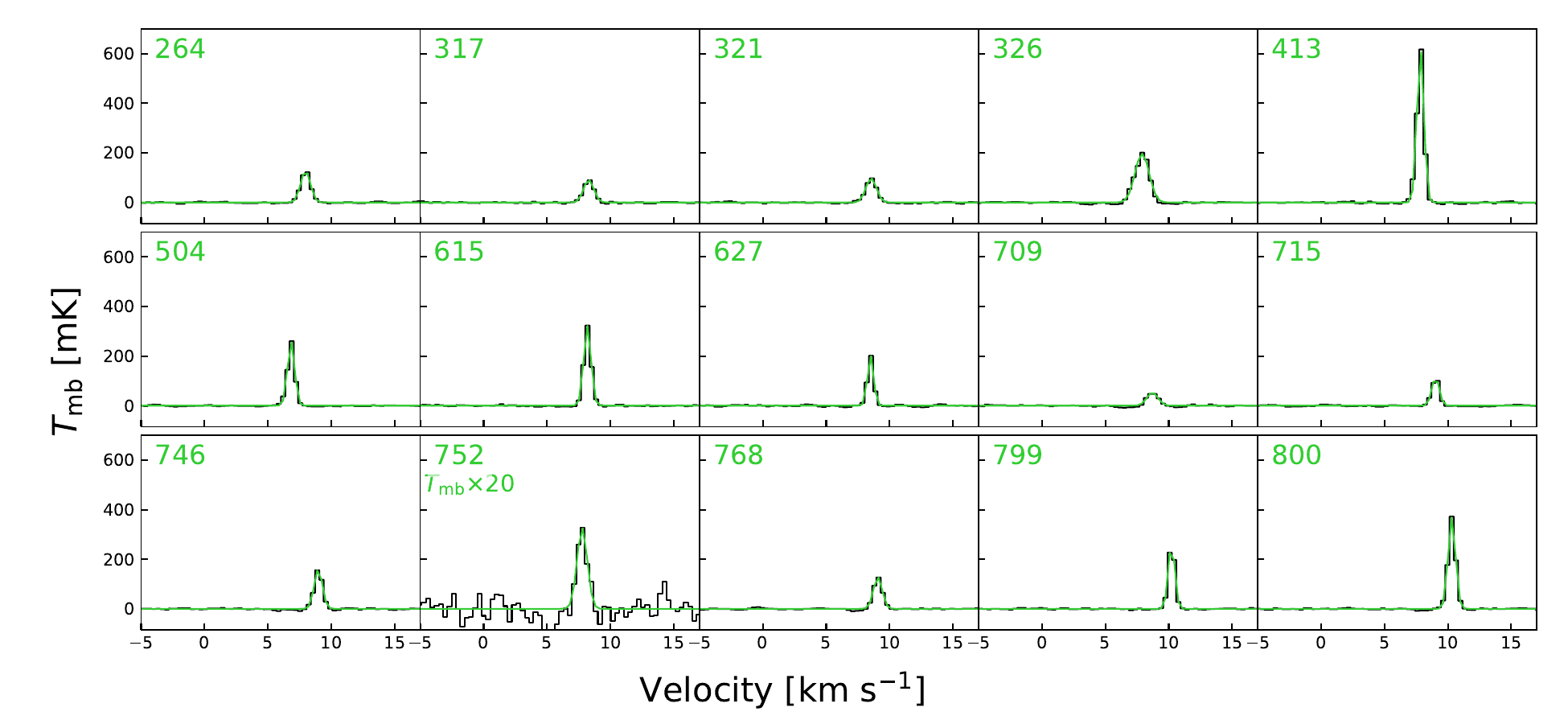}
    \caption{Diagnostic plots of CCS (N=2-1, J=3-2). The Gaussian fit detected in all 15 cores produced by our data pipeline, using Pyspeckit \citep{2011ascl.soft09001G, Ginsburg2022}, is shown in lime green. The intensity of core 752 has been magnified by 20$\times$ for visual and comparison purposes.} \label{CCS_diagnostic}
\end{figure*}

\begin{figure*}
    \centering
    \includegraphics[width=12cm]{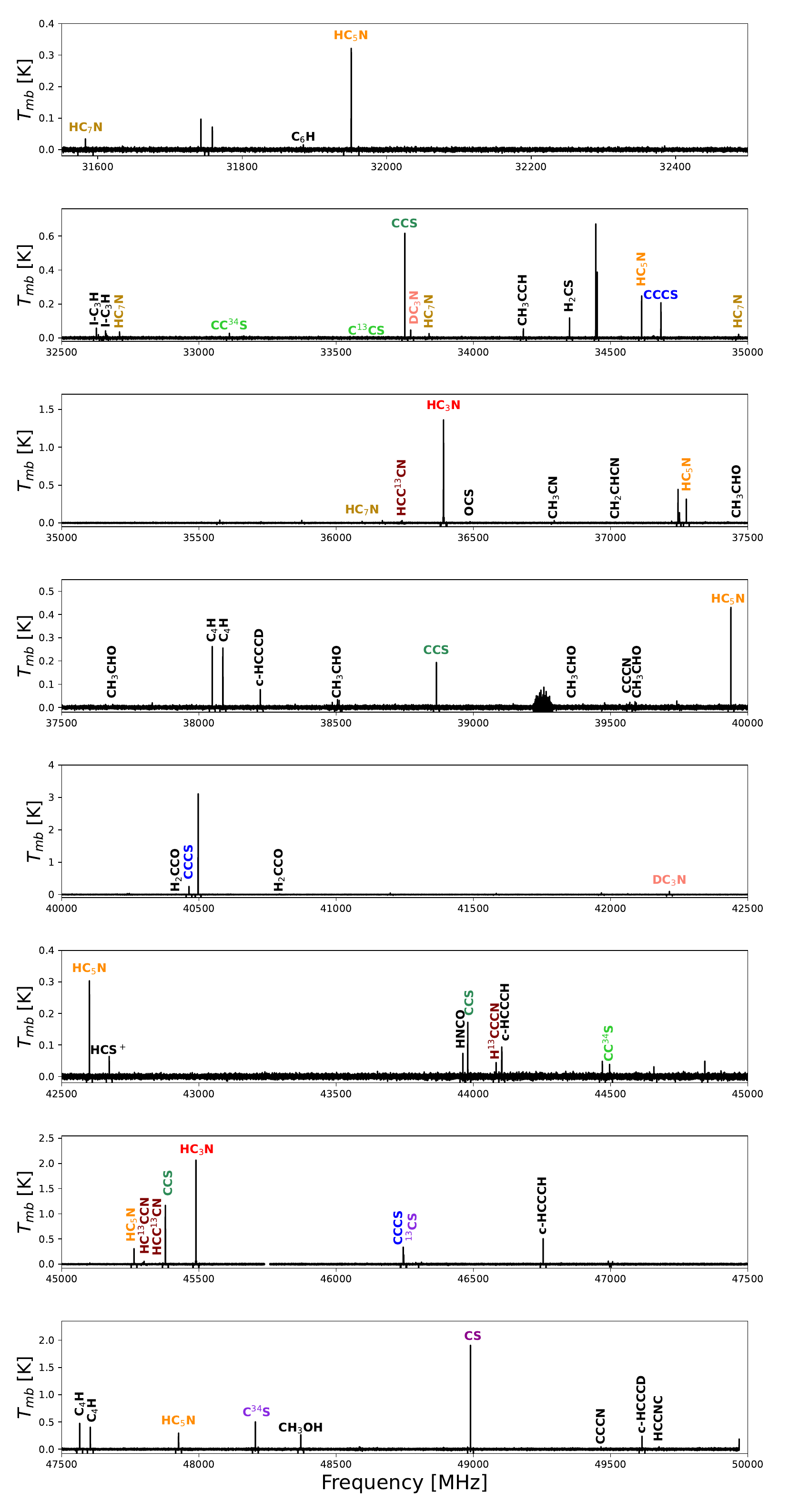}
    \caption{Yebes 40m spectrum of core 413 with carbon-chain detections focused on in this study labeled in color, and the remaining line detections denoted in black. Other bright lines seen in this data are not attributed to any energetically favorable molecular transitions.}
    \label{spectrum}
\end{figure*}

The $>$ 3$\sigma$ lines are matched with the selected \textsc{Splatalogue} transitions within 1.5 MHz to account for velocity shifts between the source and calibration process. The pipeline then performs Gaussian fits for all the detected lines and returns fit parameters that include the amplitude (peak intensity), full width half maximum (FWHM), and line-of-sight velocity ($v_{\rm lsr}$). In order to constrain the detection sample size and filter out false detections, the pipeline only registers (1) line widths larger than roughly one channel, $\Delta v$ $>$ 0.28 km s$^{-1}$, (2) positive intensities, $T_{\rm mb}$ $>$ 0, and (3) velocity shifts within the known source velocity. Since the 15 prestellar/starless cores had a broad range of $v_{\rm lsr}$ values, 6.0 $\leq v \leq$ 11.0 km s$^{-1}$, the velocity window was shifted for each core individually. For example, for core 413 we find a $v_{\rm lsr} =$  7.7 km s$^{-1}$ based on the CH$_3$OH observations, so its velocity window spans from 6.0 $\leq v \leq$ 9.0 km s$^{-1}$ to ensure we inspected all relevant possible detections. Typical line widths in Perseus cores exceed 0.3 km s$^{-1}$, broader than those observed in the Taurus Molecular Cloud ($\sim$ 0.1 -- 0.2 km s$^{-1}$). We therefore adopt a minimum line width threshold of approximately one spectral channel, consistent with our spectral resolution limitations, to reduce the likelihood of claiming false detections. 

After undergoing the filtering process, the remaining detections' Gaussian fit diagnostic plots, fit parameters, and corresponding line information from \textsc{Splatalogue} are organized into an exportable table. Figure \ref{CCS_diagnostic} displays an example of the diagnostic plots that the pipeline produces for each fit. Given that the table includes detections for all molecular species, we then extracted carbon-chain molecule transitions to analyze further. 

Despite the relatively broad velocity window, $\Delta v \sim$ 3.0 km s$^{-1}$, allowed for each core in the sample, there were a few instances in which the pipeline failed to recognize a molecular detection due to a slightly larger offset ($\sim$1.65 MHz or $\sim$$\pm$0.05 km s$^{-1}$) between the rest frequency and the frequency of the line. For example, in core 615, we detected another HC$_5$N transition at 42602.171 MHz from this discrepancy. For these scenarios, the lines were fitted individually and added to the detection list. Using the same observations for surveying COMs, the authors of \cite{Scibelli2024} acknowledge that Yebes 40m beam size ($\theta_\mathrm{beam} =$ 37$\arcsec$--57$\arcsec$) is larger than the size of the core ($\theta_\mathrm{src} =$ 22$\arcsec$--25$\arcsec$), and therefore we cannot pinpoint where each species' emission peaks spatially. Thus, small offsets in $v_{\rm lsr}$ are not significant and could instead indicate different emission peaks within the core.

In most cases, we did not have an issue with line blending due to overlapping molecular line transitions, as starless and prestellar cores are generally line-poor compared to other hotter sources. For cases in which multiple transitions were deemed a `good fit', the transition with the closest velocity shift to that of the methanol, CH$_3$OH, line was selected to claim the detection. 

\section{Results} \label{sec:3}

\subsection{Detection Statistics}
We focused on two carbon-chain families: sulfur-bearing chains (C$_n$S) and cyanopolyynes (HC$_n$N). We included CS as well as isotopologues of both families in our analysis to investigate carbon chemistry and deuterated species more closely. All of the carbon-chain spectroscopic parameters for the species of interest, including quantum numbers, rest frequencies, and upper energies, as well as the molecular detections, are summarized in Table \ref{detections}. The complete spectrum towards one of our sources, core 413, with carbon-chain classifications, is shown in Figure \ref{spectrum}. 

A molecule was considered detected if we observed (1) at least two lines with 3$\sigma$ detections or (2) one line with a 5$\sigma$ detection, and (3) the upper energies were consistent with expectations for cold core chemistry detections. In cases where only one transition is available (H$^{13}$CCCCCN, HCCC$^{13}$CCN, HCCCCC$^{15}$N, DCCCCCN), we treat a single 3$\sigma$ line as a tentative detection. Tentative detections are denoted with an asterisk `*' in Table \ref{detections}. As mentioned in Section \ref{sec:2.2}, no competing line identifications from potential line blending were found, so a single line is sufficient to claim a molecule detection for species with only one available transition. For instance, since CS is detected at $>$ 5$\sigma$ and only one transition is available for $^{13}$CS and C$^{34}$S, we claim detections of CS and its isotopologues. All of the detections listed in Table \ref{detections} have upper excitation energies $<$ 30 K, which is similar and consistent with the expected temperatures of starless and prestellar cores of $<$15 K. 

As for the main carbon-chain species, CCS, CCCS, HC$_3$N, HC$_5$N, and HC$_7$N were detected, respectively, in 15/15 (100$\%$), 14/15 (93.3$\%$), 15/15 (100$\%$), 14/15 (93.3$\%$), and 4/15 (26.7$\%$) of the Perseus cores (see Table \ref{detections}). Related molecules within the HC$_n$N and C$_n$S families, such as $^{13}$CS, C$^{34}$S, CS, C$^{13}$CS, CC$^{34}$S, H$^{13}$CCCN, HC$^{13}$CCN, HCC$^{13}$CN, HC$^{13}$CCCCN, HCC$^{13}$CCCN, HCCC$^{13}$CCN, HCCCCC$^{15}$N, and DCCCCCN were detected, respectively, in 15/15 (100\%), 15/15 (100\%), 15/15 (100\%), 4/15 (26.7\%), 7/15 (46.7\%), 

\begin{landscape}
\begin{table}
\centering
\caption{Detection Line Parameter List. Tentative detections are denoted by '\checkmark *'. We present detections for two carbon-chain families and their isotopologues: C$_n$S and HC$_n$N.} 
\label{detections}
\begin{tabular}{lclr|lllllllllllllll}
\hline
\multicolumn{4}{c|}{\textbf{Line Parameters}} & \multicolumn{15}{c}{\textbf{Perseus Core \#}} \\ \hline
\textbf{Species} & \textbf{Transition} & \textbf{$\boldsymbol{\nu}$ {{[MHz]}}} & \textbf{$E_\textbf{u}$ {{[K]}}} & \textbf{264} & \textbf{317} & \textbf{321} & \textbf{326} & \textbf{413} & \textbf{504} & \textbf{615} & \textbf{627} & \textbf{709} & \textbf{715} & \textbf{746} & \textbf{752} & \textbf{768} & \textbf{799} & \textbf{800} \\ \hline
$^{13}$CS & J=1-0 & 46247.567 & 2.220 & \checkmark & \checkmark & \checkmark & \checkmark & \checkmark & \checkmark & \checkmark & \checkmark & \checkmark & \checkmark & \checkmark & \checkmark & \checkmark & \checkmark & \checkmark \\
C$^{34}$S & J=1-0 & 48206.915 & 2.314 & \checkmark & \checkmark & \checkmark & \checkmark & \checkmark & \checkmark & \checkmark & \checkmark & \checkmark & \checkmark & \checkmark & \checkmark & \checkmark & \checkmark & \checkmark \\
CS & J=1-0 & 48990.978 & 2.351 & \checkmark & \checkmark & \checkmark & \checkmark & \checkmark & \checkmark & \checkmark & \checkmark & \checkmark & \checkmark & \checkmark & \checkmark & \checkmark & \checkmark & \checkmark \\
CCS & N=2-1, J=3-2 & 33751.374 & 3.226 & \checkmark & \checkmark & \checkmark & \checkmark & \checkmark & \checkmark & \checkmark & \checkmark & \checkmark & \checkmark & \checkmark & \checkmark & \checkmark & \checkmark & \checkmark \\
 & N=3-2, J=3-2 & 38866.423 & 12.437 & \checkmark & \checkmark & \checkmark & \checkmark & \checkmark & \checkmark & \checkmark & \checkmark &  & \checkmark & \checkmark &  & \checkmark & \checkmark & \checkmark \\
 & N=4-3, J=3-2 & 43981.027 & 12.942 &  & \checkmark & \checkmark & \checkmark & \checkmark & \checkmark & \checkmark & \checkmark & \checkmark & \checkmark & \checkmark &  & \checkmark & \checkmark & \checkmark \\
 & N=3-2, J=4-3 & 45379.033 & 5.404 & \checkmark & \checkmark & \checkmark & \checkmark & \checkmark & \checkmark & \checkmark & \checkmark & \checkmark & \checkmark & \checkmark & \checkmark & \checkmark & \checkmark & \checkmark \\
C$^{13}$CS & N=2-1, J=3-2, F=7/2-5/2 & 33613.339 & 3.213 &  &  &  &  & \checkmark * &  &  &  &  &  &  &  &  &  &  \\
 & N=3-2, J=3-2, F=7/2-5/2 & 38682.482 & 12.426 &  &  &  &  &  &  &  &  &  &  &  &  &  &  & \checkmark * \\
 & N=4-3, J=3-2, F=5/2-3/2 & 43746.043 & 4.604 &  &  &  &  &  & \checkmark * &  &  &  &  &  &  & \checkmark &  &  \\
CC$^{34}$S & N=2-1, J=3-2 & 33111.839 & 3.166 &  &  &  &  & \checkmark & \checkmark &  & \checkmark * &  &  &  &  &  & \checkmark * &  \\
 & N=3-2, J=4-3 & 44497.599 & 5.301 &  &  &  &  & \checkmark & \checkmark & \checkmark * &  &  &  &  &  & \checkmark * &  & \checkmark \\
CCCS & J=6-5 & 34684.368 & 5.826 & \checkmark &  & \checkmark & \checkmark & \checkmark & \checkmark & \checkmark & \checkmark & \checkmark & \checkmark & \checkmark &  & \checkmark & \checkmark & \checkmark \\
 & J=7-6 & 40465.014 & 7.768 & \checkmark &  & \checkmark & \checkmark & \checkmark & \checkmark & \checkmark & \checkmark & \checkmark & \checkmark & \checkmark &  & \checkmark & \checkmark & \checkmark \\
 & J=8-7 & 46245.623 & 9.987 & \checkmark & \checkmark & \checkmark & \checkmark & \checkmark & \checkmark & \checkmark & \checkmark &  & \checkmark & \checkmark &  & \checkmark & \checkmark & \checkmark \\ \hline
HC$_3$N \textit{v}=0 & J=4-3, F=3-2 & 36392.326 & 4.366 & \checkmark & \checkmark & \checkmark & \checkmark & \checkmark & \checkmark & \checkmark & \checkmark & \checkmark & \checkmark & \checkmark & \checkmark * & \checkmark & \checkmark & \checkmark \\
 & J=5-4, F=5-4 & 45490.310 & 6.550 & \checkmark & \checkmark & \checkmark & \checkmark & \checkmark & \checkmark & \checkmark & \checkmark & \checkmark & \checkmark & \checkmark &  & \checkmark & \checkmark & \checkmark \\
DC$_3$N \textit{v}=0 & J=4-3, F=3-2 & 33772.541 & 4.052 & \checkmark &  &  & \checkmark & \checkmark & \checkmark & \checkmark & \checkmark & \checkmark & \checkmark & \checkmark &  &  & \checkmark & \checkmark \\
 & J=5-4, F=4-3 & 42215.595 & 6.078 & \checkmark & \checkmark * & \checkmark & \checkmark & \checkmark & \checkmark & \checkmark & \checkmark & \checkmark & \checkmark & \checkmark &  & \checkmark * & \checkmark & \checkmark \\
H$^{13}$CCCN \textit{v}=0 & J=4-3, F=3-2 & 35267.403 & 4.231 &  &  &  &  & \checkmark & \checkmark &  &  &  &  &  &  &  &  &  \\
 & J=5-4, F=4-3 & 44084.171 & 6.347 &  &  &  &  & \checkmark & \checkmark &  &  &  &  &  &  &  &  & \checkmark * \\
HC$^{13}$CCN \textit{v}=0 & J= 4-3, F= 3-2 & 36237.945 & 4.348 &  &  &  &  & \checkmark & \checkmark &  &  &  &  &  &  &  &  & \checkmark \\
 & J=5-4, F=4-3 & 45297.347 & 6.522 &  &  &  & \checkmark * & \checkmark & \checkmark &  &  &  &  &  &  &  &  & \checkmark \\
HCC$^{13}$CN \textit{v}=0 & J=4-3, F=3-2 & 36241.435 & 4.348 & \checkmark &  &  &  & \checkmark & \checkmark &  &  &  &  &  &  &  &  & \checkmark \\
 & J=5-4, F=4-3 & 45301.711 & 6.522 & \checkmark &  &  & \checkmark * & \checkmark & \checkmark &  &  &  &  &  &  & \checkmark * &  & \checkmark \\
HC$_5$N \textit{v}=0 & J=12-11 & 31951.772 & 9.967 & \checkmark &  &  & \checkmark & \checkmark & \checkmark & \checkmark & \checkmark &  & \checkmark * & \checkmark &  & \checkmark & \checkmark & \checkmark \\
 & J=13-12 & 34614.387 & 11.629 & \checkmark &  & \checkmark & \checkmark & \checkmark & \checkmark & \checkmark & \checkmark & \checkmark * &  & \checkmark &  & \checkmark & \checkmark & \checkmark \\
 & J=14-13 & 37276.994 & 13.418 & \checkmark & \checkmark & \checkmark & \checkmark & \checkmark & \checkmark & \checkmark & \checkmark &  &  & \checkmark &  & \checkmark & \checkmark & \checkmark \\
 & J=15-14 & 39939.591 & 15.334 & \checkmark & \checkmark & \checkmark & \checkmark & \checkmark & \checkmark & \checkmark & \checkmark &  &  & \checkmark &  & \checkmark & \checkmark & \checkmark \\
 & J=16-15 & 42602.171 & 17.379 & \checkmark &  & \checkmark & \checkmark & \checkmark & \checkmark & \checkmark * & \checkmark & \checkmark  & \checkmark & \checkmark &  & \checkmark & \checkmark & \checkmark \\
 & J=17-16 & 45264.745 & 19.551 & \checkmark &  &  & \checkmark & \checkmark & \checkmark & \checkmark & \checkmark & \checkmark & \checkmark & \checkmark &  & \checkmark & \checkmark & \checkmark \\
 & J=18-17 & 47927.306 & 21.851 & \checkmark &  &  & \checkmark & \checkmark & \checkmark &  & \checkmark &  &  & \checkmark &  & \checkmark & \checkmark * & \checkmark \\
HC$^{13}$CCCCN & 12-11 & 31624.340 & 9.865 &  &  &  & \checkmark * &  &  &  &  &  &  &  &  &  &  &  \\
HCC$^{13}$CCCN & 14-13 & 37238.390 & 13.404 &  &  &  &  & \checkmark * &  &  &  &  &  &  &  & \checkmark * &  &  \\
 & 18-17 & 47877.653 & 21.829 &  &  &  &  &  & \checkmark * &  &  &  &  &  &  &  &  &  \\
HCCC$^{13}$CCN & 15-14 & 39903.082 & 15.320 &  &  &  &  & \checkmark * &  &  &  &  &  &  &  &  &  &  \\
HCCCCC$^{15}$N & 19-18 & 49347.584 & 23.683 &  &  &  &  &  &  & \checkmark * &  &  &  &  &  &  &  &  \\
DCCCCCN & 14-13 & 35589.324 & 12.810 &  &  &  &  & \checkmark * &  &  &  &  &  &  &  &  &  &  \\
HC$_7$N \textit{v}=0 & J=28-27 & 31583.704 & 21.979 & \checkmark &  &  &  & \checkmark &  &  &  &  &  &  &  &  &  & \checkmark \\
 & J=29-28 & 32711.666 & 23.549 &  &  &  &  & \checkmark &  &  &  &  &  &  &  &  &  & \checkmark \\
 & J=30-29 & 33839.628 & 25.173 & \checkmark &  &  &  & \checkmark &  &  &  &  &  &  &  &  &  &  \\
 & J=31-30 & 34967.585 & 26.851 & \checkmark &  &  &  & \checkmark &  &  &  &  &  &  &  &  &  &  \\
 & J=32-31 & 36095.541 & 28.583 & \checkmark &  &  &  & \checkmark &  &  &  &  &  &  &  & \checkmark * &  &  \\ \hline
\end{tabular}
\end{table}
\end{landscape}
\noindent
3/15 (20\%), 4/15 (26.7\%), 6/15 (40\%), 1/15 (6.7\%), 3/15 (20\%), 1/15 (6.7\%),  1/15 (6.7\%), and 1/15 (6.7\%) of the sample (see Table \ref{detections}).

These 15 Perseus cores contain a great diversity of detected carbon-chain species, with several isotopologues detected as well. However, core 413 clearly emerges as the most carbon-chain-rich prestellar/starless core in Perseus. Core 413 had overall, from the full molecular line identification pipeline, 32 unique molecular detections with 10 additional tentative unique detections. Every transition for each of the carbon-chain species we were concerned with in this study was detected in core 413, along with 16/22 isotopologue detections (tentative species included). Core 413 is situated in the B1 region of Perseus, where it appears to be relatively isolated from any protostellar outflows (see Figure \ref{cdmap}). This core's abundant reservoir of carbon-chains makes it a fascinating prestellar/starless core to be studied further. In comparison, core 752, located in the IC 348 region, had 14 total unique molecular detections and 9 tentative unique detections (Table \ref{detections}). This scarcity of carbon-chain species stands in sharp contrast to the rich inventory seen in core 413, which is the only core in \cite{Scibelli2024} with the most robust detections of CH$_2$CHCN. These results highlight that local environmental factors likely play a big role in the chemistry at this stage of star and planet formation. 

\subsection{Column Densities} \label{sec:3.2}
In order to quantify the line detections observed in molecular spectra, line-of-sight (LOS) column densities must be calculated and further analyzed. The line-of-sight column density represents the number of respective molecules per unit area along the straight path between the telescope receiver and the source. Depending on the molecule and how many available transitions were detected within the observed frequency bandwidth, different methods of calculating column densities can be applied. For our analysis, we employ the rotation diagram (RD) and fixed temperature (FT) methods, in which both assume local thermodynamical equilibrium (LTE). In the FT method, a fixed excitation temperature, $T_{\rm ex}$, is assumed for all transitions of each molecular family. When using the FT method on sulfur-bearing molecules (C$_n$S), we adopt the $T_{\rm ex}$ calculated from the detected CCS transitions. Similarly, when applying the FT method on cyanopolyyne species (HC$_n$N), we use the $T_{\rm ex}$ calculated from HC$_3$N detections. Since the RD method is performed for both CCS and HC$_3$N in all cores (except for core 752 for HC$_3$N), these serve as `reference species' to constrain $T_{\rm ex}$ in cases where insufficient transitions are available for other molecules. Throughout our analysis, we assume radio emission to be optically thin ($\tau$ $\ll$ 1). Using the following equations, we calculate the column density of each molecule. The Planck function in energy density, $u_{\nu}$, and in temperature units, $J_{\nu}$, are defined in the respective Equations \ref{eq:1} and \ref{eq:2}, where $h$ is the Planck constant, $k$ is the Boltzmann constant, and $c$ is the speed of light:

\begin{equation} \label{eq:1}
    u_\nu \equiv \frac{8\pi h\nu^3}{c^3} \frac{1}{\exp(h\nu/kT) - 1}
\end{equation}

\begin{equation} \label{eq:2}
  J_\nu \equiv \frac{h\nu}{k} \frac{1}{\exp(h\nu/kT) - 1}.
\end{equation}

The upper state column density $N_u$ can be calculated, 
\begin{equation} \label{eq:3}
    N_u = \frac{I}{hA_{ul}f} \frac{u_{\nu}(T_{\rm ex})}{[J_{\nu}(T_{\rm ex}) - J_{\nu}(T_{\rm cmb})]},
\end{equation}
where $I$ is the integrated intensity of the line, $A_{ul}$ is the spontaneous emission coefficient for a molecule transitioning from an upper to lower state (or `Einstein A'), $f$ is the frequency-dependent filling factor, and $T_{\rm cmb}$ = 2.73 K is the cosmic microwave background (CMB) temperature. We must assume a standard filling factor of $f=1$, consistent with previous studies of COMs in cold cores \citep{2020ApJ...891...73S, Scibelli2024}, as the source sizes are not currently resolved by our observations. The total column density, $N_{\rm tot}$, of a molecule can be calculated as,

\begin{equation} \label{eq:4}
    \frac{N_u}{g_u} = \frac{N_{\rm tot}}{Q(T_{\rm ex})} \exp\left(\frac{-E_u}{kT_{\rm ex}}\right),
\end{equation}
where $g_u$ is the upper state degeneracy, $Q(T_{\rm ex})$ is the partition function dependent on the excitation temperature of the molecule, and $E_u$ is the upper state energy.

\begin{figure}
    \centering
    \includegraphics[width=8.5cm]{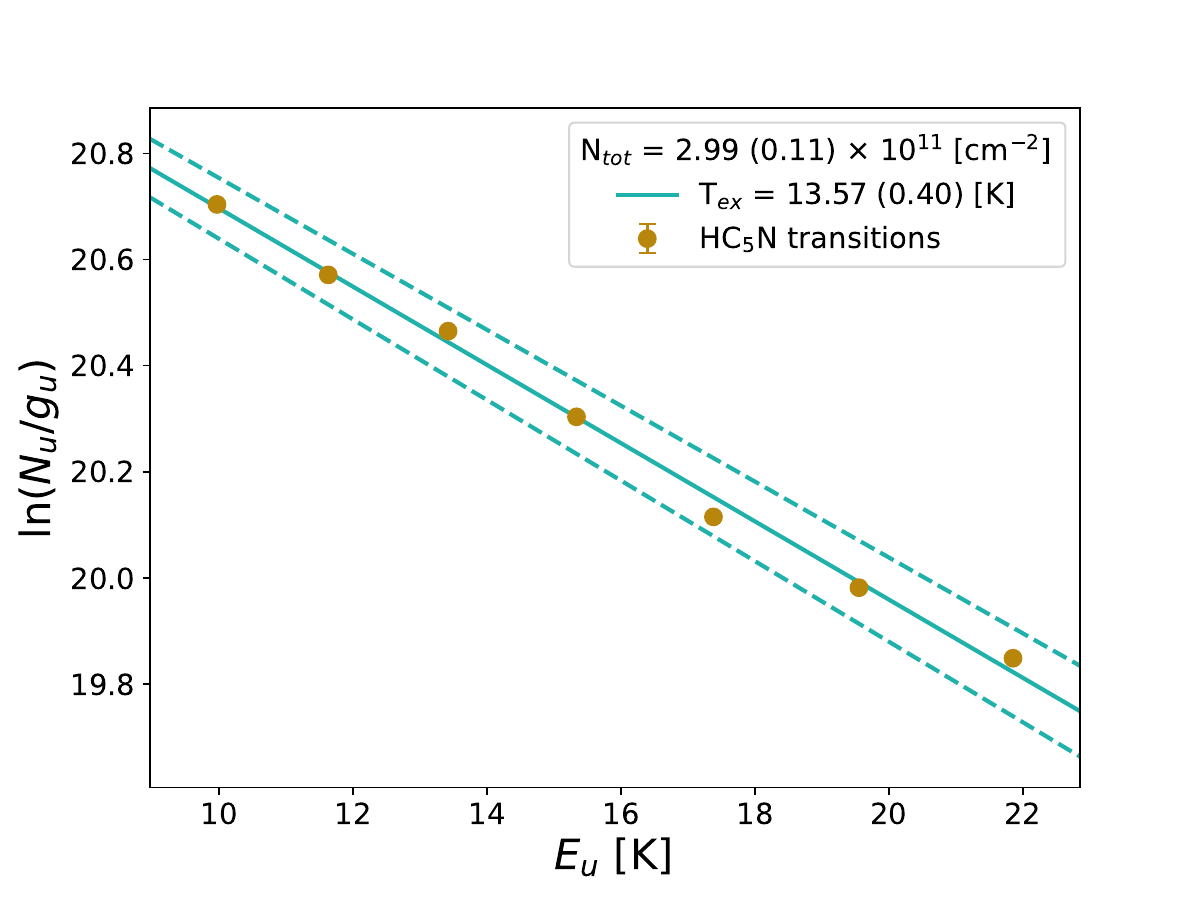}
    \caption{Rotation Diagram for HC$_5$N transitions in core 413. The 7 observed transitions of HC$_5$N are marked in gold, and the best-fit line with its 1$\sigma$ uncertainty range is indicated by solid and dashed turquoise lines, respectively. The excitation temperature T$_{\mathrm{ex}}$ is derived from the negative inverse slope of the best-fit line, as reported in the legend. Uncertainties of the data were taken into account to derive the model's parameter uncertainties. Errors are included in parentheses.}
    \label{rd}
\end{figure}

In the rotation diagram method, $E_u$ is plotted against the natural logarithm of the left-hand side of equation \ref{eq:4}, ${\rm ln}(N_u/g_u)$. Each point represents a detected transition for the molecule, and the negative inverse of the slope is the $T_{\rm ex}$. The y-intercept of the linear fit slope can be used to calculate the $N_{\rm tot}$ when equation \ref{eq:4} is rearranged \citep{1999ApJ...517..209G}. We used the RD method for molecules with at least 2 transitions and a reasonably large range of $E_u$ values (e.g., see Figure \ref{rd}). In some cases where the upper state energies were too similar (e.g., within 5 K), the total column density $N_\mathrm{{tot}}$ could not be calculated accurately, and the FT method was applied instead. 

The RD method was only employed for CCS, CCCS, HC$_3$N, HC$_5$N, and HC$_7$N, with exceptions for isotopologues CC$^{34}$S and HCC$^{13}$CN in some cores. With only one HC$_3$N line detected in core 752 (e.g., see weak HC$_3$N spectrum in Figure \ref{HC3Ndiagnostic}), we assumed a $T_{\rm ex}$ from HC$_3$N detections in the closest nearby source, core 768, to employ the LTE method. Tables \ref{tab:CS}-\ref{tab:HC7N} displays $N_{\rm tot}$, abundance relative to H$_2$, $T_{\rm ex}$, and method applied for each molecule.

\begin{figure*}
    \centering
    \includegraphics[width=150mm]{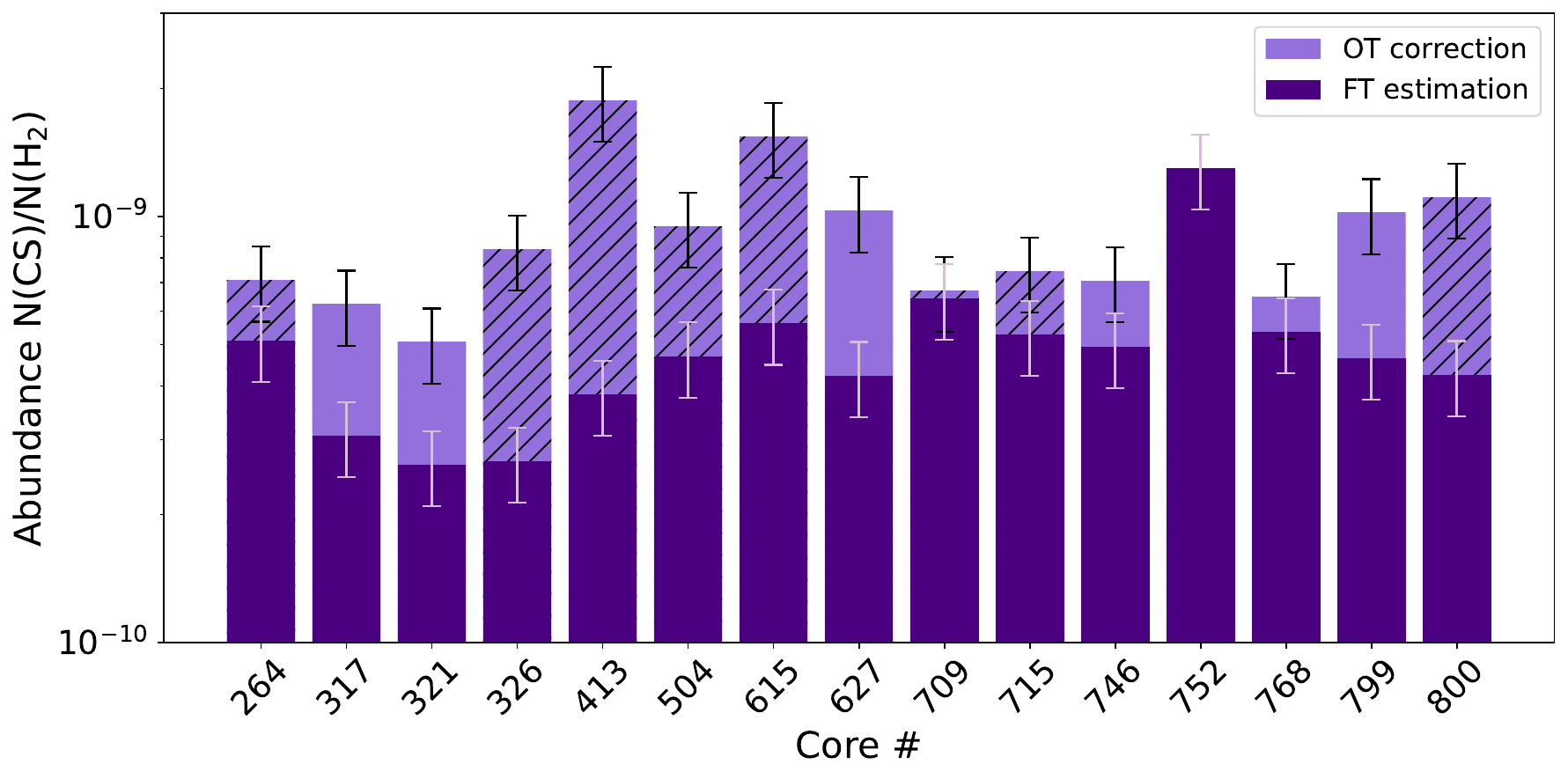}
    \caption{Comparisons between fixed temperature FT and optical thick (OT) correction factor methods for estimating the fractional abundance of CS, with respect to H$_2$. Black hatched bars for CS represent the corrected abundance values (via OT correction factor) to account for CS being optically thick in our observations. Error bars are included for OT corrections in black and FT estimations in light purple. The FT estimation is used for core 752 because CS is likely optically thin in this core.}
    \label{cs}
\end{figure*}

Because of our optically thin assumption, $\tau$ $<$ 1, our calculations of CS column densities via the FT method, and therefore abundances, were smaller than values expected. Based on previous observations of CS toward prestellar cores including L1544 \citep{Kim2020}, we expect abundances relative to H$_2$ to be on the order of $\times10^{-10}-10^{-9}$, whereas our initial calculations exhibited abundances ranging from $8.1\times10^{-12}- 2.9\times 10^{-11}$. The mean $^{12}$C/$^{13}$C fraction in our cores (36.1), calculated from the detected $^{13}$CS isotopologue, is also underestimated compared to the ISM carbon isotopic value of 68 \citep{2005ApJ...634.1126M}. Studies of sulfur chemistry in the L1544 prestellar core by \cite{Vastel2018} find similarly low $^{12}$C/$^{13}$C ratios (13.3--16.9) and resolve that CS is optically thick in their source. Therefore, we likewise conclude that CS is optically thick in our prestellar/starless cores. We thus calculate CS column densities three different ways: under the LTE optically thin limit (FT method), using optical depth correction equations (OT method), and by running the non-LTE code RADEX via the Python version \texttt{pythonradex}
\footnote{\url{https://github.com/gica3618/pythonradex}} (the comparison is displayed in Table \ref{CSmethods}; \cite{Cataldi2026}). 
For the optical depth correction method, we use the equation,
\begin{equation} \label{eq:5}
    \frac{I_{\rm CS}}{I_{^{13}\rm CS}} = \frac{1-e^{-\tau}}{1-e^{-\frac{\tau}{68}}},
\end{equation}
\noindent
by \citealt{2015PASP..127..266M} to estimate an optical depth, $\tau$, assuming a fixed $^{12}$C/$^{13}$C ratio of the ISM value 68, from the integrated intensities of CS and $^{13}\mathrm{CS}$ (I$_\mathrm{CS}$ and I$ \rm _{^{13}CS}$). Using the column density of CS in the optically thin limit, N$_{\rm tot}^{\rm thin}$, we use the following equation,
\begin{equation} \label{eq:6}
    N_{\rm tot} = N_{\rm tot}^{\rm thin} \frac{\tau}{1- \rm exp(-\tau)},
\end{equation}
\noindent
to re-calculate CS column densities with an optical depth correction factor \citep{1999ApJ...517..209G}. In core 752, we find CS is already optically thin and therefore does not need a correction factor. For this reason, we continue to use the FT method to estimate the CS column density in core 752. 

RADEX \citep{radex}, or \texttt{pythonradex}, is a non-LTE radiative transfer technique that assumes a uniform density and temperature profile but uses independent excitation temperature values to optimize column densities that predict the intensity and main beam temperature and can account for optical depth effects. We adopt a static slab geometry for the derivation of escape probabilities to remain consistent with the slab geometry described by an LTE assumption. Using line parameters (e.g., observed linewidth, or FWHM) that correspond to CS transitions, as well as physical excitation conditions ($T_{\rm kin}$ and \textit{Herschel} \textit{n}(H$_2$) densities, from Table 1 in \citealt{Scibelli2024}) of the prestellar/starless cores, we use \texttt{pythonradex} to constrain best-fit column densities based on running a grid of 500 model iterations. Compared to the LTE column densities from the optical depth correction, the \texttt{pythonradex} values for CS range on average $\sim2.6\times$ larger. However, since \texttt{pythonradex} column density calculations incorporate collisional rate coefficients, they introduce new uncertainties relative to LTE. Additionally, because the physical structure of these prestellar/starless cores is not well defined and the spatial origin of CS emission is unknown, we conclude \texttt{pythonradex} introduces minor uncertainties through its assumed escape probabilities, which depend on idealized geometry (see Appendix\,\ref{appendix:CS}). We therefore adjust the LTE column density and fractional abundance values for CS using the optical depth correction factor (Equations \ref{eq:5} and \ref{eq:6}) for all cores except core 752, reported in the remaining tables and figures of this study (shown in Figure \ref{cs}).

In general, due to the lack of collisional rate coefficient data for most of our larger species and the uncertainty brought about by incorporating \texttt{pythonradex}, we use the RD and FT methods throughout this paper for consistency. However, HC$_3$N and CCS column density calculations were cross-checked with \texttt{pythonradex} to verify our optically thin assumption. We obtain the collisional rate coefficients for CCS and HC$_3$N from the Leiden Atomic and Molecular Database
(LAMBDA; \citealt{2005A&A...432..369S}). For CCS, the collisional rates were derived by scaling H$_2$ with CCS-He rate coefficients from \cite{GodardPalluet2023}, and those for HC$_3$N were taken from \cite{2016MNRAS.460.2103F}. We find that $\tau$ $<$ 1 for both CCS and HC$_3$N and column densities vary by a factor of at most $\sim$2.3 -- 2.5, so our optically thin assumption holds.

Using the RD method, in core 264 we calculate a lower $T_{\rm ex}$ than expected for HC$_3$N, at 3.41 K. A $T_{\rm ex}$ value lower than the kinetic temperature of the source suggests a molecule that is sub-thermally populated. These transitions might have higher critical densities and are subsequently not thermalized at relatively low densities. A non-LTE calculation using \texttt{pythonradex} finds the $T_{\rm ex}$ to be closer to 10.2 K when the HC$_3$N column density is increased. In core 264, we find $\tau<$ 1, confirming that emission is optically thin. We caution that other cores with similarly low calculated $T_{\rm ex}$ ($<$ 5 K) for HC$_3$N might have actual $T_{\rm ex}$ values closer to 10 K. This discrepancy likely reflects the local excitation conditions of the gas rather than true kinetic temperatures or optical depth effects. Although non-LTE effects might impact a few cores in our sample, we have cross-checked our primarily-LTE analysis with a non-LTE analysis and found that the general trends in our abundance comparisons presented in this study remain consistent. With more transitions available and stronger gas temperature and density constraints toward these sources, future work could include a more sophisticated non-LTE study.

\section{Discussion} \label{sec:4}

\subsection{Fractional Abundances} \label{sec:4.1}
We calculate carbon-chain abundances with respect to molecular hydrogen (H$_2$), derived from the \textit{Herschel} column density values \citep{2021A&A...645A..55P} listed in column 8, Table 1 of \cite{Scibelli2024}. Given that the column density of a species is $N_X$, the fractional abundance of that species is simply the ratio, 
\begin{equation}
    f_X = \frac{N_X}{N_{{\rm H_2}}},
\end{equation}
where $N_{\rm H_{2}}$ is the column density of molecular hydrogen.
We recognize that the beam sizes (54.4" -- 36.4") of our observations are slightly smaller than the beam size used to measure \textit{Herschel} H$_{2}$ column densities (ARO 12m 62" beam; see Table 1, column 8 of \cite{Scibelli2024} for \textit{N}(H$_2$) values). Rather than re-calculate a beam-averaged \textit{N}(H$_2$), which would vary depending on the transition rather than the individual molecule, we assume the emission overfills our beam and include a 20\% error in our abundance calculations. Fractional abundances were determined for detections, and upper limits for non-detections, in order to compare abundances within carbon-chain families (see Tables \ref{tab:CS} - \ref{tab:HC7N}).

\begin{figure*} 
    \centering
    \includegraphics[width=\textwidth]{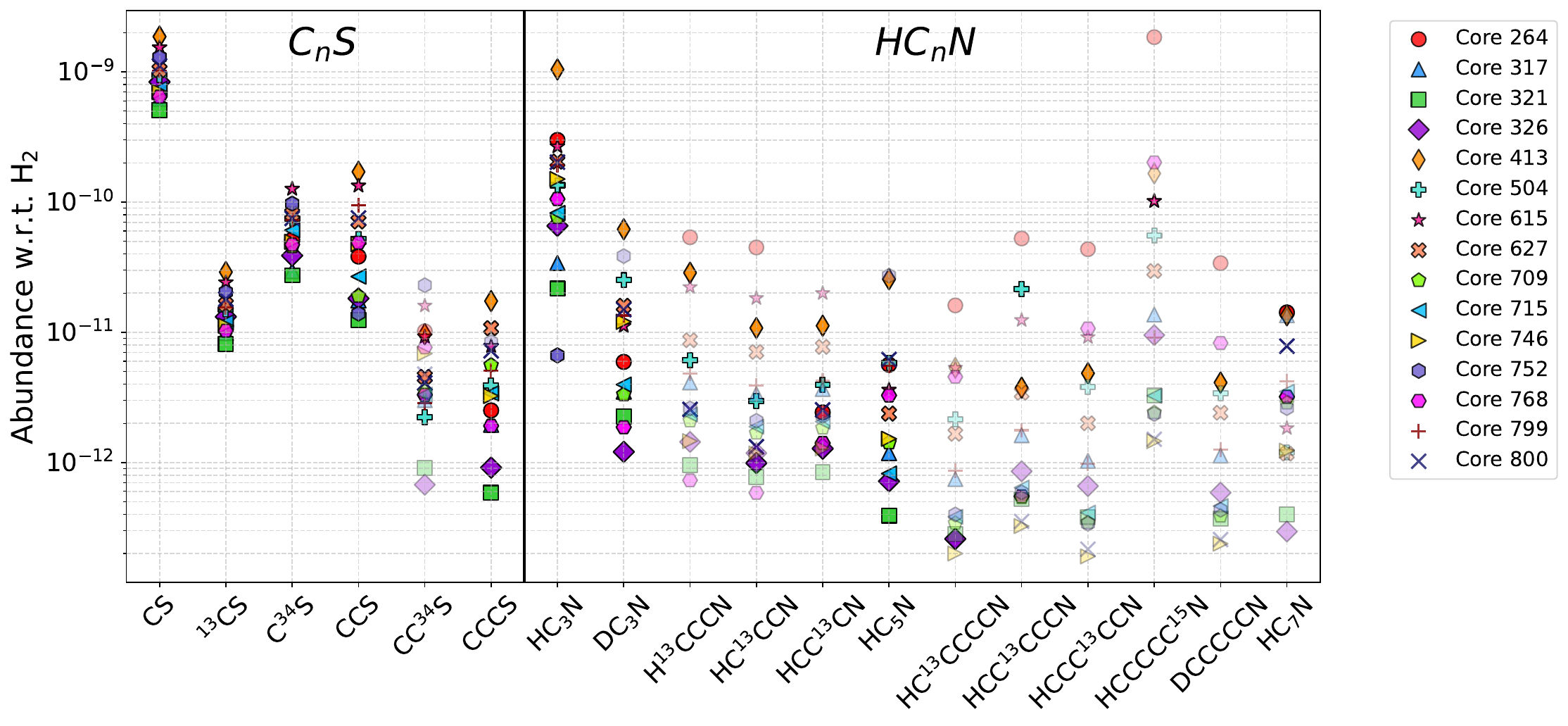}
    \caption{Observed fractional abundances relative to H$_2$ for all C$_n$S and HC$_n$N species and their isotopologues. Each core is represented by a different color and symbol. Upper limits are indicated by the faded symbols. Errors were removed in order to prioritize clarity in the figure (see Tables \ref{tab:CS}-\ref{tab:HC7N} for error values). The vertical line separates the two families, C$_n$S and HC$_n$N.} \label{abundance}
\end{figure*}

\begin{figure*} 
    \centering
    \includegraphics[width=150mm]{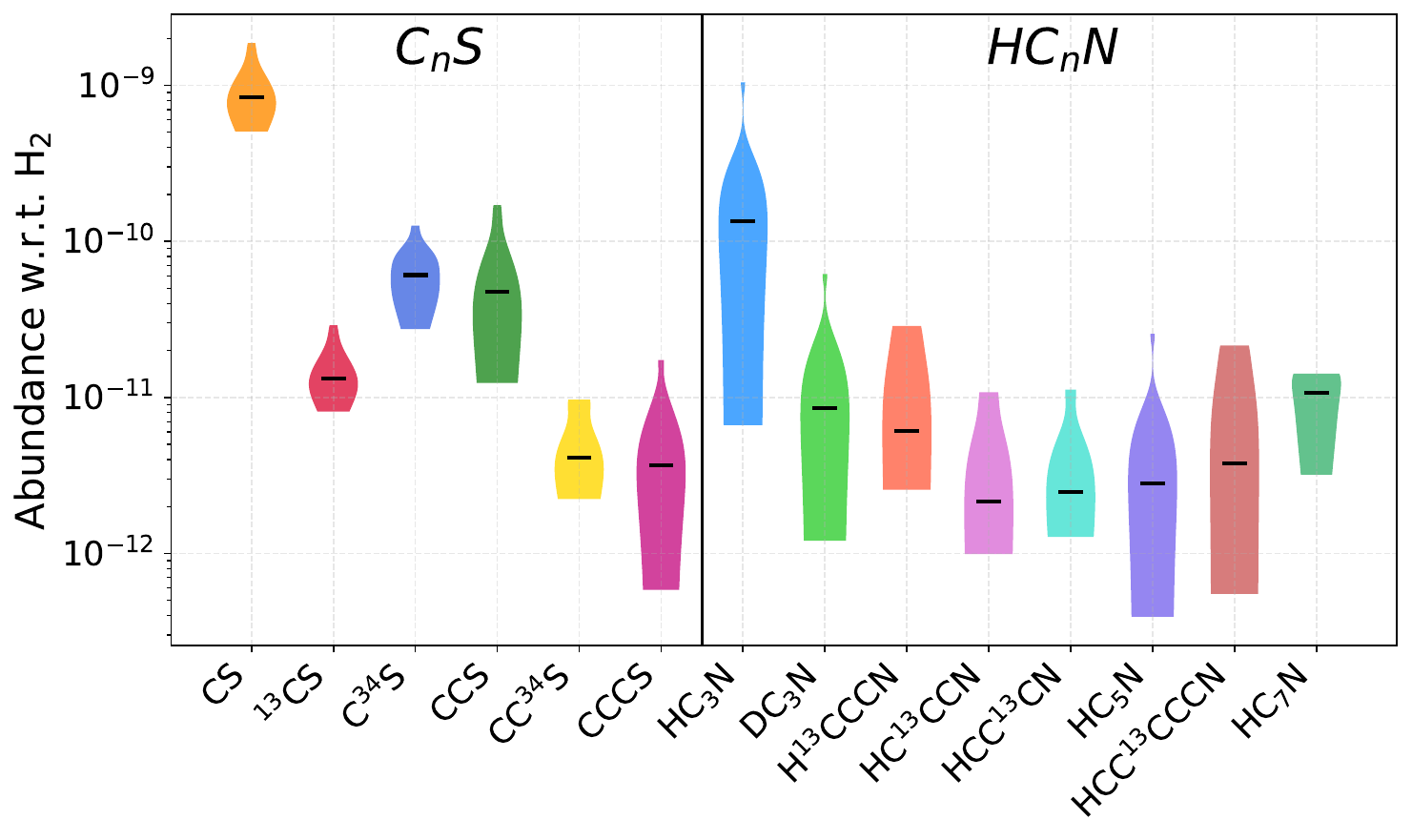}
    \caption{Violin plot of observed fractional abundances relative to H$_2$. The width of each violin represents the distribution density of data points at a given abundance, while the height spans the full range of observed values. The horizontal black line in each violin indicates the median abundance. Not all isotopologues are shown in this figure, as we required at least three detections in order to illustrate statistical information about a given species. The vertical line separates the two families, C$_n$S and HC$_n$N.}\label{violin}
\end{figure*}

In Figures \ref{abundance} and \ref{violin}, we present fractional abundances with respect to H$_2$ for all C$_n$S and HC$_n$N species, including all detected isotopologues. In the C$_n$S family, we observe at least one order magnitude decrease in abundance with each addition of a C atom (e.g., CS to CCS, CCS to CCCS). However, we do not see this same trend in the HC$_n$N family, as the median HC$_5$N and HC$_7$N abundances are quite similar. We note that HC$_7$N is detected in only 4/15 cores (264, 413, 768, 800), compared to 14/15 for HC$_5$N, so a limited sample size could explain their similarity in abundances. The HC$_7$N detections in cores 264, 413, and 800 are preferentially associated with cores that also exhibit high HC$_3$N and HC$_5$N column densities, suggesting a bias towards cyanopolyyne-rich sources. These four cores do not appear to have any unique physical properties that would result in enhanced molecular content of HC$_7$N.

The violin-style plot shown in Figure \ref{violin} illustrates the distribution density and range of abundances for each of the carbon species, in which there were at least three detections. The range of abundances spans several orders of magnitude for most species, but especially for CCCS, HC$_3$N, DC$_3$N, and HC$_5$N. The broad scatter of abundances for these carbon-chain species indicates that the cores' carbon-chain inventories are likely sensitive to their surrounding environments \citep{Law2018}. In addition to environmental factors, this spread may also reflect differences in chemical age, with younger, less chemically evolved cores typically exhibiting enhanced carbon-chain abundances with more available atomic carbon. Variations in depletion efficiency onto dust grains, as well as uncertainties in the adopted physical parameters ($H_2$ column densities and excitation conditions), may further contribute to this observed dispersion of abundances. Protostellar outflows are thought to influence the chemical structure and inventories of prestellar/starless cores \citep{2023ASPC..534..379C, Jrgensen2020, 2025A&A...702A.127S, 2025ApJ...985L..25S}, as outflow-heated gas could aid in removing COMs and carbon-chains off the dust grains. The full impact that molecular outflows have on the chemistry of their nearby environments and the evolution of star formation remains a mystery. While there does not seem to be a clear correlation between the outflow-impacted cores in the shocked regions NGC 1333 (cores 264 and 326) and IC 348 (cores 709 and 715) in our observations \citep{1999A&A...343..571G, 2013ApJ...774...22P}, it would be of great interest to study the physical conditions and spatial distribution of these cores alongside their molecular compositions. 

In contrast, $^{13}$CS and C$^{34}$S abundances are each more tightly constrained relative to the full inventory of detections, with their respective mean abundances ($\pm$ standard deviation) being 1.5$\times$10$^{-11}$ ($\pm$ 5.7$\times$10$^{-12}$) and 6.4$\times$10$^{-11}$ ($\pm$ 2.7$\times$10$^{-11}$). To further quantify the relative dispersion within the C$_n$S species, we use the coefficient of variation (also known as the normalized root-mean-square deviation), defined as the standard deviation normalized by the mean abundance. A tighter correlation has a coefficient of variation closer to 0, and a broader distribution has a value closer to 1. The coefficients of variation are 0.38 and 0.42 for $^{13}$CS and C$^{34}$S, compared to 0.83 and 0.88 for CCS and CCCS. This narrower dispersion indicates that CS isotopologues are comparatively insensitive to local environmental differences, making them less prone to the larger source-to-source variations seen in other species. Combined with the fact that isotopologues are less affected by optical depth than the main CS line \citep{Tafalla2010, Davis2013}, this suggests that $^{13}$CS and C$^{34}$S can serve as reference points for interpreting chemical differentiation with larger molecules during the prestellar phase.

The longer carbon-chain species (e.g., HC$_5$N, CCCS) are detected in fewer cores and at lower abundances than their shorter precursors (e.g., HC$_3$N, CS, and CCS), which is expected as larger molecules typically exhibit lower observed abundances \citep{1981A&A....99..239B, Suzuki1992,1996Ap&SS.240...13W, Cernicharo2020, 2024PASJ...76.1270T}. While the median abundances of our HC$_5$N and HC$_7$N detections appear similar, we note that HC$_7$N is only detected in 3/15 cores (excluding the tentative detection), which appear to be the most carbon-chain-rich cores. This limited number of detections likely biases the HC$_7$N median towards higher values and limits our ability to extend this formation build-up trend to HC$_7$N.

Some cores (413, 264, 615) show higher molecular abundances across multiple species, suggesting either earlier chemical formation timescales or more favorable conditions for carbon-chain production. As noted previously, core 413 remains to have the most diverse inventory and is among the most abundant in every species, including CS. In contrast, core 752 has among the highest CS abundances, but is the most line-poor object in our sample and has one of the lowest abundances of CCS and HC$_3$N, at only $\sim$0.25$\times$ and $\sim$0.01$\times$ their respective mean values. Interestingly, core 752 also has a \textit{n}(H$_2$) volume density 3.4$\times$ lower than the mean of the whole core sample, suggesting that it may be at an earlier stage of its formation, since higher densities generally correlate with more chemically evolved cores \citep{Crapsi2005}. However, recent work by \cite{2022MNRAS.515.5219G} shows that core density alone may not uniquely define chemical evolution, as cores may remain at similar densities for differing lengths of time.

\begin{figure}
    \centering
    \includegraphics[width=8.5cm]{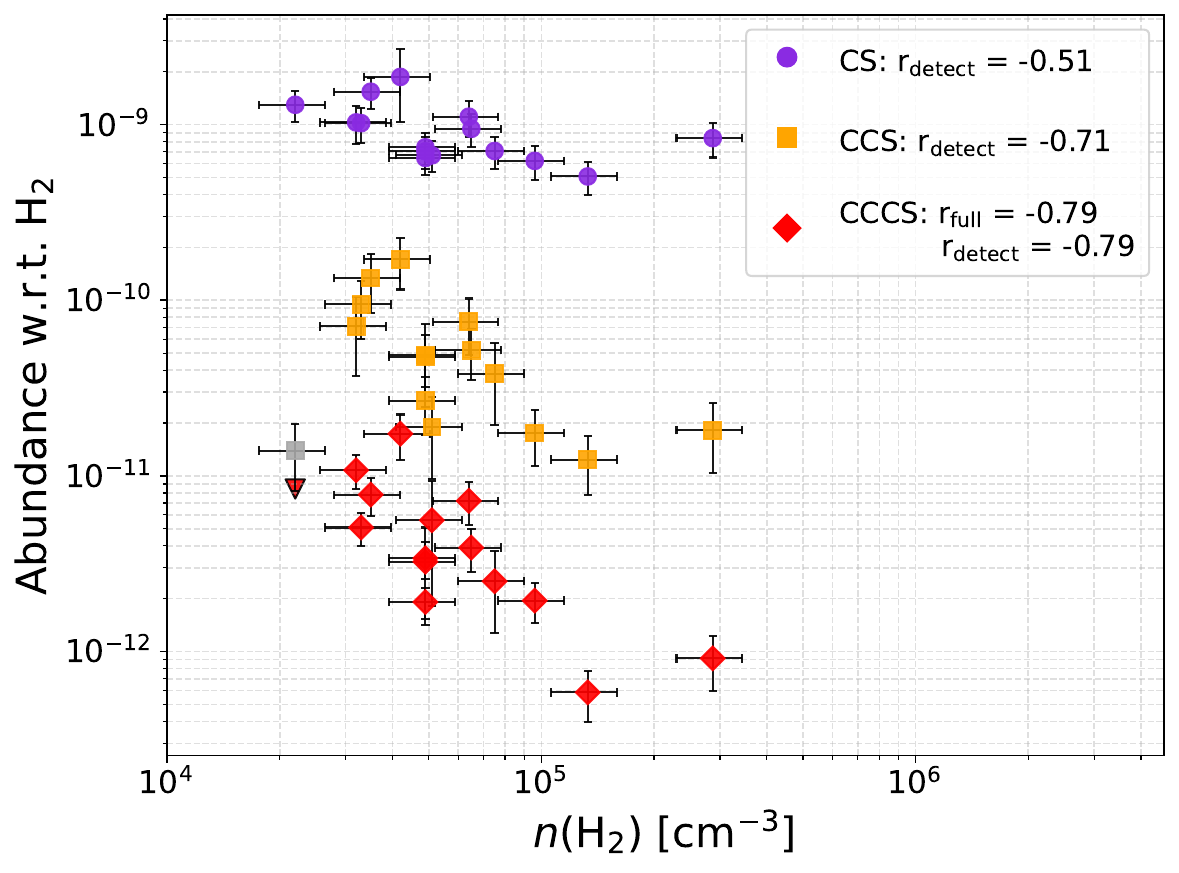}
    \caption{Comparison of C$_n$S abundances with respect to the mean H$_2$ volume density, \textit{n}(H$_2$), in units of cm$^{-3}$. The Pearson rank correlation coefficient `$r$' is calculated for each species in the C$_n$S family. The `$r$' coefficient is calculated both for detections, $r_{\mathrm{detect}}$, and the full sample, which includes upper limits, $r_{\mathrm{full}}$. We note that our $r_{\mathrm{detect}} = -0.71$ value for CCS excludes core 752 (marked in gray) due to its outlier behavior; however, when included in the sample, $r = -0.45$. We include a 20\% error for \textit{n}(H$_2$), which is plotted in black. Upper limits are denoted by a downward triangle.}
    \label{CnS_volume}
\end{figure}

\begin{figure}
    \centering
    \includegraphics[width=8.5cm]{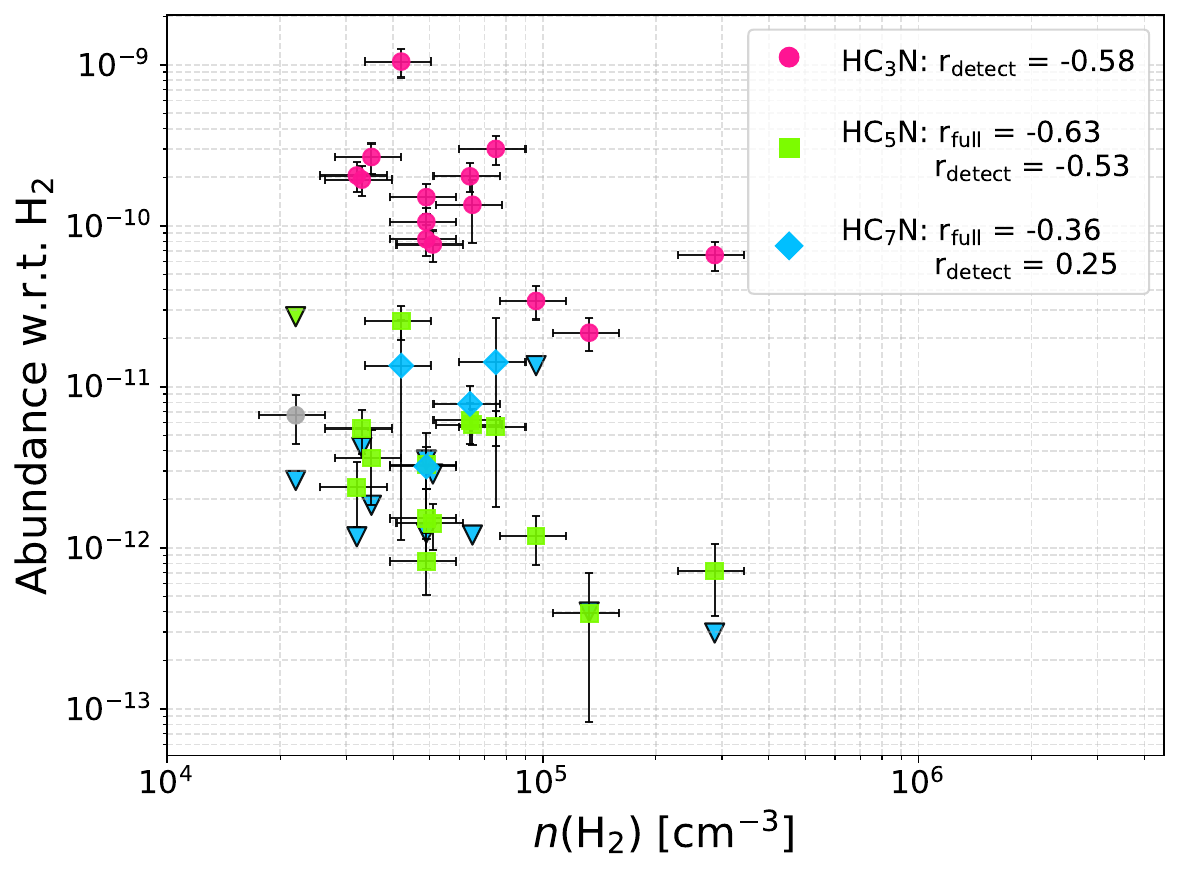}
    \caption{Comparison of HC$_n$N abundances with respect to the mean H$_2$ volume density, \textit{n}(H$_2$), in units of cm$^{-3}$. The Pearson rank correlation coefficient `$r$' is calculated for each species in the HC$_n$N family. The `$r$' coefficient is calculated both for detections, $r_{\mathrm{detect}}$, and the full sample which includes upper limits, $r_{\mathrm{full}}$. We note that our $r_{\mathrm{detect}} = -0.53$ value for HC$_3$N excludes core 752 (marked in gray) due to its outlier behavior; however, when included in the sample, $r = -0.06$. We include a 20\% error for \textit{n}(H$_2$), which is plotted in black. Upper limits are denoted by a downward triangle.}
    \label{HCnN_volume}
\end{figure}

\subsection{Abundance Correlations} \label{abundance_corr}
We first compare the cyanopolyyne (HC$_n$N) and sulfur-bearing (C$_n$S) abundances with respect to H$_2$ to the mean H$_2$ volume density, \textit{n}(H$_2$), in units of cm$^{-3}$ for each core (Figure \ref{CnS_volume}, \ref{HCnN_volume}). We do acknowledge the potential bias from comparing line-of-sight observations, as both averaged \textit{Herschel} measurements and our beam-averaged molecular line column densities both likely probe the warmer outer layers of the core.

To evaluate the relationship between two species, we apply the Pearson correlation coefficient `$r$', a measure of linear dependence (available through the scipy.stats package\footnote{\url{https://docs.scipy.org/doc/scipy/reference/stats.html}}; \citealt{Virtanen2020}). An `$r$' value approaching $-$1 or $+$1 reflects a perfectly decreasing or increasing linear trend, whereas a value near zero indicates an absence of correlation. Molecules that share chemical formation pathways or are governed by comparable physical conditions, such as temperature, are often expected to exhibit strong positive correlations.

For sulfur-bearing carbon-chains, we find a tight negative correlation with volume density for CCCS ($r = -0.79$) and a weaker correlation for CS ($r = -0.51$), as shown in Figure \ref{CnS_volume}. As for CCS, there seems to be a weaker correlation when all cores are included ($r = -0.45$), but excluding core 752 strengthens the trend ($r = -0.71$), indicating that this source is chemically distinct with an unusually low abundance of CCS (Figure \ref{CnS_volume}). Among cyanopolyynes, HC$_n$N, HC$_5$N exhibits a negative correlation (r$_\mathrm{full} = -$0.63 and r$_\mathrm{detect} = -$0.53 for the full sample that includes upper limits and the sample with only detections, respectively; see Figure \ref{HCnN_volume}). However, for HC$_3$N, the correlation is negligible when all cores are included ($r = -0.06$), but a negative correlation becomes more notable when excluding core 752 ($r = -0.58$), similar to CCS, as shown in Figure \ref{HCnN_volume}. Whereas, HC$_7$N, with only four detections, shows substantial scatter and therefore, a much weaker correlation, limiting its diagnostic capability as an evolutionary tracer.

Excluding the outlier core 752, these results suggest that both sulfur-bearing carbon-chains and cyanopolyynes are preferentially enhanced in chemically young, lower-density cores, but decline in abundance as density increases and the prestellar cores chemically evolve \citep{Ohashi2015}. This was previously noted in \cite{2021ApJ...917...44J} and \cite{2023MNRAS.519.1601M}, who suggested that N-bearing species (including HC$_3$N) are enhanced in younger starless cores. Indeed, at higher densities, carbon-chain species are either adsorbed onto dust grains or destroyed through reactions with ions like He$^+$ and H$^+$, or atomic oxygen, resulting in their depletion during the later stages of core development \citep{Lu2025}. In contrast, early-phase prestellar conditions favor the formation of unsaturated carbon-chains (HC$_5$N, CCS, CCCS; \citealt{Suzuki1992, Benson1998, deGregorioMonsalvo2006,2008ApJ...672..371S}). 

A similar density-dependent relationship is seen in the same set of Perseus cores with COMs \citep{Scibelli2024}, in which the authors report a decreasing COM abundance with increasing \textit{n}(H$_2$) (excluding CH$_3$OH). In the \cite{Scibelli2024} survey of COMs, the authors attribute the negative trend with \textit{n}(H$_2$) to the Yebes 40m beam encompassing both the dense core and its surrounding envelope, therefore concluding that COM emission in the Perseus sample is peaking away from the dust peak in higher density cores. 

Spatially resolved maps of the highly evolved and dense prestellar core L1544 reveal H$_2$CCO and CH$_3$CN peak offset from the dusk peak of the core \citep{2017A&A...606A..82S}, and larger COMs, like CH$_3$CHO, were found to peak co-spatially to the CH$_3$OH peak \citep{2016ApJ...830L...6J} from single-point observations towards the dust and CH$_3$OH peak positions. This reflects the influence of depletion, where molecules freeze out most efficiently in the highest-density cores, resulting in lower COM and carbon-chain abundances in the central regions. Given that in this study our beam is larger than the size of the core, it is probable that our detections of carbon-chains primarily trace the parts of the surrounding envelope rather than the central region, especially for the denser cores. The lower abundances of HC$_n$N and C$_n$S species observed in the denser Perseus cores are consistent with the effects of depletion, as shown in Figures \ref{CnS_volume} and \ref{HCnN_volume}. It still remains possible that COMs and carbon-chains are spatially separated, peaking in different regions of the core \citep{2017A&A...606A..82S, Spezzano_2020}.

In L1544, carbon-chain emission is concentrated in the southeast region, distinct from the methanol peak situated in the north part of the core \citep{2016A&A...592L..11S, 2017A&A...606A..82S}. The carbon-chain peak is slightly redshifted with respect to the methanol peak, indicating that COM and carbon-chain chemistry proceed in different regions and along distinct formation networks \citep{2016A&A...592L..11S, 2017A&A...606A..82S,2022arXiv220613270C, Bianchi2023}. Because carbon-chains are enhanced toward the external regions of the core, where material is more exposed to the ultraviolet (UV) interstellar radiation field, there are more gaseous carbon atoms available to drive the formation of larger cyanopolyyne chains. Consistent with this, \cite{Bianchi2023} find that HC$_n$N abundances in TMC-1 and L1544 indicate that larger cyanopolyyne ($n \geq 5$) chemistry depends primarily on gaseous carbon abundance. Similarly, maps of the L1495-B2188 filament in the Taurus Molecular Cloud exhibit that CCS and HC$_7$N peaks are offset from NH$_3$ and dust continuum peaks \citep{2019ApJ...871..134S} and CCS and CCCS emission towards the starless core L1498 are centrally depleted and offset from its NH$_3$ peak, where gas is usually shielded from UV radiation \citep{Willacy1998,2005ApJ...632..982S}. Therefore, it is believed that UV illumination likely drives the divergence between COMs and carbon-chain peaks in prestellar/starless cores \citep{2016A&A...592L..11S,2017A&A...606A..82S,2022arXiv220613270C, Bianchi2023}. 

\subsection{Column Density Ratios}

\begin{table*}
\centering
\caption{Column density ratio comparisons with Taurus, Serpens South, and protostar sources. Since HC$_5$N and CCCS were not detected in core 752, we do not calculate these column density ratios. The regions and cloud column density ratios are averaged values from detections in this work or previous literature. \textsuperscript{1}Column density ratios for Taurus prestellar/starless cores are calculated from \citealt{Suzuki1992}. \textsuperscript{2}Taurus and Perseus protostars column density ratios are calculated from \citealt{Law2018}. \textsuperscript{3}Serpens South cluster-forming region for a pre-cluster clump are calculated from \citealt{2024PASJ...76.1270T}. Errors are included in parentheses for each individual core, and the standard deviation of the sample is included for each Perseus region and source.} \label{cd_ratios}
\begin{tblr}{
  row{even} = {c},
  row{1} = {c},
  row{3} = {c},
  row{5} = {c},
  row{7} = {c},
  row{9} = {c},
  row{11} = {c},
  row{13} = {c},
  row{15} = {c},
  row{17} = {c},
  row{19} = {c},
  row{21} = {c},
  row{23} = {c},
  row{25} = {c},
  hline{2,17,21} = {-}{},
}
Core \# (or source) & N(HC$_3$N)/N(HC$_5$N) & N(CCS)/N(CCCS) & N(CCS)/N(HC$_3$N) & N(CCS)/N(HC$_5$N)\\
264 & 53.2 (2.4) & 15.2 (6.3) & 0.13 (0.04) & 6.8 (2.0)\\
317 & 28.9 (3.3) & 9.0 (1.5) & 0.52 (0.08) & 14.9 (2.8)\\
321 & 55.3 (32.6) & 21.2 (4.3) & 0.57 (0.10) & 31.6 (19.4)\\
326 & 91.7 (25.1) & 19.9 (5.4) & 0.28 (0.06) & 25.4 (9.0)\\
413 & 40.9 (1.5) & 9.8 (1.5)& 0.16 (0.02) & 6.7 (0.9)\\
504 & 47.0 (2.3) & 13.3 (1.9) & 0.19 (0.02) & 8.9 (1.2)\\
615 & 74.0 (21.5) & 17.1 (3.0) & 0.50 (0.08) & 37.0 (12.4)\\
627 & 86.4 (20.1) & 6.6 (1.8) & 0.35 (0.10) & 29.9 (10.8)\\
709 & 53.9 (5.7) & 3.4 (1.9) & 0.25 (0.07) & 13.4 (4.1)\\
715 & 100.4 (16.5) & 7.9 (1.3) & 0.33 (0.05) & 32.6 (7.5)\\
746 & 98.9 (31.3) & 14.7 (5.7) & 0.32 (0.04) & 31.4 (10.8)\\
752 & -- & -- & 2.09 (1.22) & --\\
768 & 32.2 (3.0) & 25.5 (7.6) & 0.46 (0.14) & 14.9 (4.6)\\
799 & 35.1 (3.8) & 18.7 (3.1) & 0.49 (0.08) & 17.2 (3.4)\\
800 & 32.8 (2.8) & 10.5 (1.8) & 0.37 (0.06) & 12.2 (2.2)\\
NGC 1333 & 57.3 (25.9) & 16.3 (5.5) & 0.38 (0.21) & 19.7 (11.0) \\
B1 & 44.0 (4.3) & 11.6 (2.5) & 0.18 (0.02) & 7.8 (1.6) \\
IC 348 & 74.3 (26.9) & 12.5 (8.2) & 0.61 (0.66) & 26.5 (9.9)\\
B5 & 34.0 (1.6) & 14.6 (5.8) & 0.43 (0.08) & 14.7 (3.5)\\
Perseus prestellar/starless cores & 59.3 (26.0) & 13.8 (6.3) & 0.47 (0.47) & 20.2 (10.7) \\
Taurus prestellar/starless cores\textsuperscript{1} & 3.2 (0.4) & 4.1 (0.3) & 1.10 (0.14) & 2.6 (0.2)\\
Perseus protostars\textsuperscript{2} & 5.6 & 10.3 & 1.013 & 5.7\\
Taurus protostars\textsuperscript{2} & 0.6 & 5.4 & 1.421 & 2.3 \\
Serpens South pre-cluster clump\textsuperscript{3} & 3.7 & 8.8 & 1.03 & 3.8
\end{tblr}
\end{table*}

We present the column density ratios of HC$_3$N/HC$_5$N, CCS/CCCS, CCS/HC$_3$N, and CCS/HC$_5$N in the 15 Perseus starless and prestellar cores (Table \ref{cd_ratios}). We calculated all column density ratios based on species' detections, so no upper limits are included in these values.
Among the Perseus sources, we find that HC$_3$N column densities are consistently two orders of magnitude higher than HC$_5$N column densities for each of the cores where both species are detected, with a mean HC$_3$N/HC$_5$N ratio of 57.3. This indicates that smaller cyanopolyynes dominate the chemical inventory of the cores, consistent with the expectation that HC$_3$N forms more efficiently in cold, dense conditions \citep{Seki1996,https://doi.org/10.48550/arxiv.2206.13270,Bianchi2023}. Core 715 exhibits the highest ratio (100.4), although still similar to the other HC$_3$N/HC$_5$N ratios. 

Similarly, we find consistently high CCS/CCCS ratios (mean 13.8), showing that CCS is generally 1--2 orders of magnitude higher in column density than CCCS. This result suggests that CCS might be a reliable tracer of carbon-chain chemistry in cold, dense cores, while CCCS is less efficiently formed. The CCS/HC$_3$N ratios span a narrow range across most Perseus cores (0.13 -- 0.57), implying a close chemical relationship between these species during the prestellar phase. Studies by \cite{Hirahara1992} and \cite{Petrie1996} suggest that neutral-neutral (term includes radical-neutral) pathways explain the strong correlation between C$_{n}$S and HC$_n$N species, highlighting a common gas-phase formation network for sulfur- and nitrogen-bearing carbon-chains. This potentially shared chemistry is discussed in more detail in Section \ref{sec:4.5}.

Core 752 remains an outlier, as it exhibits an interestingly high CCS/HC$_3$N ratio of 2.09. Located in IC 348, it resides farther from the HH 211 protostellar outflow where it is more isolated. Combined with its lower \textit{n}(H$_2$) density and lack of carbon-chain detections compared to the other Perseus cores, this suggests that its local environment may be influencing its carbon-chain reservoir. A study by \cite{Asensio2026} investigates the carbon-chain molecule c-C$_3$H$_2$ towards the same set of Perseus prestellar and starless cores, in which c-C$_3$H$_2$ is not detected in core 752. Since several COMs, such as CH$_3$OH and CH$_3$CHO, were detected toward core 752 by \cite{Scibelli2024}, its limited carbon-chain prevalence might indicate particularly weak carbon-chain emission rather than reflecting the core's overall chemical condition. This exception points to a potentially distinct carbon-chain chemistry or evolutionary phase relative to the rest of the Perseus cores, making core 752 an interesting source for future follow-up studies.

\subsection{Comparisons with Taurus, Serpens South, and Protostar Sources}

The Taurus Molecular Cloud is one of the most carbon-rich regions, with the dark cloud TMC-1 serving as a benchmark source for molecular line surveys and chemical modeling (e.g., \citealt{2004PASJ...56...69K, 2025ApJS..281....9X}). Its rich inventory of unsaturated carbon-chains has been extensively cataloged, making Taurus a useful comparison source for evaluating how carbon-chain inventories vary across molecular cloud environments. By contrasting abundances in Taurus prestellar/starless and protostellar sources, we can explore the extent to which carbon-chain chemistry is inherited from the prestellar to the protostellar stage in low-mass stellar evolution. We also compare abundances with Perseus protostellar sources to be able to better differentiate between Perseus and Taurus environments. In Table \ref{cd_ratios}, we include comparisons of column density ratios with Perseus protostars, Taurus prestellar/starless cores, and Taurus protostars using data from a cold cloud and carbon-chain survey conducted by \cite{Suzuki1992} and \cite{Law2018}, respectively. 

We calculated the mean column density for each species and type of source to determine the ratios. Taurus prestellar/starless cores exhibit systematically lower HC$_3$N/HC$_5$N and CCS/CCCS ratios compared to Perseus by an order of magnitude. The CCS/HC$_5$N ratio averages to 20.2 (standard deviation is 10.7) for Perseus prestellar/starless cores, which is a factor of 5 higher than Perseus protostars, Taurus protostars, and Taurus prestellar/starless cores. 

HC$_5$N was only detected in 9/16 Taurus prestellar/starless cores in the study by \cite{Suzuki1992}, where the authors calculated the HC$_5$N column density via the LTE fixed temperature (FT) method, assuming an excitation temperature of 6.5 K. This excitation temperature was based on previous observations of HC$_5$N \citep{1986A&A...155..237S, Takano1990}, which generated column densities on the order of $\times$10$^{12}$ cm$^{-2}$. In our study of Perseus cores, we calculated the column density based on multiple transitions of HC$_5$N using a rotational diagram, where excitation temperatures ranging from 6.1 -- 18.1 K and column densities spanning 1.19$\times$10$^{10}$ -- 2.99$\times$10$^{11}$ cm$^{-2}$ were calculated. When we assume the same excitation temperature of 6.5 K as \cite{Suzuki1992}, we similarly calculate column densities on the order of $\times$10$^{12}$ cm$^{-2}$. \cite{Suzuki1992} also notes that increasing the excitation temperature by 1 K, increases their column density by 30\% for HC$_5$N, which explains the significant disparity between both our HC$_3$N/HC$_5$N and CCS/HC$_5$N ratios when compared to Taurus prestellar/starless cores.

Lower Taurus HC$_3$N/HC$_5$N and CCS/CCCS ratios suggest systemic differences in carbon-chain chemistry between the two clouds, where Taurus favors a higher relative abundance of longer, more complex carbon-chains compared to Perseus. This disparity can be explained by both environmental and evolutionary mechanisms. Perseus cores may be chemically younger, with the observed molecular abundances reflecting an earlier phase of carbon-chain chemistry in which shorter chains (e.g., HC$_3$N, CCS) remain relatively more abundant. Alternatively, as cold, dense cores age, carbon-bearing species progressively freeze-out onto dust grains, reducing the reservoir of gas-phase atomic carbon available for the production of new carbon-chains \citep{Suzuki1992,2014MNRAS.437..930L}. This may cause shorter carbon-chain species to be destroyed more rapidly than their longer-chain counterparts (e.g., HC$_5$N, CCCS) in Taurus if it is generally more chemically evolved. While depletion of carbon-bearing species becomes increasingly important at later stages of core evolution, additional chemical modeling would be useful in understanding whether the observed HC$_3$N/HC$_5$N and CCS/CCCS ratios are directly related to depletion in carbon-chain growth. The lower ratios in the Taurus Molecular Cloud likely also reflect its uniquely quiescent environment and lower ambient G$_0$ interstellar radiation field relative to the Perseus Molecular Cloud. While an elevated G$_0$ field in Perseus can increase the availability of atomic carbon through CO photodissociation processes \citep{1988ApJ...334..771V}, the accompanying high-energy UV photons can also create a more destructive environment for longer carbon-chains. The weaker radiation field in Taurus might be effectively shielding these carbon-chain molecules from destruction via photodissociation, allowing them to accumulate relative to shorter chains. 

Evolutionary differences between Perseus and Taurus might also be evident when comparing the prestellar and protostellar stages. The CCS/HC$_3$N ratio is rather consistent around unity from prestellar cores to protostars for both Perseus (0.47 $\rightarrow$ 1.013) and Taurus (1.10 $\rightarrow$ 1.421) sources. This trend further suggests that HC$_3$N and CCS form similarly, or their timescales align at this stage of their prestellar phase. Time-dependent models likewise predict that both species peak at $\sim10^5$–$10^6$ yr before declining as carbon is locked into CO \citep{Suzuki1992, 1990A&A...231..466M, 2019ApJ...871..134S}. However, once the protostellar stage begins, there are two possible scenarios. Either HC$_3$N is destroyed at a faster rate and CCS persists longer. Or, CCS forms more efficiently than HC$_3$N, and the results of their slightly differing formation timescales are revealed at the start of the protostellar stage.

We also compare Perseus cores with the Serpens South cluster-forming region, in which \cite{2024PASJ...76.1270T} conducted a line-survey within the same 30 -- 50 GHz bandwidth toward the carbon-chain emission peak, corresponding to a pre-cluster clump. As shown in Table \ref{cd_ratios}, the Serpens South clump exhibits slightly lower column density ratios relative to Perseus prestellar/starless cores, similar to those in Taurus prestellar/starless cores. \cite{2024PASJ...76.1270T} concludes that the Serpens South pre-cluster clump is more chemically evolved but not yet forming protostars, consistent with the interpretation that Perseus cores are likely generally chemically younger compared to those in Taurus.

These formation timelines are consistent with the scenarios in \cite{Taniguchi2024}, where carbon-chains such as HC$_3$N and CCS are considered `early-type species' because they are prevalent in the initial evolutionary stages of prestellar cores \citep{2021ApJ...917...44J, 2023MNRAS.519.1601M}. Since sulfur-bearing species form on faster timescales than cyanopolyyne species \citep{Bergin1997}, this further supports the scenario in which HC$_3$N and CCS both form early but evolve at different rates. Alternatively, it is possible that HC$_3$N is destroyed by He$^+$ and H$^+$ ions, transformed, or impacted by freeze-out as the cores evolve \citep{Suzuki1992}. CCS is known to be susceptible to freeze-out and is likely among the first molecules to do so. As CO also freezes out, this creates space for molecular species like N$_2$H$^+$ to increase in abundance, allowing other nitrogen-containing carbon chains to form, evolve, and subsequently freeze out \citep{Bergin2007}. In this scenario, if CCS freezes out first, then the destruction of HC$_3$N might be occurring at a faster rate, so the CCS/HC$_3$N ratio still increases through the transition from a prestellar to protostellar core. However, in either case, higher-resolution observations, as well as continuum mapping images, are necessary to make stronger conclusions about the spatial distribution, inheritance, and depletion of carbon chains.

To illustrate the range of carbon-chain chemistry within Perseus and to understand the level of sensitivity that carbon-chain species have to their local conditions, we provide a source-by-source comparison of the most extreme cases presented in Table \ref{cd_ratios}. Cores 715 and 746 exhibit the largest HC$_3$N/HC$_5$N ratios in Perseus (100.4 and 98.9, respectively), while core 317 shows the smallest ratio in Perseus (28.9). Although all Perseus values are substantially larger than those reported for the Taurus prestellar core L1544 (2.3; \citealt{Suzuki1992}) and Serpens South pre-cluster clump (3.7; \citealt{2024PASJ...76.1270T}), they span more than a factor within Perseus itself, indicating significant source-to-source chemical variation.

Core 768 has the largest CCS/CCCS ratio in Perseus (25.5), whereas core 709 has the smallest (3.4). The latter is more comparable to the value measured toward L1544 in Taurus (4.3; \citealt{Suzuki1992}), known to be isolated from neighboring activity, while the ratio in core 768 exceeds those reported for both Taurus prestellar/starless cores and the Serpens South clump (8.8). Notably, both cores 709 and 768 are located within the IC 348 region, demonstrating that substantial chemical diversity can exist even among sources within the same larger-scale environment. Higher resolution maps of this region would be useful in understanding how these two prestellar/starless cores are affected by known protostellar activity in IC 348.

Among Perseus prestellar and starless cores, core 752 exhibits the largest CCS/HC$_3$N ratio (2.09), while core 264 shows the lowest ratio (0.13). Core 752's ratio is certainly a unique case, being more than a factor of 3.7$\times$ larger than other Perseus cores and about 2$\times$ larger than Taurus and Serpens South prestellar sources. The smallest CCS/HC$_3$N observed in core 264 is approximately a factor of 7$\times$ larger than L1544 in Taurus. Notably, core 264 is located in the active NGC 1333 environment and is exposed to nearby protostellar activity, whereas L1544 is a relatively isolated prestellar core. This contrast suggests that local environmental conditions may contribute to the observed differences, although the difference in ratios does not allow protostellar influence to be uniquely determined.

The CCS/HC$_5$N ratio also exhibits considerable variation. Cores 264 and 413 show the smallest ratios (6.8 and 6.7, respectively), comparable to the values measured towards Perseus protostars (5.7; \citealt{Law2018}), the Serpens South clump (3.8; \citealt{2024PASJ...76.1270T}), and the Taurus prestellar core L1498 (3.8; \citealt{Suzuki1992}). In contrast, core 615 exhibits the largest CCS/HC$_5$N ratio in Perseus (37.0), differing substantially from values reported in Taurus and Serpens South prestellar cores. Core 615 is situated in an isolated environment, away from protostellar activity, while core 264 is likely much more exposed to protostellar outflow activity in the NGC 1333 region. However, core 413 resides in the B1 region, where it is relatively isolated from the influence of nearby protostars. The wide range of ratios observed among Perseus cores, including sources located in both clustered and relatively isolated environments, suggests that local physical conditions and chemical histories may both contribute to the observed carbon-chain abundance patterns.

\begin{figure} 
    \centering
    \includegraphics[width=8.5cm]{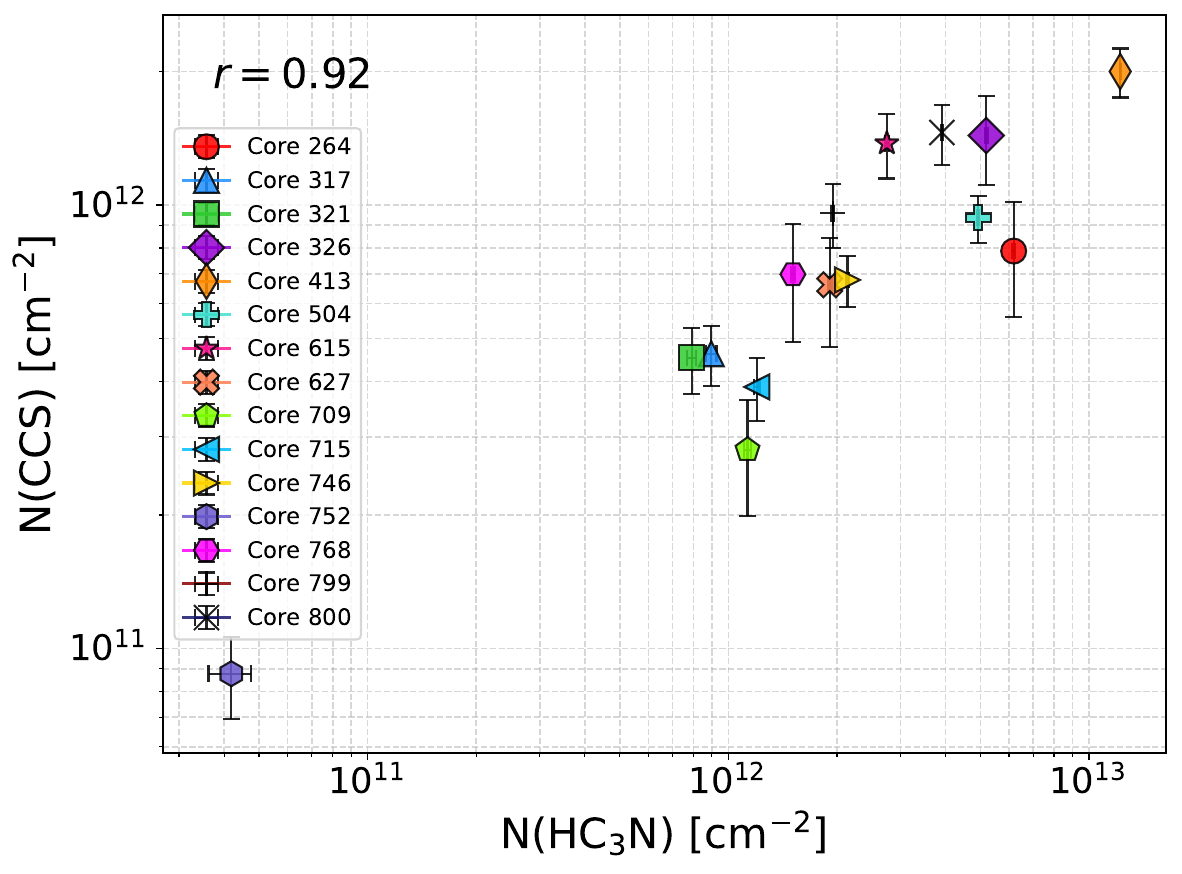}
    \caption{Comparison of HC$_3$N and CCS column densities, in which we calculate the Pearson rank correlation coefficient `$r$' between these two species. HC$_3$N and CCS column density errors range between 0.3-13\% and 12.2-29.7\%, respectively.} \label{fig:ccs_hc3n}
\end{figure}

\begin{figure} 
    \centering
    \includegraphics[width=8.5cm]{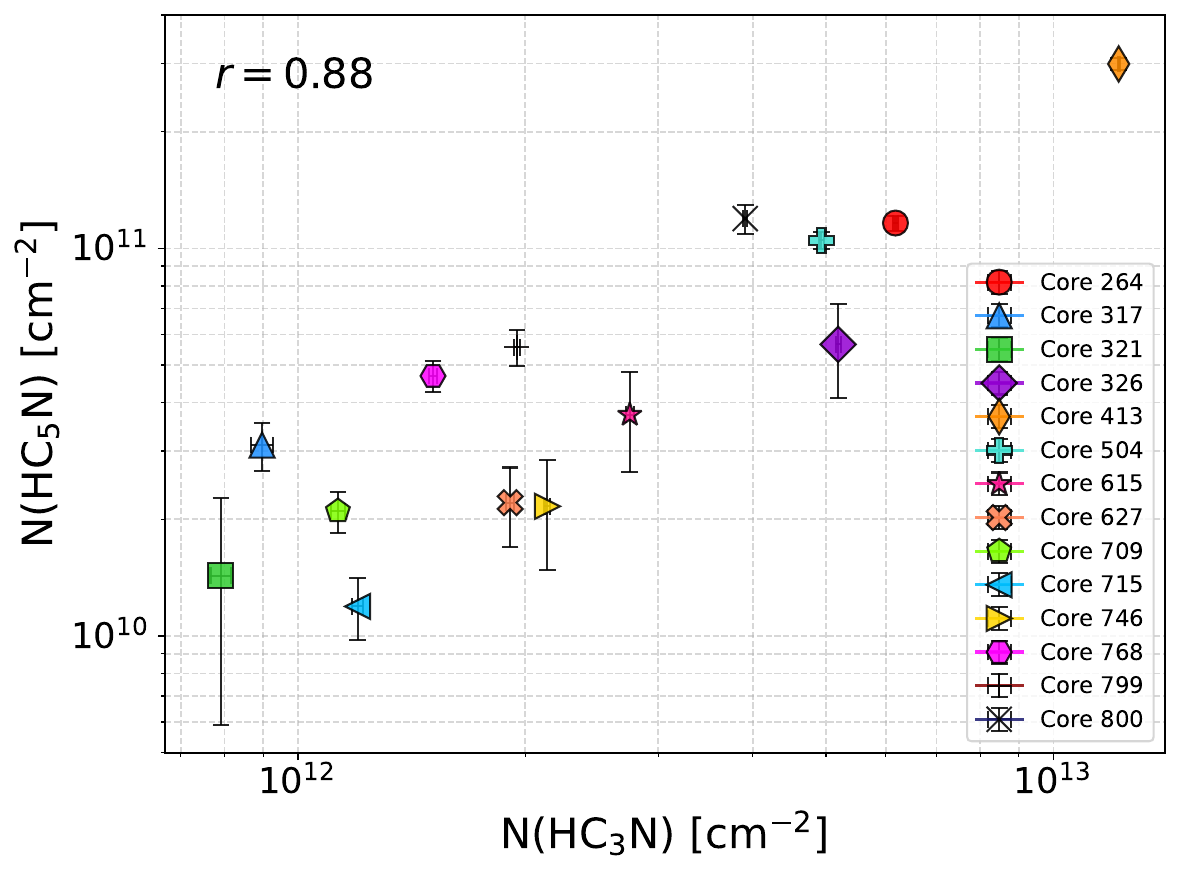}
    \caption{Comparison of HC$_3$N and HC$_5$N column densities, in which we calculate the Pearson rank correlation coefficient `$r$' between these two species. HC$_3$N and HC$_5$N column density errors range between 0.3-13\% and 3.5-31.7\%, respectively.} \label{fig:hc3n_hc5n}
\end{figure}

\subsection{Carbon-Chain Formation Pathways} \label{sec:4.5}
In Figure \ref{fig:ccs_hc3n}, we plot the column densities of HC$_3$N and CCS to provide a comparison and identify the correlation between these two species. We use the same `$r$' coefficient correlation technique, described in Section \ref{abundance_corr}. We observe a tight, positive correlation ($r = 0.92$) between CCS and HC$_3$N column densities in our Perseus sample (Figure \ref{fig:ccs_hc3n}), indicating that these two species might share closely linked chemical pathways \citep{Petrie1996, Suzuki1992, Hirahara1992}. Although neither CCS ($r = -0.71$) nor HC$_3$N ($r = -0.63$) shows as tight of a correlation with \textit{n}(H$_2$) (Figures \ref{CnS_volume}, \ref{HCnN_volume}), their covariation suggests they may be enhanced or depleted together throughout prestellar evolution \citep{Suzuki1992}. This agrees with previous chemical model predictions that show that both CCS and HC$_3$N are most prominent in the early stages of core evolution and decline as cores evolve into protostars, with CCS generally depleting more rapidly than HC$_3$N \citep{Suzuki1992, Herbst1989, 1989ApJS...69..271H}. Although a statistical correlation does not inherently imply a common chemical origin \citep{Belloche2020}, the potentially linked behavior of these two molecules reinforces their role as early-stage tracers, while also highlighting that their relative timescales of depletion may provide information about prestellar core age and surrounding physical conditions.

Cyanopolyynes can form efficiently at low temperatures ($<$ 10 K; \citealt{Seki1996}) via reactions between hydrocarbon ions and a nitrogen atom, followed by electron recombination reactions for HC$_5$N and HC$_7$N and neutral-neutral reactions for HC$_3$N \citep{1998A&A...329.1156T, Taniguchi2016}. In cold, dark cloud conditions, carbon-chain growth is further driven by reactions involving atomic carbon C and C$^+$ with unsaturated hydrocarbons, while barrierless neutral-neutral reactions involving radicals such as CN also play a dominant role at $\sim 10$ K \citep{2014MNRAS.437..930L}. Observations of single-substituted HC$_3$N isotopologues in TMC-1 reveal unequal relative abundances of H$^{13}$CCCN, HC$^{13}$CCN, and HCC$^{13}$CN (1.0:1.1:1.6) in \cite{2024A&A...682L..12T} and similar ratios in \citep{1998A&A...329.1156T}, demonstrating that $^{13}$C fractionation occurs during the molecule’s formation rather than after \citep{Taniguchi2024}. In this work, we derive different ratios for individual cores where all three single-substituted isotopologues are detected: 2.7:1.0:1.0 for core 413, 2.1:1.0:1.3 for core 504, and 1.9:1.0:1.9 for core 800. Here, H$^{13}$CCCN, HC$^{13}$CCN, and HCC$^{13}$CN have varying relative abundances from each other, which is different from what recent observations have found towards TMC-1 \citep{1998A&A...329.1156T, 2024A&A...682L..12T}, L1527 \citep{2016ApJ...833..291A}, and L1521B \citep{2017ApJ...846...46T}; therefore, $^{13}$C enrichment might not be as sensitive to the positional fractionation of HC$_3$N. However, it is important to note the caveat that our ratios are based on only two transitions, whereas ratios from \cite{2024A&A...682L..12T} are based on 12 different transitions. \cite{Giani2025} detected one transition from each of the three HC$_3$N isotopologues toward the prestellar core L1544 in the Taurus Molecular Cloud, where they find a relative intensity ratio of 0.8:1.0:1.2 for H$^{13}$CCCN, HC$^{13}$CCN, and HCC$^{13}$CN, respectively. The observations toward L1544 show more uniform isotopologue ratios with no strong $^{13}$C enrichment of HC$_3$N, suggesting that significant $^{13}$C fractionation is likely dependent on the chemical and physical conditions of the source's environment \citep{Giani2025}. This still supports a neutral-neutral chemistry pathway, since the carbon atoms from the precursor molecules are preserved into the product of HC$_3$N. The reaction C$_2$H$_2$ + CN $\to$ HC$_3$N + H has been proposed as the dominant formation route \citep{1990A&A...231..466M}, and \textit{ab initio} calculations confirm that it proceeds without an activation barrier at $\sim$ 10 K \citep{1996ApJ...465..795W, 1997ApJ...477..204W, Fukuzawa1997, VidalVidal2017}. 

Laboratory experiments and chemical models further demonstrate that several reactions of CN with hydrocarbon radicals exhibit increasing rate coefficients with decreasing temperature, supporting the efficiency of longer cyanopolyyne growth and formation under cold prestellar conditions \citep{Sims1993, Cherchneff1993, Balucani2000}. Although, updated reaction networks suggest that the formation of longer chains becomes progressively less efficient and more susceptible to destruction \citep{2014MNRAS.437..930L}. In particular, cyanopolyyne abundances exhibit a strong time dependence, increasing and reaching peak values at $\sim 10^3 -10^5$ yr before declining as destruction pathways become more efficient. This timing is sensitive to the available reservoir of atomic carbon, which plays a dual role in both the growth and destruction of cyanopolyynes \citep{2014MNRAS.437..930L}. Early in the core's chemical evolution, atomic carbon facilitates the rapid build-up of cyanopolyynes through formation reactions. However, once HC$_n$N species become abundant, atomic carbon transitions into a primary destruction pathway \citep{2014MNRAS.437..930L}. Updated reaction networks from these studies suggest that as CO freeze-out progresses, the resulting depletion of gas-phase atomic carbon may help the survival of long chains by removing this major destruction pathway. This allows species like HC$_3$N and HC$_5$N to persist and peak at later times. The observed abundances are not just a reflection of formation efficiency, but a balance governed by the chemical age of the gas and the elemental C/O ratio.

Two broad classes of pathways have been proposed for the formation of CCS and CCCS: ion-molecule and neutral-neutral reaction chemistry. Early work emphasizes ion-molecule reactions of S$^+$  with small hydrocarbons such as C$_2$H, c-C$_3$H, and c-C$_3$H$_2$. These pathways remain viable and efficient for producing CCS and are consistent with the detection of these hydrocarbons in TMC-1 \citep{Suzuki1988,1988A&A...200..191S, 1990A&A...231..466M}. However, extending the same ion-molecule chemistry to CCCS becomes problematic as the key reactions either have activation barriers that are prohibitive at $\sim$10 K or depend on precursor species that are too scarce \citep{Petrie1996, Sakai2007}. Further investigation of C$_n$S formation routes were accomplished by \cite{Sakai2007}, where the authors dismiss ion-molecule pathways that include reactions with C$_2$H$_2$, C$_2$H, C$_2$S$^+$, and C$_2$. Their study reports different abundances of $^{13}$CCS and C$^{13}$CS, indicating that the two carbon atoms in CCS are nonequivalent in the main production pathway and therefore need to originate from different precursor species. They follow this same reasoning for the production of CCCS, which leads to the discussion of neutral-neutral pathways as the more probable main production pathway for both CCS and CCCS. More recent laboratory and theoretical studies \citep{Yamada2002, Sakai2007} demonstrate that reactions involving radicals, in particular CH + CS $\to$ CCS + H and CH + CCS $\to$ CCCS + H, are exothermic, activation energy barrier-free \citep{1988A&A...200..191S, Smith1995_IJMSIP_149_231,Clary1993, Rowe1993}, and efficient in cold prestellar conditions. Although these pathways are often grouped under the umbrella term `neutral-neutral' chemistry, they are chemically described as radical-neutral since CH is a radical. The current consensus is that CCS and CCCS are primarily formed through the `neutral-neutral' reactions, while ion-molecule chemistry may still play a secondary role in producing CCS. However, further observations of hydrocarbons and CCCS are necessary to make a stronger statement on the primary production pathway for CCS.

Although we do not detect $^{13}$CCS in our study, we calculate the CCS/C$^{13}$CS abundance ratio using column densities since we lack sufficient C$^{13}$CS transitions to derive an intensity ratio corrected for hyperfine components. We find CCS/C$^{13}$CS ratios of 32.7, 1.7, 1.2, and 7.2 for cores 413, 504, 768, and 800, respectively. While the ratios in cores 504, 768, and 800 are a couple of factors smaller than TMC-1 literature values, the ratio found in core 413 is similar to standard interstellar values derived for TMC-1 ($54 \pm 5$) and L1521E ($51 \pm 13$) by \cite{Sakai2007}.

The strong correlation observed between CCS and HC$_3$N column densities and abundances in this study of prestellar/starless cores is similarly seen both in observations of TMC-1 and other Taurus prestellar/starless sources \citep{Hirahara1992, Suzuki1992} and pseudo time-dependent chemical model calculations \citep{Suzuki1992}. This observed correlation likely points to a shared reliance on gas-phase precursor species rather than a direct chemical reaction between the two families. Both C$_n$S and HC$_n$N formation are fueled by the same reservoir of unsaturated hydrocarbons (e.g., C$_2$H and C$_2$H$_2$), reactive radicals (e.g., CH and CN), and atomic carbon \citep{1990A&A...231..466M,Yamada2002, Sakai2007, Giani2025}. Because grain-surface formation for these species is considered unlikely due to the rapid hydrogenation of carbon-chains by accreting hydrogen atoms \citep{Giani2025}, their covariation is likely a signature of gas-phase chemistry regulated by the local C/O ratio. 

The potential chemical link is further evidenced by the similarity in their dominant neutral-neutral (or radical-neutral) pathways  \citep{Petrie1996}. For example, the primary route for HC$_3$N (C$_2$H$_2$ + CN $\to$ HC$_3$N + H) might be analogous to the proposed CCS formation route (CH + CS $\to$ CCS + H), where a simple radical reacts with a neutral species to extend the carbon-chain without an activation barrier. This shared chemistry offers a rationale for the consistent column density ratios throughout Perseus and Taurus prestellar/starless and protostellar cores (Table \ref{cd_ratios}) and their strongly correlated column densities (Figure \ref{fig:ccs_hc3n}). While these statistical trends suggest that C$_n$S and HC$_n$N are chemically coupled through their precursor species, the primary formation routes need to be further constrained, particularly for CCS. Higher spatial resolution observations of the distribution of the C$_n$S family and its specific hydrocarbon precursors would be of great interest to further illuminate these pathways \citep{Taniguchi2024}.

In Figure \ref{fig:hc3n_hc5n}, we observe a tight positive correlation between HC$_3$N and HC$_5$N column densities ($r = 0.88$). The persistence of HC$_3$N to higher volume densities, coupled with the correlated but lower abundances of HC$_5$N, suggests that HC$_5$N production is regulated by the availability of HC$_3$N and C$_2$-bearing precursors. The relationship of HC$_3$N/HC$_5$N $>$ 1 (see Table \ref{cd_ratios}) is consistent with expectations from carbon-chain chemistry, in which longer cyanopolyynes such as HC$_5$N are generally less abundant than shorter ones like HC$_3$N \citep{1981A&A....99..239B, Suzuki1992,1996Ap&SS.240...13W,Cernicharo2020}. However, the HC$_3$N/HC$_5$N ratios are still similar between cores, so we may not be able to gain much insight into the evolutionary stage of individual prestellar cores using this ratio.

\subsection{Deuterium Fractionation of HC$_3$N}

\begin{figure*} 
    \centering
    \includegraphics[width=\textwidth]{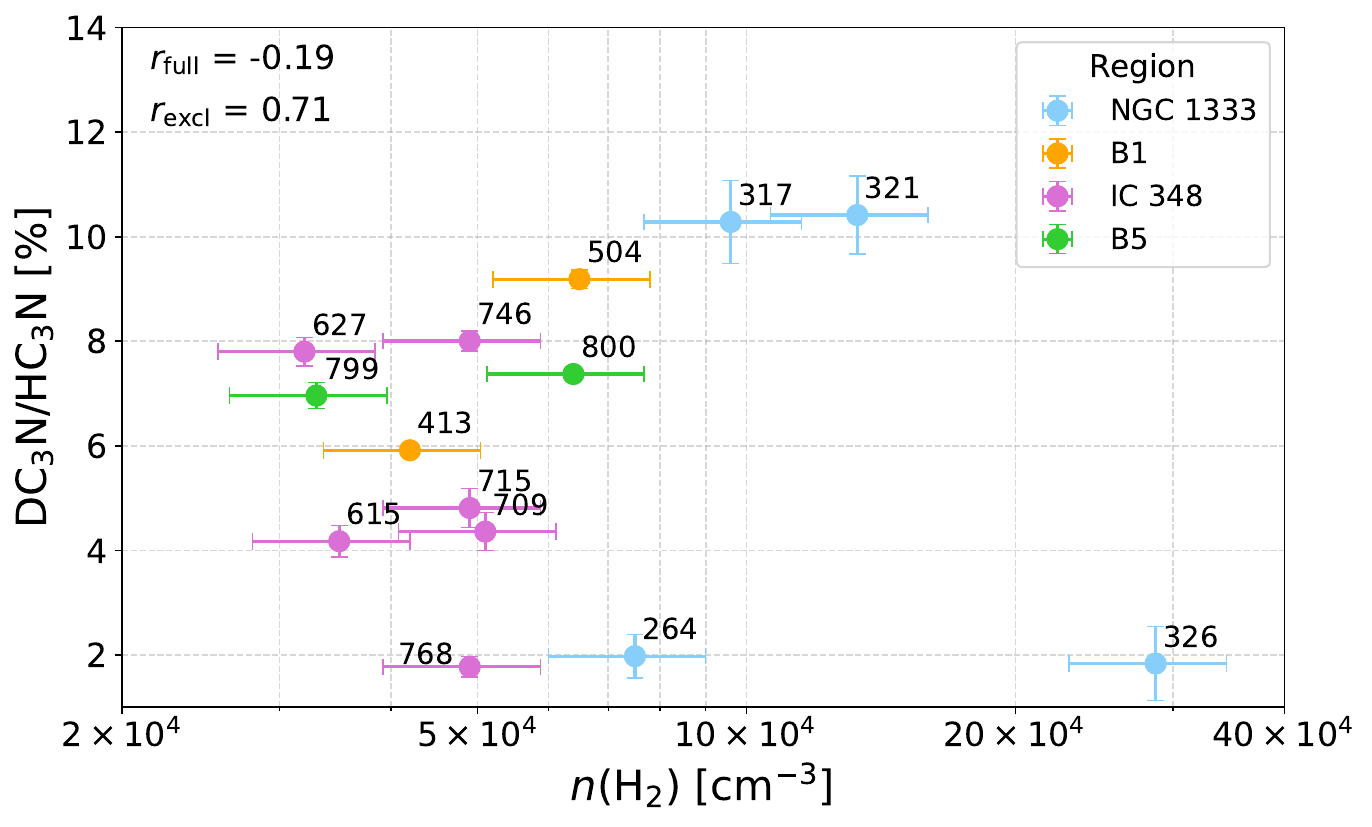}
    \caption{Comparison of deuterium fractionation (DC$_3$N/HC$_3$N) to the average $H_2$ volume density, \textit{n}($H_2$), for each of the cores in our sample, except core 752, where DC$_3$N was not detected. The Pearson rank correlation coefficient $r_{\rm full} = -0.19$ for the full sample of cores. However, when excluding outflow-impacted cores 264, 326, and 768, $r_{\rm excl} =$ 0.71, and the remaining cores appear to follow a positive trend where deuteration increases with \textit{n}($H_2$).}
    \label{deuterium_volume}
\end{figure*}

Astrochemists often use deuterated species and deuterium fractions to examine chemical inheritance from starless cores to protostellar evolution \citep{2014prpl.conf..859C}, as deuterated species are enhanced in cold and dense regions. At the cold temperatures of dense molecular clouds, neutral molecules like CO tend to freeze out, or deplete, onto the surfaces of dust grains. The depletion of CO and other neutral species provides less competition for reactions involving H$_2$D$^+$, allowing deuterated species to form more efficiently \citep{Caselli2002, Walmsley2004, Roueff2005, Crapsi2005, Crapsi2007}. Additionally, since deuterium is heavier, it gets preferentially incorporated into molecules at cold temperatures. Deuterium fractionation is thus quantified as the ratio of a deuterated isotopologue to its hydrogenated counterpart (e.g., DC$_3$N/HC$_3$N). However, it is worth noting that the observed deuterium fraction is not merely a function of current physical conditions, but it is sensitive to the gas density and temperature of a specific epoch of icy mantle formation \citep{Aikawa2012,Taquet2013}. Therefore, the D/H ratio of species like HC$_3$N might also reflect past thermal, density, and CO depletion histories, as chemistry is time-dependent.

As shown in Table \ref{deuterium_table}, the deuterium fraction ratios across Perseus cores span nearly an order of magnitude, ranging from as low as 1.8\% (core 326, 768) to as high as 10.4\% (core 321). Most cores fall within the 4 -- 9\% range, consistent with earlier observations of enhanced deuteration in cold, quiescent cores \citep{Howe1994}. The correlation between higher volume density and increased deuterium enrichment (Figure \ref{deuterium_volume}) supports the picture that deuteration is enhanced as cores evolve and CO depletion progresses. However, the notably low values in cores 264, 326, and 768 suggest that the local environment influences deuterium fractionation as 2/3 of these cores lie along the path of protostellar outflows from SVS13, IRAS 4A/B, and IRAS 2A/B (Figure \ref{cdmap}). Core 326 is a particularly unique case, located on an extended post-shocked SiO ridge of gas \citep{1998ApJ...504L.109L}, it is also the only core in the sample with phosphorus-bearing molecule detections (PN, PO, and PO$^{+}$; \citealt{2025ApJ...985L..25S}).
However, core 321 is also in close proximity to protostar SVS13, but does not seem to be in the direct path of its outflows \citep{2013ApJ...774...22P}, and its high deuterium fraction supports this. Core 768 similarly exhibits a lower D/H ratio than expected and is situated close to active protostars in the IC 348 region \citep{1999A&A...343..571G}. However, it is not as clear how directly the outflows and associated shocked gas impact this core specifically. In IC 348, cores 709 and 715 are also in close proximity to the path of the outflow jets from protostars HH 211 \citep{1999A&A...343..571G}, which seem to have slightly lower deuterium fractions than expected (Figure \ref{deuterium_volume}). Overall, the 5 potentially outflow-affected sources (cores 264, 326, 709, 715, and 768) are among the 5 lowest DC$_3$N/HC$_3$N percentages in Figure \ref{deuterium_volume}. These outflow-impacted regions, where the gas and dust in these cores are reheated, may suppress deuterium enrichment by driving CO back into the gas-phase, likely destroying the deuterium present \citep{Petrashkevich2024}. By contrast, cores 317, 321, and 504 exhibit unusually high deuterium fractions ($\geq$ 9$\%$), indicative of cold, dense conditions where CO depletion is advanced, and deuterium chemistry is enhanced. 

\cite{Asensio2026} study the singly- and doubly-deuterated forms of c-C$_3$H$_2$ towards the same subset of 15 prestellar/starless cores in the Perseus Molecular Cloud, reporting a statistically corrected average D/H ratio of 2.4\%. c-C$_3$HD was detected in all cores, except core 752, similar to our detection results of DC$_3$N. However, \cite{Asensio2026} find a tight correlation of $r=0.94$ between \textit{n}($H_2$) volume density and the c-C$_3$H$_2$ D/H ratio for all cores where c-C$_3$H$_2$ is detected. The prestellar/starless cores 264 and 326 situated near protostellar outflow activity exhibit corrected D/H ratios of 1.2\% and 9.2\%, respectively, suggesting that outflow-shocked gas does not appear to influence c-C$_3$H$_2$ deuteration chemistry to the same degree as HC$_3$N chemistry in this work. The divergence of deuteration enhancement levels in c-C$_3$H$_2$ and HC$_3$N likely indicates differing chemical formation pathways and reactivity across these particular carbon-chain species. However, this may also be influenced by the different timescales upon which these molecules respond to changes in the local physical environment, such as the sublimation of ices or variations in grain surface chemistry during the core's earlier, more diffuse stages \citep{Aikawa2012,Taquet2013}.

\begin{table*}
\centering
\caption{Deuterium fractionation of DC$_3$N/HC$_3$N compared between this work (Perseus prestellar/starless cores), Taurus prestellar/starless and protostellar sources, and a pre-cluster clump in the Serpens South cluster-forming region \citep{Howe1994, Turner2001, Saito2002, Sakai2009, Cordiner2011, 2020A&A...633A.118L, 2024PASJ...76.1270T}. The mean DC$_3$N/HC$_3$N ratios do not include non-detections. Errors are included in parentheses for each of the individual cores, Taurus sources, and Serpens South. The standard deviation of the sample is included instead of error values for each Perseus region (NGC 1333, B1, IC 348, B5).} \label{deuterium_table}
\begin{tblr}{
  cells = {c},
  hline{2,17} = {-}{},
}
\textbf{Cloud} & \textbf{Source} & $\mathbf{DC_3N/HC_3N}$ & \textbf{Stage} & \textbf{Reference}\\
\hline
Perseus & 264 & 1.98 (0.41) \% & prestellar/starless core & this work\\
 & 317 & 10.28 (0.72) \% & prestellar/starless core & this work\\
 & 321 & 10.42 (0.81) \% & prestellar/starless core & this work\\
 & 326 & 1.83 (0.71) \% & prestellar/starless core & this work\\
 & 413 & 5.91 (0.06) \% & prestellar/starless core & this work\\
 & 504 & 9.19 (0.14) \% & prestellar/starless core & this work\\
 & 615 & 4.18 (0.25) \% & prestellar/starless core & this work\\
 & 627 & 7.80 (0.24) \% & prestellar/starless core & this work\\
 & 709 & 4.36 (0.34) \% & prestellar/starless core & this work\\
 & 715 & 4.81 (0.34) \% & prestellar/starless core & this work\\
 & 746 & 8.01 (0.18) \% & prestellar/starless core & this work\\
 & 752 & -- & prestellar/starless core & this work\\
 & 768 & 1.77 (0.45) \% & prestellar/starless core & this work\\
 & 799 & 6.96 (0.22) \% & prestellar/starless core & this work\\
 & 800 & 7.37 (0.10) \% & prestellar/starless core & this work\\
Perseus & NGC 1333 & 6.13 (4.88) \% & prestellar/starless core & this work\\
 & B1 & 6.69 (2.32) \% & prestellar/starless core & this work\\
 & IC 348 & 5.16 (2.38) \% & prestellar/starless core & this work\\
 & B5 & 7.17 (0.29) \% & prestellar/starless core & this work\\
Taurus & L1544 & 6 \%; 1.5 -- 9.3 \% & prestellar core & \cite{Howe1994, 2020A&A...633A.118L} \\
 & L1527 & 3.1 (1.1) -- 3.5 (1.0) \% & Class 0 protostar & \cite{Sakai2009}\\
 & Cha-MMS1 & 3.6 (1.5) \% & Class 0 protostar & \cite{Cordiner2011}\\
 & TMC-1 & 1 -- 6 \% & star-forming region & \cite{Turner2001, Saito2002}\\
Serpens & Serpens South & 1.72 (0.12) \% & pre-cluster clump & \cite{2024PASJ...76.1270T}
\end{tblr}
\end{table*}

When comparing to other regions, the Perseus prestellar/starless cores (1.8 -- 2.4\%) show deuterium fractions that are consistent with values measured in well-studied prestellar cores elsewhere. For example, the prestellar core L1544 in Taurus exhibits a DC$_3$N/HC$_3$N ratio of $\sim$6\% \citep{Howe1994} and 1.5 -- 9.3\% (depending on assumed T$_\mathrm{ex}$ of 5 K and 10 K; \citealt{2020A&A...633A.118L}), similar to the Perseus median value (6.44\%), whereas the protostellar sources L1527 \citep{Sakai2009} and Cha-MMS1 \citep{Cordiner2011} show slightly lower ratios (3 -- 3.6\%), reflecting partial destruction of deuterated species once heating and active accretion begin. The range of deuterium fractions reported for TMC-1 (1 -- 6\%; \citealt{Langer1980, Howe1994,Turner2001, Saito2002}) also falls within the spread observed in Perseus cores, although the highest Perseus values exceed those typically found in TMC-1 (see Table \ref{deuterium_table}). The DC$_3$N/HC$_3$N ratio for the Serpens South pre-cluster clump is 1.72\% \citep{2024PASJ...76.1270T}, comparable to the lower bound reported for TMC-1 and L1544. Overall, Perseus prestellar/starless cores exhibit the highest levels of deuteration among these three molecular clouds, and the Serpens clump shows the lowest degree of deuterium fractionation (Table \ref{deuterium_table}). Collectively, these comparisons suggest that DC$_3$N/HC$_3$N is a sensitive probe of the local physical conditions associated with different core evolutionary stages. While the D/H ratio is often used as an evolutionary tracer, the observed degree of deuteration may also reflect the thermal and chemical histories of the core \citep{Taquet2013}.

Ultimately, the variation in deuterium fractionation observed in these 15 Perseus prestellar/starless cores reinforces the strong influence of local physical conditions, such as outflow feedback, density, and temperature, on molecular chemistry. The suppressed ratios in outflow-impacted cores demonstrate how local feedback processes can disrupt the positive correlation between volume density and deuterium fractionation, while the high fractions in the cold, quiescent cores reveal that deuteration is otherwise a sensitive indicator of its local physical environment and possibly its evolutionary state. Together, these results emphasize the usefulness of DC$_3$N as a probe of prestellar evolution and the chemical inheritance pathways that may be imprinted on future planetary systems.

\section{Conclusions} \label{sec:5}
We find a myriad of carbon-chain species in starless and prestellar cores in the Perseus Molecular Cloud. From a Yebes 40m line survey, we detect CCS, CCCS, HC$_3$N, DC$_3$N, HC$_5$N, and HC$_7$N in 15/15 (100\%), 14/15 (93.3\%), 15/15 (100\%), 14/15 (93.3\%), 14/15 (93.3\%), and 4/15 (26.7\%) of our sample of 15 cores. We also detect a variety of isotopologues and related species within the HC$_n$N and C$_n$S families: $^{13}$CS, C$^{34}$S, CS, C$^{13}$CS, CC$^{34}$S, H$^{13}$CCCN, HC$^{13}$CCN, HCC$^{13}$CN, HC$^{13}$CCCCN, HCC$^{13}$CCCN, HCCC$^{13}$CCN, HCCCCC$^{15}$N, and DCCCCCN are detected in 15/15 (100\%), 15/15 (100\%), 15/15 (100\%), 4/15 (26.7\%), 7/15 (46.7\%), 3/15 (20\%), 4/15 (26.7\%), 6/15 (40\%), 1/15 (6.7\%), 3/15 (20\%), 1/15 (6.7\%),  1/15 (6.7\%), and 1/15 (6.7\%) of the sample. From the analysis of these results, we conclude the following:
\begin{enumerate}
    \item Core 413 contains a diverse and rich inventory of complex carbon-chain chemistry with 32 unique molecular detections, making it a chemically interesting source for future study. In sharp contrast, core 752's sparse budget of carbon-chains along with its lower volume density (\textit{n}(H$_2$) $=$ 0.22$\times$10$^5$ cm$^{-3}$), indicates that its carbon-chain inventory is likely governed by its distinct surrounding physical environment or its evolutionary stage.
    \item The abundances for most species, especially CCCS, HC$_3$N, DC$_3$N, and HC$_5$N, span several orders of magnitude, suggesting carbon-chain chemistry's sensitivity to each core's unique surrounding environments. However, we observe a narrower dispersion of abundances between CS isotopologues, indicating that $^{13}$CS and C$^{34}$S species are comparatively independent from their surrounding environmental conditions. This reinforces the concept that local environments impact the complex chemistry and carbon-chain reservoirs of prestellar and starless cores.
    \item Sulfur-bearing carbon-chains and cyanopolyynes (specifically CS, CCS, CCCS, HC$_3$N, and HC$_5$N) are preferentially enhanced in early-prestellar phase conditions and decline in abundance with core density. These results likely reflect that the Yebes 40m beam encompasses both the dense core and its surrounding envelope, such that chemical depletion in the central regions of the higher-density cores reduces abundances relative to early-phase cores. 
    \item A narrow span of CCS/HC$_3$N ratios across the core sample and a tight positive correlation between CCS and HC$_3$N column densities unveil the potentially linked chemical formation networks between these species. The CCS/HC$_3$N ratio remains close to unity across both Taurus and Perseus sources from the prestellar to protostellar stage, further supporting the rationale of similar formation histories.
    \item Perseus cores 264, 326, and likely also core 768, are impacted by nearby protostellar outflows and exhibit suppressed deuteration, which emphasizes the significant influence of local environmental conditions on deuterium chemistry. Shocked regions may reduce deuterium fractionation of HC$_3$N by reheating gas and dust, which could influence the carbon-chain inheritance of deuterated species into planetary systems. 
\end{enumerate}

\vspace{-6mm}
\section*{Acknowledgments}
We thank the anonymous reviewer for their constructive comments and feedback. We thank the Spanish Geographic Institute (IGN) for their operation in the Yebes 40m radio telescope (22A022 and 23A025), which provided our observations. Paula Tarrío and Alba Vidal García carried out the observations and the first inspection of the data quality. The 40m radio telescope at Yebes Observatory
is operated by the Spanish Geographic Institute (IGN; Ministerio
de Transportes, Movilidad y Agenda Urbana). The Yebes Observatory
thanks the European Union’s Horizon 2020 research and innovation
program for funding support to ORP project under grant agreement No 101004719.

AP-Y thanks the National Radio Astronomy Observatory (NRAO) for the opportunity and resources to conduct research under the NRAO Research Experience for Undergraduates (REU) 2024 Summer Program. AP-Y and SS acknowledge that the National Radio Astronomy Observatory is a facility of the National Science Foundation operated under cooperative agreement by Associated Universities, Inc. 

JFA thanks RIKEN Special Postdoctoral Researcher Program (Fellowships) for financial support. Support for YS was provided by National Science Foundation Astronomy and Astrophysics Grant (AAG) AST-2205474. 

IJ-S and AM acknowledge funding from the European Research Council (ERC) grant OPENS (GA No. 101125858) funded by the European Union, and from grant PID2022-136814NB-I00 funded by the Spanish Ministry of Science, Innovation and Universities/State Agency of Research MICIU/AEI/ 10.13039/501100011033 and by ERDF/EU. Views and opinions expressed are however those of the authors only and do not necessarily reflect those of the European Union or the European Research Council Executive Agency. Neither the European Union nor the granting authority can be held responsible for them.

Software: APLpy \citep{2012ascl.soft08017R}, astropy \citep{2013A&A...558A..33A}, Matplotlib \citep{2007CSE.....9...90H}, NumPy \citep{2020arXiv200610256H}, Pandas \citep{reback2020pandas}, Pyspeckit \citep{2011ascl.soft09001G, Ginsburg2022}, SciPy \citep{Virtanen2020}.

\vspace{-6mm}

\section*{Data Availability}
The reduced spectra in the $T_{\rm mb}$ scale presented in this paper can be accessed through the Harvard Dataverse at the following link: \url{https://doi.org/10.7910/DVN/AN3EFQ}. The GILDAS-CLASS pipeline used for data reduction by Andrés Megías is available on GitHub: \url{https://github.com/andresmegias/gildas-class-pipeline/} \citep{2023MNRAS.519.1601M}.
The Python cold core molecular detection pipeline is also available on GitHub at the following link: \url{https://github.com/anissapy/Cold-Core-Detection-Pipeline}. 

\vspace{-6mm}


\bibliographystyle{mnras}
\bibliography{references_paper} 

@ARTICLE{1963JChPh..39.2856K,
       author = {{Kewley}, Roger and {Sastry}, K.~V.~L.~N. and {Winnewisser}, Manfred and {Gordy}, Walter},
        title = "{Millimeter Wave Spectroscopy of Unstable Molecular Species. I. Carbon Monosulfide}",
      journal = {\jcp},
         year = 1963,
        month = dec,
       volume = {39},
       number = {11},
        pages = {2856-2860},
          doi = {10.1063/1.1734116},
       adsurl = {https://ui.adsabs.harvard.edu/abs/1963JChPh..39.2856K}
}

@article{Cataldi2026, doi = {10.21105/joss.09233}, url = {https://doi.org/10.21105/joss.09233}, year = {2026}, publisher = {The Open Journal}, volume = {11}, number = {118}, pages = {9233}, author = {Cataldi, Gianni}, title = {pythonradex: a fast Python re-implementation of RADEX with extended functionality}, journal = {Journal of Open Source Software} }

@article{Aikawa2012,
  title = {FROM PRESTELLAR TO PROTOSTELLAR CORES. II. TIME DEPENDENCE AND DEUTERIUM FRACTIONATION},
  volume = {760},
  ISSN = {1538-4357},
  url = {http://dx.doi.org/10.1088/0004-637X/760/1/40},
  DOI = {10.1088/0004-637x/760/1/40},
  number = {1},
  journal = {The Astrophysical Journal},
  publisher = {American Astronomical Society},
  author = {Aikawa,  Y. and Wakelam,  V. and Hersant,  F. and Garrod,  R. T. and Herbst,  E.},
  year = {2012},
  month = Nov,
  pages = {40}
}

@article{Taquet2013,
  title = {WATER DEUTERIUM FRACTIONATION IN THE INNER REGIONS OF TWO SOLAR-TYPE PROTOSTARS},
  volume = {768},
  ISSN = {2041-8213},
  url = {http://dx.doi.org/10.1088/2041-8205/768/2/L29},
  DOI = {10.1088/2041-8205/768/2/l29},
  number = {2},
  journal = {The Astrophysical Journal},
  publisher = {American Astronomical Society},
  author = {Taquet,  V. and López-Sepulcre,  A. and Ceccarelli,  C. and Neri,  R. and Kahane,  C. and Coutens,  A. and Vastel,  C.},
  year = {2013},
  month = Apr,
  pages = {L29}
}

@article{Cernicharo2020,
  title = {Discovery of HC<sub>4</sub>NC in TMC-1: A study of the isomers of HC<sub>3</sub>N,  HC<sub>5</sub>N,  and HC<sub>7</sub>N},
  volume = {642},
  ISSN = {1432-0746},
  url = {http://dx.doi.org/10.1051/0004-6361/202039274},
  DOI = {10.1051/0004-6361/202039274},
  journal = {Astronomy &amp; Astrophysics},
  publisher = {EDP Sciences},
  author = {Cernicharo,  J. and Marcelino,  N. and Agúndez,  M. and Bermúdez,  C. and Cabezas,  C. and Tercero,  B. and Pardo,  J. R.},
  year = {2020},
  month = Oct,
  pages = {L8}
}

@article{Balucani2000,
  title = {Formation of Nitriles in the Interstellar Medium via Reactions of Cyano Radicals,  CN(<i>X</i> <sup>2</sup>Σ<sup>+</sup>),  with Unsaturated Hydrocarbons},
  volume = {545},
  ISSN = {1538-4357},
  url = {http://dx.doi.org/10.1086/317848},
  DOI = {10.1086/317848},
  number = {2},
  journal = {The Astrophysical Journal},
  publisher = {American Astronomical Society},
  author = {Balucani,  N. and Asvany,  O. and Huang,  L. C. L. and Lee,  Y. T. and Kaiser,  R. I. and Osamura,  Y. and Bettinger,  H. F.},
  year = {2000},
  month = Dec,
  pages = {892–906}
}

@article{Giani2025,
  title = {A comprehensive study of the gas-phase formation network of HC5N: theory,  experiments,  observations,  and models},
  volume = {537},
  ISSN = {1365-2966},
  url = {http://dx.doi.org/10.1093/mnras/staf189},
  DOI = {10.1093/mnras/staf189},
  number = {4},
  journal = {Monthly Notices of the Royal Astronomical Society},
  publisher = {Oxford University Press (OUP)},
  author = {Giani,  Lisa and Bianchi,  Eleonora and Fournier,  Martin and Cheikh Sid Ely,  Sidaty and Ceccarelli,  Cecilia and Rosi,  Marzio and Guillemin,  Jean-Claude and Sims,  Ian R and Balucani,  Nadia},
  year = {2025},
  month = Jan,
  pages = {3861–3883}
}

@ARTICLE{2014MNRAS.437..930L,
       author = {{Loison}, Jean-Christophe and {Wakelam}, Valentine and {Hickson}, Kevin M. and {Bergeat}, Astrid and {Mereau}, Raphael},
        title = "{The gas-phase chemistry of carbon chains in dark cloud chemical models}",
      journal = {\mnras},
         year = 2014,
        month = jan,
       volume = {437},
       number = {1},
        pages = {930-945},
          doi = {10.1093/mnras/stt1956},
archivePrefix = {arXiv},
       eprint = {1311.3865},
 primaryClass = {astro-ph.GA},
       adsurl = {https://ui.adsabs.harvard.edu/abs/2014MNRAS.437..930L}
}

@article{Redaelli2025,
  title = {Cosmic-ray ionisation rate in low-mass cores: The role of the environment},
  volume = {702},
  ISSN = {1432-0746},
  url = {http://dx.doi.org/10.1051/0004-6361/202453198},
  DOI = {10.1051/0004-6361/202453198},
  journal = {Astronomy &amp; Astrophysics},
  publisher = {EDP Sciences},
  author = {Redaelli,  E. and Bovino,  S. and Sabatini,  G. and Arzoumanian,  D. and Padovani,  M. and Caselli,  P. and Wyrowski,  F. and Pineda,  J. E. and Latrille,  G.},
  year = {2025},
  month = Oct,
  pages = {A210}
}

@article{Sims1993,
  title = {Rate constants for the reactions of CN with hydrocarbons at low and ultra-low temperatures},
  volume = {211},
  ISSN = {0009-2614},
  url = {http://dx.doi.org/10.1016/0009-2614(93)87091-G},
  DOI = {10.1016/0009-2614(93)87091-g},
  number = {4-5},
  journal = {Chemical Physics Letters},
  publisher = {Elsevier BV},
  author = {Sims,  Ian R. and Queffelec,  Jean-Louis and Travers,  Daniel and Rowe,  Bertrand R. and Herbert,  Lee B. and Karth\"{a}user,  Joachim and Smith,  Ian W.M.},
  year = {1993},
  month = Aug,
  pages = {461–468}
}

@article{VidalVidal2017,
  title = {Computational study of the hydrolysis of carbonyl sulphide: Thermodynamics and kinetic constants estimation using ab initio calculations},
  volume = {110},
  ISSN = {0021-9614},
  url = {http://dx.doi.org/10.1016/j.jct.2017.03.003},
  DOI = {10.1016/j.jct.2017.03.003},
  journal = {The Journal of Chemical Thermodynamics},
  publisher = {Elsevier BV},
  author = {Vidal-Vidal,  A. and Pérez-Rodríguez,  M. and Piñeiro,  M.M.},
  year = {2017},
  month = July,
  pages = {154–161}
}

@article{Sakai2013,
  title = {Warm Carbon-Chain Chemistry},
  volume = {113},
  ISSN = {1520-6890},
  url = {http://dx.doi.org/10.1021/cr4001308},
  DOI = {10.1021/cr4001308},
  number = {12},
  journal = {Chemical Reviews},
  publisher = {American Chemical Society (ACS)},
  author = {Sakai,  Nami and Yamamoto,  Satoshi},
  year = {2013},
  month = Oct,
  pages = {8981–9015}
}

@article{Sipil2013,
  title = {HD depletion in starless cores},
  volume = {554},
  ISSN = {1432-0746},
  url = {http://dx.doi.org/10.1051/0004-6361/201220922},
  DOI = {10.1051/0004-6361/201220922},
  journal = {Astronomy &amp; Astrophysics},
  publisher = {EDP Sciences},
  author = {Sipil\"{a},  O. and Caselli,  P. and Harju,  J.},
  year = {2013},
  month = june,
  pages = {A92}
}

@ARTICLE{1987ApJ...317L.111K,
       author = {{Kaifu}, Norio and {Suzuki}, Hiroko and {Ohishi}, Masatoshi and {Miyaji}, Takeshi and {Ishikawa}, Shin-Ichi and {Kasuga}, Takashi and {Morimoto}, Masaki and {Saito}, Shuji},
        title = "{Detection of Intense Unidentified Lines in TMC-1}",
      journal = {\apjl},
         year = 1987,
        month = jun,
       volume = {317},
        pages = {L111},
          doi = {10.1086/184922},
       adsurl = {https://ui.adsabs.harvard.edu/abs/1987ApJ...317L.111K}
}

@ARTICLE{1987ApJ...317L.115S,
       author = {{Saito}, Shuji and {Kawaguchi}, Kentarou and {Yamamoto}, Satoshi and {Ohishi}, Masatoshi and {Suzuki}, Hiroko and {Kaifu}, N.},
        title = "{Laboratory Detection and Astronomical Identification of a New Free Radical, CCS( 3 Sigma -)}",
      journal = {\apjl},
         year = 1987,
        month = jun,
       volume = {317},
        pages = {L115},
          doi = {10.1086/184923},
       adsurl = {https://ui.adsabs.harvard.edu/abs/1987ApJ...317L.115S}
}

@article{Willacy1998,
  title = {Dust Emission and Molecular Depletion in L1498},
  volume = {507},
  ISSN = {0004-637X},
  url = {http://dx.doi.org/10.1086/311695},
  DOI = {10.1086/311695},
  number = {2},
  journal = {The Astrophysical Journal},
  publisher = {American Astronomical Society},
  author = {Willacy,  K. and Langer,  W. D. and Velusamy,  T.},
  year = {1998},
  month = nov,
  pages = {L171–L175}
}

@ARTICLE{2016ApJ...833..291A,
       author = {{Araki}, Mitsunori and {Takano}, Shuro and {Sakai}, Nami and {Yamamoto}, Satoshi and {Oyama}, Takahiro and {Kuze}, Nobuhiko and {Tsukiyama}, Koichi},
        title = "{Precise Observations of the $^{12}$C/$^{13}$C Ratios of HC$_{3}$N in the Low-mass Star-forming Region L1527}",
      journal = {\apj},
         year = 2016,
        month = dec,
       volume = {833},
       number = {2},
          eid = {291},
        pages = {291},
          doi = {10.3847/1538-4357/833/2/291},
archivePrefix = {arXiv},
       eprint = {1610.02793},
 primaryClass = {astro-ph.SR},
       adsurl = {https://ui.adsabs.harvard.edu/abs/2016ApJ...833..291A}
}

@ARTICLE{2017ApJ...846...46T,
       author = {{Taniguchi}, Kotomi and {Ozeki}, Hiroyuki and {Saito}, Masao},
        title = "{$^{13}$C Isotopic Fractionation of HC$_{3}$N in Two Starless Cores: L1521B and L134N (L183)}",
      journal = {\apj},
         year = 2017,
        month = sep,
       volume = {846},
       number = {1},
          eid = {46},
        pages = {46},
          doi = {10.3847/1538-4357/aa82ba},
archivePrefix = {arXiv},
       eprint = {1707.08267},
 primaryClass = {astro-ph.GA},
       adsurl = {https://ui.adsabs.harvard.edu/abs/2017ApJ...846...46T}
}

@ARTICLE{2024A&A...682L..12T,
       author = {{Tercero}, B. and {Marcelino}, N. and {Roueff}, E. and {Ag{\'u}ndez}, M. and {Cabezas}, C. and {Fuentetaja}, R. and {de Vicente}, P. and {Cernicharo}, J.},
        title = "{Doubly substituted isotopologues of Hcccn in TMC-1: Detection of D$^{13}$CCCN, DC$^{13}$CCN, DCC$^{13}$CN, DCCC$^{15}$N, H$^{13}$C$^{13}$CCN, H$^{13}$CC$^{13}$CN, HC$^{13}$C$^{13}$CN, HCC$^{13}$C$^{15}$N, and HC$^{13}$CC$^{15}$N}",
      journal = {\aap},
         year = 2024,
        month = feb,
       volume = {682},
          eid = {L12},
        pages = {L12},
          doi = {10.1051/0004-6361/202348929},
archivePrefix = {arXiv},
       eprint = {2402.01318},
 primaryClass = {astro-ph.GA},
       adsurl = {https://ui.adsabs.harvard.edu/abs/2024A&A...682L..12T}
}

@ARTICLE{1955PhRv...98.1837M,
       author = {{Mockler}, Richard C. and {Bird}, George R.},
        title = "{Microwave Spectrum of Carbon Monsulfide}",
      journal = {Physical Review},
         year = 1955,
        month = jun,
       volume = {98},
       number = {6},
        pages = {1837-1839},
          doi = {10.1103/PhysRev.98.1837},
       adsurl = {https://ui.adsabs.harvard.edu/abs/1955PhRv...98.1837M}
}

@ARTICLE{1982JMoSp..95...35B,
       author = {{Bogey}, M. and {Demuynck}, C. and {Destombes}, J.~L.},
        title = "{Millimeter and submillimeter wave spectrum of CS $^{1}${\ensuremath{\Sigma}} in high vibrational states: Isotopic dependence of Dunham coefficients}",
      journal = {Journal of Molecular Spectroscopy},
         year = 1982,
        month = sep,
       volume = {95},
       number = {1},
        pages = {35-42},
          doi = {10.1016/0022-2852(82)90234-X},
       adsurl = {https://ui.adsabs.harvard.edu/abs/1982JMoSp..95...35B}
}

@ARTICLE{1968JMoSp..28..266W,
       author = {{Winnewisser}, Gisbert and {Cook}, Robert L.},
        title = "{The dipole moment of carbon monosulfide}",
      journal = {Journal of Molecular Spectroscopy},
         year = 1968,
        month = oct,
       volume = {28},
       number = {2},
        pages = {266-268},
          doi = {10.1016/0022-2852(68)90011-8},
       adsurl = {https://ui.adsabs.harvard.edu/abs/1968JMoSp..28..266W}
}

@ARTICLE{1990ApJ...361..318Y,
       author = {{Yamamoto}, Satoshi and {Saito}, Shuji and {Kawaguchi}, Kentarou and {Chikada}, Yoshihiro and {Suzuki}, Hiroko and {Kaifu}, Norio and {Ishikawa}, Shin-Ichi and {Ohishi}, Masatoshi},
        title = "{Rotational Spectrum of the CCS Radical Studied by Laboratory Microwave Spectroscopy and Radio-astronomical Observations}",
      journal = {\apj},
         year = 1990,
        month = sep,
       volume = {361},
        pages = {318},
          doi = {10.1086/169197},
       adsurl = {https://ui.adsabs.harvard.edu/abs/1990ApJ...361..318Y}
}

@ARTICLE{1987ApJ...317L.119Y,
       author = {{Yamamoto}, Satoshi and {Saito}, Shuji and {Kawaguchi}, Kentarou and {Kaifu}, Norio and {Suzuki}, Hiroko and {Ohishi}, Masatoshi},
        title = "{Laboratory Detection of a New Carbon-Chain Molecule C 3S and Its Astronomical Identification}",
      journal = {\apjl},
         year = 1987,
        month = jun,
       volume = {317},
        pages = {L119},
          doi = {10.1086/184924},
       adsurl = {https://ui.adsabs.harvard.edu/abs/1987ApJ...317L.119Y}
}

@ARTICLE{1976JMoSp..62..175A,
       author = {{Alexander}, A.~J. and {Kroto}, H.~W. and {Walton}, D.~R.~M.},
        title = "{The microwave spectrum, substitution structure and dipole moment of cyanobutadiyne, HCCCCCN}",
      journal = {Journal of Molecular Spectroscopy},
         year = 1976,
        month = aug,
       volume = {62},
       number = {2},
        pages = {175-180},
          doi = {10.1016/0022-2852(76)90347-7},
       adsurl = {https://ui.adsabs.harvard.edu/abs/1976JMoSp..62..175A}
}

@ARTICLE{1980MNRAS.192..651R,
       author = {{Rodriguez}, L.~F. and {Chaisson}, E.~J.},
        title = "{Observations of HC5N and HC7N in SGR B2 and Cloud 2}",
      journal = {\mnras},
         year = 1980,
        month = sep,
       volume = {192},
        pages = {651-658},
          doi = {10.1093/mnras/192.4.651},
       adsurl = {https://ui.adsabs.harvard.edu/abs/1980MNRAS.192..651R}
}

@ARTICLE{1981A&A....95..143T,
       author = {{Toelle}, F. and {Ungerechts}, H. and {Walmsley}, C.~M. and {Winnewisser}, G. and {Churchwell}, E.},
        title = "{A molecular line study of the elongated daark dust cloud TMC 1.}",
      journal = {\aap},
         year = 1981,
        month = feb,
       volume = {95},
        pages = {143-155},
       adsurl = {https://ui.adsabs.harvard.edu/abs/1981A&A....95..143T}
}

@ARTICLE{1988A&A...196..194O,
       author = {{Olano}, C.~A. and {Walmsley}, C.~M. and {Wilson}, T.~L.},
        title = "{The relative distribution of NH3, HC7N and C4H in the Taurus Molecular Cloud 1 (TMC 1).}",
      journal = {\aap},
         year = 1988,
        month = may,
       volume = {196},
        pages = {194-200},
       adsurl = {https://ui.adsabs.harvard.edu/abs/1988A&A...196..194O}
}

@ARTICLE{1988A&A...199..291T,
       author = {{Truong-Bach} and {Graham}, D. and {Rieu}, N.~Q.},
        title = "{Observations of NH3, HC5N and HC7N toward AFGL 2688.}",
      journal = {\aap},
         year = 1988,
        month = jun,
       volume = {199},
        pages = {291-298},
       adsurl = {https://ui.adsabs.harvard.edu/abs/1988A&A...199..291T}
}

@ARTICLE{1988ApJ...334..771V,
       author = {{van Dishoeck}, Ewine F. and {Black}, John H.},
        title = "{The Photodissociation and Chemistry of Interstellar CO}",
      journal = {\apj},
         year = 1988,
        month = nov,
       volume = {334},
        pages = {771},
          doi = {10.1086/166877},
       adsurl = {https://ui.adsabs.harvard.edu/abs/1988ApJ...334..771V}
}

@article{Kim2020,
  title = {CS Depletion in Prestellar Cores},
  volume = {891},
  ISSN = {1538-4357},
  url = {http://dx.doi.org/10.3847/1538-4357/ab774d},
  DOI = {10.3847/1538-4357/ab774d},
  number = {2},
  journal = {The Astrophysical Journal},
  publisher = {American Astronomical Society},
  author = {Kim,  Shinyoung and Lee,  Chang Won and Gopinathan,  Maheswar and Tafalla,  Mario and Sohn,  Jungjoo and Kim,  Gwanjeong and Kim,  Mi-Ryang and Soam,  Archana and Myers,  Philip C.},
  year = {2020},
  month = mar,
  pages = {169}
}

@article{DenisAlpizar2018,
  title = {New ratecoefficients of CS in collision with para- and ortho-H2 and astrophysical implications},
  volume = {478},
  ISSN = {1365-2966},
  url = {http://dx.doi.org/10.1093/mnras/sty1177},
  DOI = {10.1093/mnras/sty1177},
  number = {2},
  journal = {Monthly Notices of the Royal Astronomical Society},
  publisher = {Oxford University Press (OUP)},
  author = {Denis-Alpizar,  Otoniel and Stoecklin,  Thierry and Guilloteau,  Stéphane and Dutrey,  Anne},
  year = {2018},
  month = may,
  pages = {1811–1817}
}

@ARTICLE{2016MNRAS.460.2103F,
       author = {{Faure}, Alexandre and {Lique}, Fran{\c{c}}ois and {Wiesenfeld}, Laurent},
        title = "{Collisional excitation of HC$_{3}$N by para- and ortho-H$_{2}$}",
      journal = {\mnras},
         year = 2016,
        month = aug,
       volume = {460},
       number = {2},
        pages = {2103-2109},
          doi = {10.1093/mnras/stw1156},
archivePrefix = {arXiv},
       eprint = {1605.03786},
 primaryClass = {astro-ph.GA},
       adsurl = {https://ui.adsabs.harvard.edu/abs/2016MNRAS.460.2103F}
}

@article{GodardPalluet2023,
  title = {Fine-structure excitation of CCS by He: Potential energy surface and scattering calculations},
  volume = {158},
  ISSN = {1089-7690},
  url = {http://dx.doi.org/10.1063/5.0138470},
  DOI = {10.1063/5.0138470},
  number = {4},
  journal = {The Journal of Chemical Physics},
  publisher = {AIP Publishing},
  author = {Godard Palluet,  A. and Lique,  F.},
  year = {2023},
  month = jan 
}

@ARTICLE{2005A&A...432..369S,
       author = {{Sch{\"o}ier}, F.~L. and {van der Tak}, F.~F.~S. and {van Dishoeck}, E.~F. and {Black}, J.~H.},
        title = "{An atomic and molecular database for analysis of submillimetre line observations}",
      journal = {\aap},
         year = 2005,
        month = mar,
       volume = {432},
       number = {1},
        pages = {369-379},
          doi = {10.1051/0004-6361:20041729},
archivePrefix = {arXiv},
       eprint = {astro-ph/0411110},
 primaryClass = {astro-ph},
       adsurl = {https://ui.adsabs.harvard.edu/abs/2005A&A...432..369S}
}

@misc{Asensio2026,
  doi = {10.48550/ARXIV.2601.13495},
  url = {https://arxiv.org/abs/2601.13495},
  author = {Ferrer Asensio,  J. and Scibelli,  S. and Steffes,  L. and Kulterer,  B. and Pokorny-Yadav,  A. and Shirley,  Y. and Megías,  A. and Jiménez-Serra,  I. and Taillard,  A.},
  title = {c-C3H2 deuteration towards prestellar and starless cores in the Perseus Molecular Cloud},
  publisher = {arXiv},
  year = {2026},
  copyright = {Creative Commons Attribution 4.0 International}
}

@ARTICLE{2021ApJ...917...44J,
       author = {{Jim{\'e}nez-Serra}, Izaskun and {Vasyunin}, Anton I. and {Spezzano}, Silvia and {Caselli}, Paola and {Cosentino}, Giuliana and {Viti}, Serena},
        title = "{The Complex Organic Molecular Content in the L1498 Starless Core}",
      journal = {\apj},
         year = 2021,
        month = aug,
       volume = {917},
       number = {1},
          eid = {44},
        pages = {44},
          doi = {10.3847/1538-4357/ac024c},
archivePrefix = {arXiv},
       eprint = {2105.08363},
 primaryClass = {astro-ph.SR},
       adsurl = {https://ui.adsabs.harvard.edu/abs/2021ApJ...917...44J}
}

@ARTICLE{1986A&A...160..181C,
       author = {{Cernicharo}, J. and {Bachiller}, R. and {Duvert}, G.},
        title = "{New HC7N cloudlets in Taurus.}",
      journal = {\aap},
         year = 1986,
        month = may,
       volume = {160},
        pages = {181-184},
       adsurl = {https://ui.adsabs.harvard.edu/abs/1986A&A...160..181C}
}

@ARTICLE{1978ApJ...219L.133K,
       author = {{Kroto}, H.~W. and {Kirby}, C. and {Walton}, D.~R.~M. and {Avery}, L.~W. and {Broten}, N.~W. and {MacLeod}, J.~M. and {Oka}, T.},
        title = "{The detection of cyanohexatriyne, H(C{\ensuremath{\equiv}} C)$_{3}$CN, in Heile's Cloud 2.}",
      journal = {\apjl},
         year = 1978,
        month = feb,
       volume = {219},
        pages = {L133-L137},
          doi = {10.1086/182623},
       adsurl = {https://ui.adsabs.harvard.edu/abs/1978ApJ...219L.133K}
}

@ARTICLE{2018Sci...359..202M,
       author = {{McGuire}, Brett A. and {Burkhardt}, Andrew M. and {Kalenskii}, Sergei and {Shingledecker}, Christopher N. and {Remijan}, Anthony J. and {Herbst}, Eric and {McCarthy}, Michael C.},
        title = "{Detection of the aromatic molecule benzonitrile (c-C$_{6}$H$_{5}$CN) in the interstellar medium}",
      journal = {Science},
         year = 2018,
        month = jan,
       volume = {359},
       number = {6372},
        pages = {202-205},
          doi = {10.1126/science.aao4890},
archivePrefix = {arXiv},
       eprint = {1801.04228},
 primaryClass = {astro-ph.GA},
       adsurl = {https://ui.adsabs.harvard.edu/abs/2018Sci...359..202M}
}

@article{Laas2019,
  title = {Modeling sulfur depletion in interstellar clouds},
  volume = {624},
  ISSN = {1432-0746},
  url = {http://dx.doi.org/10.1051/0004-6361/201834446},
  DOI = {10.1051/0004-6361/201834446},
  journal = {Astronomy &amp; Astrophysics},
  publisher = {EDP Sciences},
  author = {Laas,  Jacob C. and Caselli,  Paola},
  year = {2019},
  month = apr,
  pages = {A108}
}

@software{2011ascl.soft09001G,
       author = {{Ginsburg}, Adam and {Mirocha}, Jordan},
        title = "{PySpecKit: Python Spectroscopic Toolkit}",
 howpublished = {Astrophysics Source Code Library, record ascl:1109.001},
         year = 2011,
        month = sep,
          eid = {ascl:1109.001},
archivePrefix = {ascl},
       eprint = {1109.001},
       adsurl = {https://ui.adsabs.harvard.edu/abs/2011ascl.soft09001G}
}

@ARTICLE{2024PASJ...76.1270T,
       author = {{Taniguchi}, Kotomi and {Nakamura}, Fumitaka and {Liu}, Sheng-Yuan and {Shimoikura}, Tomomi and {Chiong}, Chau-Ching and {Dobashi}, Kazuhito and {Hirano}, Naomi and {Yonekura}, Yoshinori and {Nomura}, Hideko and {Nishimura}, Atsushi and {Ogawa}, Hideo and {Chien}, Chen and {Ho}, Chin-Ting and {Hwang}, Yuh-Jing and {Yeh}, You-Ting and {Lai}, Shih-Ping and {Fujii}, Yasunori and {Yamasaki}, Yasumasa and {Nguyen-Luong}, Quang and {Kawabe}, Ryohei},
        title = "{Q-band line survey observations toward a carbon-chain-rich clump in the Serpens South region}",
      journal = {\pasj},
         year = 2024,
        month = dec,
       volume = {76},
       number = {6},
        pages = {1270-1301},
          doi = {10.1093/pasj/psae088},
archivePrefix = {arXiv},
       eprint = {2409.16492},
 primaryClass = {astro-ph.GA},
       adsurl = {https://ui.adsabs.harvard.edu/abs/2024PASJ...76.1270T}
}

@article{McCarthy2021,
  title = {Aromatics and Cyclic Molecules in Molecular Clouds: A New Dimension of Interstellar Organic Chemistry},
  volume = {125},
  ISSN = {1520-5215},
  url = {http://dx.doi.org/10.1021/acs.jpca.1c00129},
  DOI = {10.1021/acs.jpca.1c00129},
  number = {16},
  journal = {The Journal of Physical Chemistry A},
  publisher = {American Chemical Society (ACS)},
  author = {McCarthy,  Michael C. and McGuire,  Brett A.},
  year = {2021},
  month = mar,
  pages = {3231–3243}
}

@article{Herbst2009,
  title = {Complex Organic Interstellar Molecules},
  volume = {47},
  ISSN = {1545-4282},
  url = {http://dx.doi.org/10.1146/annurev-astro-082708-101654},
  DOI = {10.1146/annurev-astro-082708-101654},
  number = {1},
  journal = {Annual Review of Astronomy and Astrophysics},
  publisher = {Annual Reviews},
  author = {Herbst,  Eric and van Dishoeck,  Ewine F.},
  year = {2009},
  month = sep,
  pages = {427–480}
}

@article{Law2018,
  title = {Carbon Chain Molecules toward Embedded Low-mass Protostars∗},
  volume = {863},
  ISSN = {1538-4357},
  url = {http://dx.doi.org/10.3847/1538-4357/aacf9d},
  DOI = {10.3847/1538-4357/aacf9d},
  number = {1},
  journal = {The Astrophysical Journal},
  publisher = {American Astronomical Society},
  author = {Law,  Charles J. and \"{O}berg,  Karin I. and Bergner,  Jennifer B. and Graninger,  Dawn},
  year = {2018},
  month = aug,
  pages = {88}
}

@article{Choe2021,
  title = {Can Cytosine,  Uracil,  and Thymine Be Formed from HC3N and H2NCO+ in Interstellar Space?},
  volume = {914},
  ISSN = {1538-4357},
  url = {http://dx.doi.org/10.3847/1538-4357/abfd34},
  DOI = {10.3847/1538-4357/abfd34},
  number = {2},
  journal = {The Astrophysical Journal},
  publisher = {American Astronomical Society},
  author = {Choe,  Joong Chul},
  year = {2021},
  month = jun,
  pages = {136}
}

@article{Herbst1989,
  title = {Gas-phase production of complex hydrocarbons,  cyanopolyynes,  and related compounds in dense interstellar clouds},
  volume = {69},
  ISSN = {1538-4365},
  url = {http://dx.doi.org/10.1086/191314},
  DOI = {10.1086/191314},
  journal = {The Astrophysical Journal Supplement Series},
  publisher = {American Astronomical Society},
  author = {Herbst,  Eric and Leung,  Chun Ming},
  year = {1989},
  month = feb,
  pages = {271}
}

@article{Ohishi1998,
  title = {Chemical and physical evolution of dark clouds Molecular spectral line survey toward TMC-1},
  volume = {109},
  ISSN = {1364-5498},
  url = {http://dx.doi.org/10.1039/A801058G},
  DOI = {10.1039/a801058g},
  journal = {Faraday Discussions},
  publisher = {Royal Society of Chemistry (RSC)},
  author = {Ohishi,  Masatoshi and Kaifu,  Norio},
  year = {1998},
  pages = {205–216}
}

@ARTICLE{1973ApJ...185..505H,
       author = {{Herbst}, Eric and {Klemperer}, William},
        title = "{The Formation and Depletion of Molecules in Dense Interstellar Clouds}",
      journal = {\apj},
         year = 1973,
        month = oct,
       volume = {185},
        pages = {505-534},
          doi = {10.1086/152436},
       adsurl = {https://ui.adsabs.harvard.edu/abs/1973ApJ...185..505H}
}

@article{Suzuki1992,
  title = {A survey of CCS,  HC3N,  HC5N,  and NH3 toward dark cloud cores and their production chemistry},
  volume = {392},
  ISSN = {1538-4357},
  url = {http://dx.doi.org/10.1086/171456},
  DOI = {10.1086/171456},
  journal = {The Astrophysical Journal},
  publisher = {American Astronomical Society},
  author = {Suzuki,  Hiroko and Yamamoto,  Satoshi and Ohishi,  Masatoshi and Kaifu,  Norio and Ishikawa,  Shin-Ichi and Hirahara,  Yasuhiro and Takano,  Shuro},
  year = {1992},
  month = jun,
  pages = {551}
}

@article{Hassel2008,
  title = {Modeling the Lukewarm Corino Phase: Is L1527 Unique?},
  volume = {681},
  ISSN = {1538-4357},
  url = {http://dx.doi.org/10.1086/588185},
  DOI = {10.1086/588185},
  number = {2},
  journal = {The Astrophysical Journal},
  publisher = {American Astronomical Society},
  author = {Hassel,  George E. and Herbst,  Eric and Garrod,  Robin T.},
  year = {2008},
  month = jul,
  pages = {1385–1395}
}

@article{Yang2021,
  title = {The Perseus ALMA Chemistry Survey (PEACHES). I. The Complex Organic Molecules in Perseus Embedded Protostars},
  volume = {910},
  ISSN = {1538-4357},
  url = {http://dx.doi.org/10.3847/1538-4357/abdfd6},
  DOI = {10.3847/1538-4357/abdfd6},
  number = {1},
  journal = {The Astrophysical Journal},
  publisher = {American Astronomical Society},
  author = {Yang,  Yao-Lun and Sakai,  Nami and Zhang,  Yichen and Murillo,  Nadia M. and Zhang,  Ziwei E. and Higuchi,  Aya E. and Zeng,  Shaoshan and López-Sepulcre,  Ana and Yamamoto,  Satoshi and Lefloch,  Bertrand and Bouvier,  Mathilde and Ceccarelli,  Cecilia and Hirota,  Tomoya and Imai,  Muneaki and Oya,  Yoko and Sakai,  Takeshi and Watanabe,  Yoshimasa},
  year = {2021},
  month = mar,
  pages = {20}
}

@article{Belloche2020,
  title = {Questioning the spatial origin of complex organic molecules in young protostars with the CALYPSO survey},
  volume = {635},
  ISSN = {1432-0746},
  url = {http://dx.doi.org/10.1051/0004-6361/201937352},
  DOI = {10.1051/0004-6361/201937352},
  journal = {Astronomy &amp; Astrophysics},
  publisher = {EDP Sciences},
  author = {Belloche,  A. and Maury,  A. J. and Maret,  S. and Anderl,  S. and Bacmann,  A. and André,  Ph. and Bontemps,  S. and Cabrit,  S. and Codella,  C. and Gaudel,  M. and Gueth,  F. and Lefèvre,  C. and Lefloch,  B. and Podio,  L. and Testi,  L.},
  year = {2020},
  month = mar,
  pages = {A198}
}

@article{Little1977,
  title = {Detection of the Formula transition of interstellar cyanodiacetylene},
  volume = {181},
  ISSN = {1365-2966},
  url = {http://dx.doi.org/10.1093/mnras/181.1.33P},
  DOI = {10.1093/mnras/181.1.33p},
  number = {1},
  journal = {Monthly Notices of the Royal Astronomical Society},
  publisher = {Oxford University Press (OUP)},
  author = {Little,  L. T. and Riley,  P. W. and Matheson,  D. N.},
  year = {1977},
  month = nov,
  pages = {33P--35P}
}

@article{Broten1978,
  title = {The detection of HC9N in interstellar space},
  volume = {223},
  ISSN = {1538-4357},
  url = {http://dx.doi.org/10.1086/182739},
  DOI = {10.1086/182739},
  journal = {The Astrophysical Journal},
  publisher = {American Astronomical Society},
  author = {Broten,  N. W. and Oka,  T. and Avery,  L. W. and MacLeod,  J. M. and Kroto,  H. W.},
  year = {1978},
  month = jul,
  pages = {L105}
}

@article{Kroto1978,
  title = {The detection of cyanohexatriyne in Heiles’s cloud 2.},
  volume = {219},
  ISSN = {1538-4357},
  url = {http://dx.doi.org/10.1086/182623},
  DOI = {10.1086/182623},
  journal = {The Astrophysical Journal},
  publisher = {American Astronomical Society},
  author = {Kroto,  H. W. and Kirby,  C. and Walton,  D. R. M. and Avery,  L. W. and Broten,  N. W. and MacLeod,  J. M. and Oka,  T.},
  year = {1978},
  month = feb,
  pages = {L133}
}

@article{Bergin2007,
  title = {Cold Dark Clouds: The Initial Conditions for Star Formation},
  volume = {45},
  ISSN = {1545-4282},
  url = {http://dx.doi.org/10.1146/annurev.astro.45.071206.100404},
  DOI = {10.1146/annurev.astro.45.071206.100404},
  number = {1},
  journal = {Annual Review of Astronomy and Astrophysics},
  publisher = {Annual Reviews},
  author = {Bergin,  Edwin A. and Tafalla,  Mario},
  year = {2007},
  month = sep,
  pages = {339–396}
}

@article{Ginsburg2022,
  title = {Pyspeckit: A Spectroscopic Analysis and Plotting Package},
  volume = {163},
  ISSN = {1538-3881},
  url = {http://dx.doi.org/10.3847/1538-3881/ac695a},
  DOI = {10.3847/1538-3881/ac695a},
  number = {6},
  journal = {The Astronomical Journal},
  publisher = {American Astronomical Society},
  author = {Ginsburg,  Adam and Sokolov,  Vlas and de Val-Borro,  Miguel and Rosolowsky,  Erik and Pineda,  Jaime E. and Sipőcz,  Brigitta M. and Henshaw,  Jonathan D.},
  year = {2022},
  month = may,
  pages = {291}
}

@ARTICLE{2023MNRAS.519.1601M,
       author = {{Meg{\'\i}as}, A. and {Jim{\'e}nez-Serra}, I. and {Mart{\'\i}n-Pintado}, J. and {Vasyunin}, A.~I. and {Spezzano}, S. and {Caselli}, P. and {Cosentino}, G. and {Viti}, S.},
        title = "{The complex organic molecular content in the L1517B starless core}",
      journal = {\mnras},
         year = 2023,
        month = feb,
       volume = {519},
       number = {2},
        pages = {1601-1617},
          doi = {10.1093/mnras/stac3449},
archivePrefix = {arXiv},
       eprint = {2211.16119},
 primaryClass = {astro-ph.GA},
       adsurl = {https://ui.adsabs.harvard.edu/abs/2023MNRAS.519.1601M}
}

@ARTICLE{1999ApJ...517..209G,
       author = {{Goldsmith}, Paul F. and {Langer}, William D.},
        title = "{Population Diagram Analysis of Molecular Line Emission}",
      journal = {\apj},
         year = 1999,
        month = may,
       volume = {517},
       number = {1},
        pages = {209-225},
          doi = {10.1086/307195},
       adsurl = {https://ui.adsabs.harvard.edu/abs/1999ApJ...517..209G}
}

@ARTICLE{Scibelli2024,
       author = {{Scibelli}, Samantha and {Shirley}, Yancy and {Meg{\'\i}as}, Andr{\'e}s and {Jim{\'e}nez-Serra}, Izaskun},
        title = "{Survey of complex organic molecules in starless and pre-stellar cores in the Perseus molecular cloud}",
      journal = {\mnras},
         year = 2024,
        month = oct,
       volume = {533},
       number = {4},
        pages = {4104-4149},
          doi = {10.1093/mnras/stae2017},
archivePrefix = {arXiv},
       eprint = {2408.11613},
 primaryClass = {astro-ph.GA},
       adsurl = {https://ui.adsabs.harvard.edu/abs/2024MNRAS.533.4104S}
}

@ARTICLE{Taniguchi2024,
       author = {{Taniguchi}, Kotomi and {Gorai}, Prasanta and {Tan}, Jonathan C.},
        title = "{Carbon-chain chemistry in the interstellar medium}",
      journal = {\apss},
         year = 2024,
        month = apr,
       volume = {369},
       number = {4},
          eid = {34},
        pages = {34},
          doi = {10.1007/s10509-024-04292-9},
archivePrefix = {arXiv},
       eprint = {2303.15769},
 primaryClass = {astro-ph.GA},
       adsurl = {https://ui.adsabs.harvard.edu/abs/2024Ap&SS.369...34T}
}

@article{Benson1998,
  title = {Dense Cores in Dark Clouds. XI. A Survey for N2H+,  C3H2,  and CCS},
  volume = {506},
  ISSN = {1538-4357},
  url = {http://dx.doi.org/10.1086/306276},
  DOI = {10.1086/306276},
  number = {2},
  journal = {The Astrophysical Journal},
  publisher = {American Astronomical Society},
  author = {Benson,  Priscilla J. and Caselli,  Paola and Myers,  Philip C.},
  year = {1998},
  month = oct,
  pages = {743–757}
}

@article{Seki1996,
  title = {Reaction rates of the CN radical with diacetylene and dicyanoacetylene},
  volume = {258},
  ISSN = {0009-2614},
  url = {http://dx.doi.org/10.1016/0009-2614(96)00697-5},
  DOI = {10.1016/0009-2614(96)00697-5},
  number = {5–6},
  journal = {Chemical Physics Letters},
  publisher = {Elsevier BV},
  author = {Seki,  Kanekazu and Yagi,  Mikio and He,  Maoqi and Halpern,  Joshua B. and Okabe,  Hideo},
  year = {1996},
  month = aug,
  pages = {657–662}
}

@ARTICLE{1998A&A...329.1156T,
       author = {{Takano}, Shuro and {Masuda}, Akimasa and {Hirahara}, Yasuhiro and {Suzuki}, Hiroko and {Ohishi}, Masatoshi and {Ishikawa}, Shin-Ichi and {Kaifu}, Norio and {Kasai}, Yasuko and {Kawaguchi}, Kentarou and {Wilson}, T.~L.},
        title = "{Observations of \^(13)C isotopomers of HC\_(3)N and HC\_(5)N in TMC-1: evidence for isotopic fractionation}",
      journal = {\aap},
         year = 1998,
        month = jan,
       volume = {329},
        pages = {1156-1169},
       adsurl = {https://ui.adsabs.harvard.edu/abs/1998A&A...329.1156T}
}

@article{Petrashkevich2024,
  title = {Deuterium fractionation in cold dense cores in the low-mass star-forming region L1688},
  volume = {528},
  ISSN = {1365-2966},
  url = {http://dx.doi.org/10.1093/mnras/stae116},
  DOI = {10.1093/mnras/stae116},
  number = {2},
  journal = {Monthly Notices of the Royal Astronomical Society},
  publisher = {Oxford University Press (OUP)},
  author = {Petrashkevich,  I V and Punanova,  A F and Caselli,  P and Sipil\"{a},  O and Pineda,  J E and Friesen,  R K and Korotaeva,  M G and Vasyunin,  A I},
  year = {2024},
  month = jan,
  pages = {1327–1353}
}

@article{Crapsi2005,
  title = {Probing the Evolutionary Status of Starless Cores through N2H+and N2D+Observations},
  volume = {619},
  ISSN = {1538-4357},
  url = {http://dx.doi.org/10.1086/426472},
  DOI = {10.1086/426472},
  number = {1},
  journal = {The Astrophysical Journal},
  publisher = {American Astronomical Society},
  author = {Crapsi,  A. and Caselli,  P. and Walmsley,  C. M. and Myers,  P. C. and Tafalla,  M. and Lee,  C. W. and Bourke,  T. L.},
  year = {2005},
  month = jan,
  pages = {379–406}
}

@article{Sakai2007,
  title = {Production Pathways of CCS and CCCS Inferred from Their13C Isotopic Species},
  volume = {663},
  ISSN = {1538-4357},
  url = {http://dx.doi.org/10.1086/518595},
  DOI = {10.1086/518595},
  number = {2},
  journal = {The Astrophysical Journal},
  publisher = {American Astronomical Society},
  author = {Sakai,  Nami and Ikeda,  Masafumi and Morita,  Masaru and Sakai,  Takeshi and Takano,  Shuro and Osamura,  Yoshihiro and Yamamoto,  Satoshi},
  year = {2007},
  month = jul,
  pages = {1174–1179}
}

@article{Yamada2002,
  title = {A comprehensive investigation on the formation of organo-sulfur molecules in dark clouds via neutral-neutral reactions},
  volume = {395},
  ISSN = {1432-0746},
  url = {http://dx.doi.org/10.1051/0004-6361:20021328},
  DOI = {10.1051/0004-6361:20021328},
  number = {3},
  journal = {Astronomy & Astrophysics},
  publisher = {EDP Sciences},
  author = {Yamada,  M. and Osamura,  Y. and Kaiser,  R. I.},
  year = {2002},
  month = nov,
  pages = {1031–1044}
}

@article{Bergin1997,
  title = {Chemical Evolution in Preprotostellar and Protostellar Cores},
  volume = {486},
  ISSN = {1538-4357},
  url = {http://dx.doi.org/10.1086/304510},
  DOI = {10.1086/304510},
  number = {1},
  journal = {The Astrophysical Journal},
  publisher = {American Astronomical Society},
  author = {Bergin,  E. A. and Langer,  W. D.},
  year = {1997},
  month = sep,
  pages = {316–328}
}

@ARTICLE{1995ZNatA..50.1179Y,
       author = {{Yamada}, K.~M.~T. and {Moravec}, A. and {Winnewisser}, G.},
        title = "{Sub-millimeter Wave Spectra of Cyanoacetylene and Revised Ground State Constants}",
      journal = {Zeitschrift Naturforschung Teil A},
         year = 1995,
        month = dec,
       volume = {50},
       number = {12},
        pages = {1179-1181},
          doi = {10.1515/zna-1995-1222},
       adsurl = {https://ui.adsabs.harvard.edu/abs/1995ZNatA..50.1179Y}
}

@article{Lafferty1978,
  title = {Microwave spectra of molecules of astrophysical interest XIII. Cyanoacetylene},
  volume = {7},
  ISSN = {1529-7845},
  url = {http://dx.doi.org/10.1063/1.555575},
  DOI = {10.1063/1.555575},
  number = {2},
  journal = {Journal of Physical and Chemical Reference Data},
  publisher = {AIP Publishing},
  author = {Lafferty,  W. J. and Lovas,  F. J.},
  year = {1978},
  month = apr,
  pages = {441–493}
}

@ARTICLE{2013A&A...559A..47B,
       author = {{Belloche}, A. and {M{\"u}ller}, H.~S.~P. and {Menten}, K.~M. and {Schilke}, P. and {Comito}, C.},
        title = "{Complex organic molecules in the interstellar medium: IRAM 30 m line survey of Sagittarius B2(N) and (M)}",
      journal = {\aap},
         year = 2013,
        month = nov,
       volume = {559},
          eid = {A47},
        pages = {A47},
          doi = {10.1051/0004-6361/201321096},
archivePrefix = {arXiv},
       eprint = {1308.5062},
 primaryClass = {astro-ph.GA},
       adsurl = {https://ui.adsabs.harvard.edu/abs/2013A&A...559A..47B}
}

@article{PICKETT1998,
  title = {SUBMILLIMETER,  MILLIMETER,  AND MICROWAVE SPECTRAL LINE CATALOG},
  volume = {60},
  ISSN = {0022-4073},
  url = {http://dx.doi.org/10.1016/S0022-4073(98)00091-0},
  DOI = {10.1016/s0022-4073(98)00091-0},
  number = {5},
  journal = {Journal of Quantitative Spectroscopy and Radiative Transfer},
  publisher = {Elsevier BV},
  author = {Pickett,  H.M. and Poynter,  R.L. and Cohen,  E.A. and Delistky,  M.L. and Pearson,  J.C. and M\"{u}ller,  H.S.P.},
  year = {1998},
  month = nov,
  pages = {883–890}
}

@ARTICLE{2008ApJ...684.1240E,
       author = {{Enoch}, Melissa L. and {Evans}, Neal J., II and {Sargent}, Anneila I. and {Glenn}, Jason and {Rosolowsky}, Erik and {Myers}, Philip},
        title = "{The Mass Distribution and Lifetime of Prestellar Cores in Perseus, Serpens, and Ophiuchus}",
      journal = {\apj},
         year = 2008,
        month = sep,
       volume = {684},
       number = {2},
        pages = {1240-1259},
          doi = {10.1086/589963},
archivePrefix = {arXiv},
       eprint = {0805.1075},
 primaryClass = {astro-ph},
       adsurl = {https://ui.adsabs.harvard.edu/abs/2008ApJ...684.1240E}
}

@ARTICLE{2008ApJS..175..509R,
       author = {{Rosolowsky}, E.~W. and {Pineda}, J.~E. and {Foster}, J.~B. and {Borkin}, M.~A. and {Kauffmann}, J. and {Caselli}, P. and {Myers}, P.~C. and {Goodman}, A.~A.},
        title = "{An Ammonia Spectral Atlas of Dense Cores in Perseus}",
      journal = {\apjs},
         year = 2008,
        month = apr,
       volume = {175},
       number = {2},
        pages = {509-521},
          doi = {10.1086/524299},
archivePrefix = {arXiv},
       eprint = {0711.0231},
 primaryClass = {astro-ph},
       adsurl = {https://ui.adsabs.harvard.edu/abs/2008ApJS..175..509R}
}

@article{radex,
    author = {{Van der Tak} and F.F.S., Black and J.H., Schöier and F.L., Jansen and D.J., van Dishoeck and E.F.},
    title = "{online RADEX}",
    journal = {\aap},
    year = 2007,
    volume = {468},
    pages = {627-635},
    adsurl = {https://var.sron.nl/radex/radex.php}
}

@ARTICLE{2012A&A...540A..10S,
       author = {{Sadavoy}, S.~I. and {di Francesco}, J. and {Andr{\'e}}, Ph. and {Pezzuto}, S. and {Bernard}, J. -P. and {Bontemps}, S. and {Bressert}, E. and {Chitsazzadeh}, S. and {Fallscheer}, C. and {Hennemann}, M. and {Hill}, T. and {Martin}, P. and {Motte}, F. and {Nguyen Luong}, Q. and {Peretto}, N. and {Reid}, M. and {Schneider}, N. and {Testi}, L. and {White}, G.~J. and {Wilson}, C.},
        title = "{Herschel observations of a potential core-forming clump: Perseus B1-E}",
      journal = {\aap},
         year = 2012,
        month = apr,
       volume = {540},
          eid = {A10},
        pages = {A10},
          doi = {10.1051/0004-6361/201117934},
archivePrefix = {arXiv},
       eprint = {1111.7021},
 primaryClass = {astro-ph.GA},
       adsurl = {https://ui.adsabs.harvard.edu/abs/2012A&A...540A..10S}
}

@ARTICLE{2023A&A...674L...4C,
       author = {{Cernicharo}, J. and {Tercero}, B. and {Marcelino}, N. and {Ag{\'u}ndez}, M. and {de Vicente}, P.},
        title = "{The spatial distribution of an aromatic molecule, C$_{6}$H$_{5}$CN, in the cold dark cloud TMC-1}",
      journal = {\aap},
         year = 2023,
        month = jun,
       volume = {674},
          eid = {L4},
        pages = {L4},
          doi = {10.1051/0004-6361/202346722},
archivePrefix = {arXiv},
       eprint = {2305.15315},
 primaryClass = {astro-ph.GA},
       adsurl = {https://ui.adsabs.harvard.edu/abs/2023A&A...674L...4C}
}

@ARTICLE{1998ApJ...504L.109L,
       author = {{Lefloch}, B. and {Castets}, A. and {Cernicharo}, J. and {Loinard}, L.},
        title = "{Widespread SiO Emission in NGC 1333}",
      journal = {\apjl},
         year = 1998,
        month = sep,
       volume = {504},
       number = {2},
        pages = {L109-L112},
          doi = {10.1086/311581},
       adsurl = {https://ui.adsabs.harvard.edu/abs/1998ApJ...504L.109L}
}

@ARTICLE{2004PASJ...56...69K,
       author = {{Kaifu}, Norio and {Ohishi}, Masatoshi and {Kawaguchi}, Kentarou and {Saito}, Shuji and {Yamamoto}, Satoshi and {Miyaji}, Takeshi and {Miyazawa}, Keisuke and {Ishikawa}, Shin-Ichi and {Noumaru}, Chiaki and {Harasawa}, Sumiko and {Okuda}, Michiko and {Suzuki}, Hiroko},
        title = "{A 8.8--50GHz Complete Spectral Line Survey toward TMC-1 I. Survey Data}",
      journal = {\pasj},
         year = 2004,
        month = feb,
       volume = {56},
        pages = {69-173},
          doi = {10.1093/pasj/56.1.69},
       adsurl = {https://ui.adsabs.harvard.edu/abs/2004PASJ...56...69K}
}

@ARTICLE{2019ApJ...871..134S,
       author = {{Seo}, Young Min and {Majumdar}, Liton and {Goldsmith}, Paul F. and {Shirley}, Yancy L. and {Willacy}, Karen and {Ward-Thompson}, Derek and {Friesen}, Rachel and {Frayer}, David and {Church}, Sarah E. and {Chung}, Dongwoo and {Cleary}, Kieran and {Cunningham}, Nichol and {Devaraj}, Kiruthika and {Egan}, Dennis and {Gaier}, Todd and {Gawande}, Rohit and {Gundersen}, Joshua O. and {Harris}, Andrew I. and {Kangaslahti}, Pekka and {Readhead}, Anthony C.~S. and {Samoska}, Lorene and {Sieth}, Matthew and {Stennes}, Michael and {Voll}, Patricia and {White}, Steve},
        title = "{An Ammonia Spectral Map of the L1495-B218 Filaments in the Taurus Molecular Cloud. II. CCS and HC$_{7}$N Chemistry and Three Modes of Star Formation in the Filaments}",
      journal = {\apj},
         year = 2019,
        month = feb,
       volume = {871},
       number = {2},
          eid = {134},
        pages = {134},
          doi = {10.3847/1538-4357/aaf887},
archivePrefix = {arXiv},
       eprint = {1812.06121},
 primaryClass = {astro-ph.GA},
       adsurl = {https://ui.adsabs.harvard.edu/abs/2019ApJ...871..134S}
}

@INPROCEEDINGS{2005sf2a.conf..721P,
       author = {{Pety}, J.},
        title = "{Successes of and Challenges to GILDAS, a State-of-the-Art Radioastronomy Toolkit}",
    booktitle = {SF2A-2005: Semaine de l'Astrophysique Francaise},
         year = 2005,
       editor = {{Casoli}, F. and {Contini}, T. and {Hameury}, J.~M. and {Pagani}, L.},
        month = dec,
        pages = {721},
       adsurl = {https://ui.adsabs.harvard.edu/abs/2005sf2a.conf..721P}
}

@MISC{2013ascl.soft05010G,
       author = {{Gildas Team}},
        title = "{GILDAS: Grenoble Image and Line Data Analysis Software}",
         year = 2013,
        month = may,
          eid = {ascl:1305.010},
        pages = {ascl:1305.010},
archivePrefix = {ascl},
       eprint = {1305.010},
       adsurl = {https://ui.adsabs.harvard.edu/abs/2013ascl.soft05010G}
}

@ARTICLE{2021A&A...645A..37T,
       author = {{Tercero}, F. and {L{\'o}pez-P{\'e}rez}, J.~A. and {Gallego}, J.~D. and {Beltr{\'a}n}, F. and {Garc{\'\i}a}, O. and {Patino-Esteban}, M. and {L{\'o}pez-Fern{\'a}ndez}, I. and {G{\'o}mez-Molina}, G. and {Diez}, M. and {Garc{\'\i}a-Carre{\~n}o}, P. and {Malo}, I. and {Amils}, R. and {Serna}, J.~M. and {Albo}, C. and {Hern{\'a}ndez}, J.~M. and {Vaquero}, B. and {Gonz{\'a}lez-Garc{\'\i}a}, J. and {Barbas}, L. and {L{\'o}pez-Fern{\'a}ndez}, J.~A. and {Bujarrabal}, V. and {G{\'o}mez-Garrido}, M. and {Pardo}, J.~R. and {Santander-Garc{\'\i}a}, M. and {Tercero}, B. and {Cernicharo}, J. and {de Vicente}, P.},
        title = "{Yebes 40 m radio telescope and the broad band Nanocosmos receivers at 7 mm and 3 mm for line surveys}",
      journal = {\aap},
         year = 2021,
        month = jan,
       volume = {645},
          eid = {A37},
        pages = {A37},
          doi = {10.1051/0004-6361/202038701},
archivePrefix = {arXiv},
       eprint = {2010.16224},
 primaryClass = {astro-ph.IM},
       adsurl = {https://ui.adsabs.harvard.edu/abs/2021A&A...645A..37T}
}

@ARTICLE{2021A&A...645A..55P,
       author = {{Pezzuto}, S. and {Benedettini}, M. and {Di Francesco}, J. and {Palmeirim}, P. and {Sadavoy}, S. and {Schisano}, E. and {Li Causi}, G. and {Andr{\'e}}, Ph. and {Arzoumanian}, D. and {Bernard}, J. -Ph. and {Bontemps}, S. and {Elia}, D. and {Fiorellino}, E. and {Kirk}, J.~M. and {K{\"o}nyves}, V. and {Ladjelate}, B. and {Men'shchikov}, A. and {Motte}, F. and {Piccotti}, L. and {Schneider}, N. and {Spinoglio}, L. and {Ward-Thompson}, D. and {Wilson}, C.~D.},
        title = "{Physical properties of the ambient medium and of dense cores in the Perseus star-forming region derived from Herschel Gould Belt Survey observations}",
      journal = {\aap},
         year = 2021,
        month = jan,
       volume = {645},
          eid = {A55},
        pages = {A55},
          doi = {10.1051/0004-6361/201936534},
archivePrefix = {arXiv},
       eprint = {2010.00006},
 primaryClass = {astro-ph.GA},
       adsurl = {https://ui.adsabs.harvard.edu/abs/2021A&A...645A..55P}
}

@INPROCEEDINGS{2023ASPC..534..379C,
       author = {{Ceccarelli}, C. and {Codella}, C. and {Balucani}, N. and {Bockelee-Morvan}, D. and {Herbst}, E. and {Vastel}, C. and {Caselli}, P. and {Favre}, C. and {Lefloch}, B. and {Oberg}, K. and {Yamamoto}, S.},
        title = "{Organic Chemistry in the First Phases of Solar-Type Protostars}",
    booktitle = {Protostars and Planets VII},
         year = 2023,
       editor = {{Inutsuka}, S. and {Aikawa}, Y. and {Muto}, T. and {Tomida}, K. and {Tamura}, M.},
       series = {Astronomical Society of the Pacific Conference Series},
       volume = {534},
        month = jul,
        pages = {379},
       adsurl = {https://ui.adsabs.harvard.edu/abs/2023ASPC..534..379C}
}

@article{Ladd1994,
  title = {Dense cores in dark clouds. 10: Ammonia emission in the Perseus molecular cloud complex},
  volume = {433},
  ISSN = {1538-4357},
  url = {http://dx.doi.org/10.1086/174629},
  DOI = {10.1086/174629},
  journal = {The Astrophysical Journal},
  publisher = {American Astronomical Society},
  author = {Ladd,  E. F. and Myers,  P. C. and Goodman,  A. A.},
  year = {1994},
  month = sep,
  pages = {117}
}

@article{Jrgensen2020,
  title = {Astrochemistry During the Formation of Stars},
  volume = {58},
  ISSN = {1545-4282},
  url = {http://dx.doi.org/10.1146/annurev-astro-032620-021927},
  DOI = {10.1146/annurev-astro-032620-021927},
  number = {1},
  journal = {Annual Review of Astronomy and Astrophysics},
  publisher = {Annual Reviews},
  author = {Jørgensen,  Jes K. and Belloche,  Arnaud and Garrod,  Robin T.},
  year = {2020},
  month = aug,
  pages = {727–778}
}

@INPROCEEDINGS{2014prpl.conf..859C,
       author = {{Ceccarelli}, C. and {Caselli}, P. and {Bockel{\'e}e-Morvan}, D. and {Mousis}, O. and {Pizzarello}, S. and {Robert}, F. and {Semenov}, D.},
        title = "{Deuterium Fractionation: The Ariadne's Thread from the Precollapse Phase to Meteorites and Comets Today}",
    booktitle = {Protostars and Planets VI},
         year = 2014,
       editor = {{Beuther}, Henrik and {Klessen}, Ralf S. and {Dullemond}, Cornelis P. and {Henning}, Thomas},
        month = jan,
        pages = {859-882},
          doi = {10.2458/azu_uapress_9780816531240-ch037},
archivePrefix = {arXiv},
       eprint = {1403.7143},
 primaryClass = {astro-ph.EP},
       adsurl = {https://ui.adsabs.harvard.edu/abs/2014prpl.conf..859C}
}

@ARTICLE{2022MNRAS.515.5219G,
       author = {{Galloway-Sprietsma}, Maria and {Shirley}, Yancy L. and {Di Francesco}, James and {Keown}, Jared and {Scibelli}, Samantha and {Sipil{\"a}}, Olli and {Smullen}, Rachel},
        title = "{A survey of deuterated ammonia in the Cepheus star-forming region L1251}",
      journal = {\mnras},
         year = 2022,
        month = oct,
       volume = {515},
       number = {4},
        pages = {5219-5234},
          doi = {10.1093/mnras/stac2084},
archivePrefix = {arXiv},
       eprint = {2207.10718},
 primaryClass = {astro-ph.SR},
       adsurl = {https://ui.adsabs.harvard.edu/abs/2022MNRAS.515.5219G}
}

@ARTICLE{2025ApJ...985L..25S,
       author = {{Scibelli}, Samantha and {Meg{\'\i}as}, Andr{\'e}s and {Jim{\'e}nez-Serra}, Izaskun and {Shirley}, Yancy and {Bergner}, Jennifer and {Ferrer Asensio}, Judit and {Garrod}, Robin T. and {Bonfand}, M{\'e}lisse and {Pokorny-Yadav}, Anissa},
        title = "{First Detections of PN, PO, and PO$^{+}$ toward a Shocked Low-mass Starless Core}",
      journal = {\apjl},
         year = 2025,
        month = jun,
       volume = {985},
       number = {2},
          eid = {L25},
        pages = {L25},
          doi = {10.3847/2041-8213/add344},
archivePrefix = {arXiv},
       eprint = {2504.17849},
 primaryClass = {astro-ph.GA},
       adsurl = {https://ui.adsabs.harvard.edu/abs/2025ApJ...985L..25S}
}

@ARTICLE{2025A&A...702A.127S,
       author = {{Scibelli}, S. and {Drozdovskaya}, M.~N. and {Caselli}, P. and {Ferrer Asensio}, J. and {Kulterer}, B. and {Spezzano}, S. and {Lin}, Y. and {Shirley}, Y.},
        title = "{Nascent chemical complexity in the prestellar core IRAS 16293E: Complex organics and deuterated methanol}",
      journal = {\aap},
         year = 2025,
        month = oct,
       volume = {702},
          eid = {A127},
        pages = {A127},
          doi = {10.1051/0004-6361/202553696},
archivePrefix = {arXiv},
       eprint = {2508.04762},
 primaryClass = {astro-ph.GA},
       adsurl = {https://ui.adsabs.harvard.edu/abs/2025A&A...702A.127S}
}

@article{Tafalla2010,
  title = {A molecular survey of outflow gas: velocity-dependent shock chemistry and the peculiar composition of the EHV gas},
  volume = {522},
  ISSN = {1432-0746},
  url = {http://dx.doi.org/10.1051/0004-6361/201015158},
  DOI = {10.1051/0004-6361/201015158},
  journal = {Astronomy &amp; Astrophysics},
  publisher = {EDP Sciences},
  author = {Tafalla,  M. and Santiago-García,  J. and Hacar,  A. and Bachiller,  R.},
  year = {2010},
  month = nov,
  pages = {A91}
}

@article{Davis2013,
  title = {ISM chemistry in metal-rich environments: molecular tracers of metallicity},
  volume = {433},
  ISSN = {0035-8711},
  url = {http://dx.doi.org/10.1093/mnras/stt842},
  DOI = {10.1093/mnras/stt842},
  number = {2},
  journal = {Monthly Notices of the Royal Astronomical Society},
  publisher = {Oxford University Press (OUP)},
  author = {Davis,  Timothy A. and Bayet,  Estelle and Crocker,  Alison and Topal,  Sel\c{c}uk and Bureau,  Martin},
  year = {2013},
  month = jun,
  pages = {1659–1674}
}

@article{Caselli2002,
  title = {Deuterated molecules as a probe of ionization fraction in dense interstellar clouds},
  volume = {50},
  ISSN = {0032-0633},
  url = {http://dx.doi.org/10.1016/S0032-0633(02)00074-0},
  DOI = {10.1016/s0032-0633(02)00074-0},
  number = {12–13},
  journal = {Planetary and Space Science},
  publisher = {Elsevier BV},
  author = {Caselli,  P},
  year = {2002},
  month = oct,
  pages = {1133–1144}
}

@article{Walmsley2004,
  title = {Complete depletion in prestellar cores},
  volume = {418},
  ISSN = {1432-0746},
  url = {http://dx.doi.org/10.1051/0004-6361:20035718},
  DOI = {10.1051/0004-6361:20035718},
  number = {3},
  journal = {Astronomy &amp; Astrophysics},
  publisher = {EDP Sciences},
  author = {Walmsley,  C. M. and Flower,  D. R. and G. Pineau des For\^ets},
  year = {2004},
  month = apr,
  pages = {1035–1043}
}

@article{Roueff2005,
  title = {Interstellar deuterated ammonia: from NH3 to ND3},
  volume = {438},
  ISSN = {1432-0746},
  url = {http://dx.doi.org/10.1051/0004-6361:20052724},
  DOI = {10.1051/0004-6361:20052724},
  number = {2},
  journal = {Astronomy &amp; Astrophysics},
  publisher = {EDP Sciences},
  author = {Roueff,  E. and Lis,  D. C. and van der Tak,  F. F. S. and Gerin,  M. and Goldsmith,  P. F.},
  year = {2005},
  month = jul,
  pages = {585–598}
}

@ARTICLE{1999A&A...343..571G,
       author = {{Gueth}, F. and {Guilloteau}, S.},
        title = "{The jet-driven molecular outflow of HH 211}",
      journal = {\aap},
         year = 1999,
        month = mar,
       volume = {343},
        pages = {571-584},
       adsurl = {https://ui.adsabs.harvard.edu/abs/1999A&A...343..571G}
}

@article{Cordiner2011,
  title = {ORGANIC CHEMISTRY OF LOW-MASS STAR-FORMING CORES. I. 7 mm SPECTROSCOPY OF CHAMAELEON MMS1},
  volume = {744},
  ISSN = {1538-4357},
  url = {http://dx.doi.org/10.1088/0004-637X/744/2/131},
  DOI = {10.1088/0004-637x/744/2/131},
  number = {2},
  journal = {The Astrophysical Journal},
  publisher = {American Astronomical Society},
  author = {Cordiner,  Martin A. and Charnley,  Steven B. and Wirstr\"{o}m,  Eva S. and Smith,  Robert G.},
  year = {2011},
  month = dec,
  pages = {131}
}

@article{Howe1994,
  title = {Observations of deuterated cyanoacetylene in dark clouds},
  volume = {267},
  ISSN = {1365-2966},
  url = {http://dx.doi.org/10.1093/mnras/267.1.59},
  DOI = {10.1093/mnras/267.1.59},
  number = {1},
  journal = {Monthly Notices of the Royal Astronomical Society},
  publisher = {Oxford University Press (OUP)},
  author = {Howe,  D. A. and Millar,  T. J. and Schilke,  P. and Walmsley,  C. M.},
  year = {1994},
  month = mar,
  pages = {59–68}
}

@article{Saito2002,
  title = {Chemical Timescale of Dark Cloud Cores Estimated from Deuterium Fractionation},
  volume = {569},
  ISSN = {1538-4357},
  url = {http://dx.doi.org/10.1086/339474},
  DOI = {10.1086/339474},
  number = {2},
  journal = {The Astrophysical Journal},
  publisher = {American Astronomical Society},
  author = {Saito,  Shuji and Aikawa,  Yuri and Herbst,  Eric and Ohishi,  Masatoshi and Hirota,  Tomoya and Yamamoto,  Satoshi and Kaifu,  Norio},
  year = {2002},
  month = apr,
  pages = {836–840}
}

@article{Turner2001,
  title = {Deuterated Molecules in Translucent and Dark Clouds},
  volume = {136},
  ISSN = {1538-4365},
  url = {http://dx.doi.org/10.1086/322536},
  DOI = {10.1086/322536},
  number = {2},
  journal = {The Astrophysical Journal Supplement Series},
  publisher = {American Astronomical Society},
  author = {Turner,  B. E.},
  year = {2001},
  month = oct,
  pages = {579–629}
}

@article{Sakai2009,
  title = {DEUTERATED MOLECULES IN WARM CARBON CHAIN CHEMISTRY: THE L1527 CASE},
  volume = {702},
  ISSN = {1538-4357},
  url = {http://dx.doi.org/10.1088/0004-637X/702/2/1025},
  DOI = {10.1088/0004-637x/702/2/1025},
  number = {2},
  journal = {The Astrophysical Journal},
  publisher = {American Astronomical Society},
  author = {Sakai,  Nami and Sakai,  Takeshi and Hirota,  Tomoya and Yamamoto,  Satoshi},
  year = {2009},
  month = aug,
  pages = {1025–1035}
}

@ARTICLE{1986A&A...155..237S,
       author = {{Sorochenko}, R.~L. and {Tolmachev}, A.~M. and {Winnewisser}, G.},
        title = "{High resolution measurements of cyanoacetylene in dark clouds.}",
      journal = {\aap},
         year = 1986,
        month = feb,
       volume = {155},
        pages = {237-241},
       adsurl = {https://ui.adsabs.harvard.edu/abs/1986A&A...155..237S}
}

@article{Takano1990,
  title = {Detection of five C-13 isotopic species of HC5N in TMC-1},
  volume = {361},
  ISSN = {1538-4357},
  url = {http://dx.doi.org/10.1086/185816},
  DOI = {10.1086/185816},
  journal = {The Astrophysical Journal},
  publisher = {American Astronomical Society},
  author = {Takano,  Shuro and Suzuki,  Hiroko and Ohishi,  Masatoshi and Ishikawa,  Shin-Ichi and Kaifu,  Norio and Hirahara,  Yasuhiro and Masuda,  Akimasa},
  year = {1990},
  month = sep,
  pages = {L15}
}

@article{Langer1980,
  title = {Detection of deuterated cyanoacetylene in the interstellar cloud TMC 1},
  volume = {239},
  ISSN = {1538-4357},
  url = {http://dx.doi.org/10.1086/183307},
  DOI = {10.1086/183307},
  journal = {The Astrophysical Journal},
  publisher = {American Astronomical Society},
  author = {Langer,  W. D. and Schloerb,  F. P. and Snell,  R. L. and Young,  J. S.},
  year = {1980},
  month = aug,
  pages = {L125}
}

@article{Bianchi2023,
  title = {Cyanopolyyne Chemistry in the L1544 Prestellar Core: New Insights from GBT Observations},
  volume = {944},
  ISSN = {1538-4357},
  url = {http://dx.doi.org/10.3847/1538-4357/acb5e8},
  DOI = {10.3847/1538-4357/acb5e8},
  number = {2},
  journal = {The Astrophysical Journal},
  publisher = {American Astronomical Society},
  author = {Bianchi,  Eleonora and Remijan,  Anthony and Codella,  Claudio and Ceccarelli,  Cecilia and Lique,  Francois and Spezzano,  Silvia and Balucani,  Nadia and Caselli,  Paola and Herbst,  Eric and Podio,  Linda and Vastel,  Charlotte and McGuire,  Brett},
  year = {2023},
  month = feb,
  pages = {208}
}

@misc{https://doi.org/10.48550/arxiv.2206.13270,
  doi = {10.48550/ARXIV.2206.13270},
  url = {https://arxiv.org/abs/2206.13270},
  author = {Ceccarelli,  C. and Codella,  C. and Balucani,  N. and Bockelée-Morvan,  D. and Herbst,  E. and Vastel,  C. and Caselli,  P. and Favre,  C. and Lefloch,  B. and \"{O}berg,  K.},
  title = {Organic chemistry in the first phases of Solar-type protostars},
  publisher = {arXiv},
  year = {2022},
  copyright = {Creative Commons Attribution 4.0 International}
}

@article{Petrie1996,
  title = {Formation of interstellar CCS and CCCS: a case for radical/neutral chemistry?},
  volume = {281},
  ISSN = {1365-2966},
  url = {http://dx.doi.org/10.1093/mnras/281.2.666},
  DOI = {10.1093/mnras/281.2.666},
  number = {2},
  journal = {Monthly Notices of the Royal Astronomical Society},
  publisher = {Oxford University Press (OUP)},
  author = {Petrie,  S.},
  year = {1996},
  month = jul,
  pages = {666–672}
}

@ARTICLE{1989ApJS...69..271H,
       author = {{Herbst}, Eric and {Leung}, Chun Ming},
        title = "{Gas Phase Production of Complex Hydrocarbons, Cyanopolyynes, and Related Compounds in Dense Interstellar Clouds}",
      journal = {APJS},
         year = 1989,
        month = feb,
       volume = {69},
        pages = {271},
          doi = {10.1086/191314},
       adsurl = {https://ui.adsabs.harvard.edu/abs/1989ApJS...69..271H}
}

@software{reback2020pandas,
    author       = {The pandas development team},
    title        = {pandas-dev/pandas: Pandas},
    month        = feb,
    year         = 2020,
    publisher    = {Zenodo},
    version      = {latest},
    doi          = {10.5281/zenodo.3509134},
    url          = {https://doi.org/10.5281/zenodo.3509134}
}

@software{2012ascl.soft08017R,
       author = {{Robitaille}, Thomas and {Bressert}, Eli},
        title = "{APLpy: Astronomical Plotting Library in Python}",
 howpublished = {Astrophysics Source Code Library, record ascl:1208.017},
         year = 2012,
        month = aug,
          eid = {ascl:1208.017},
archivePrefix = {ascl},
       eprint = {1208.017},
       adsurl = {https://ui.adsabs.harvard.edu/abs/2012ascl.soft08017R}
}

@article{Spezzano_2020,
   title={Distribution of methanol and cyclopropenylidene around starless cores},
   volume={643},
   ISSN={1432-0746},
   url={http://dx.doi.org/10.1051/0004-6361/201936598},
   DOI={10.1051/0004-6361/201936598},
   journal={Astronomy &amp; Astrophysics},
   publisher={EDP Sciences},
   author={Spezzano, S. and Caselli, P. and Pineda, J. E. and Bizzocchi, L. and Prudenzano, D. and Nagy, Z.},
   year={2020},
   month=nov, pages={A60} }

@ARTICLE{2017A&A...606A..82S,
       author = {{Spezzano}, S. and {Caselli}, P. and {Bizzocchi}, L. and {Giuliano}, B.~M. and {Lattanzi}, V.},
        title = "{The observed chemical structure of L1544}",
      journal = {\aap},
         year = 2017,
        month = oct,
       volume = {606},
          eid = {A82},
        pages = {A82},
          doi = {10.1051/0004-6361/201731262},
archivePrefix = {arXiv},
       eprint = {1707.06015},
 primaryClass = {astro-ph.GA},
       adsurl = {https://ui.adsabs.harvard.edu/abs/2017A&A...606A..82S}
}

@article{Taniguchi2016,
  title = {13C ISOTOPIC FRACTIONATION OF HC3N IN STAR-FORMING REGIONS: LOW-MASS STAR-FORMING REGION L1527 AND HIGH-MASS STAR-FORMING REGION G28.28-0.36},
  volume = {830},
  ISSN = {1538-4357},
  url = {http://dx.doi.org/10.3847/0004-637X/830/2/106},
  DOI = {10.3847/0004-637x/830/2/106},
  number = {2},
  journal = {The Astrophysical Journal},
  publisher = {American Astronomical Society},
  author = {Taniguchi,  Kotomi and Saito,  Masao and Ozeki,  Hiroyuki},
  year = {2016},
  month = oct,
  pages = {106}
}

@ARTICLE{1981A&A....99..239B,
       author = {{Bujarrabal}, V. and {Guelin}, M. and {Morris}, M. and {Thaddeus}, P.},
        title = "{The abundance and excitation of the carbon chains in interstellar molecular clouds.}",
      journal = {\aap},
         year = 1981,
        month = jun,
       volume = {99},
        pages = {239-247},
       adsurl = {https://ui.adsabs.harvard.edu/abs/1981A&A....99..239B}
}

@ARTICLE{1996Ap&SS.240...13W,
       author = {{Winstanley}, N. and {Nejad}, L.~A.~M.},
        title = "{Cyanopolyyne Chemistry in TMC-1}",
      journal = {\apss},
         year = 1996,
        month = mar,
       volume = {240},
       number = {1},
        pages = {13-37},
          doi = {10.1007/BF00640193},
       adsurl = {https://ui.adsabs.harvard.edu/abs/1996Ap&SS.240...13W}
}

@ARTICLE{1971ApJ...163L..35T,
       author = {{Turner}, B.~E.},
        title = "{Detection of Interstellar Cyanoacetylene}",
      journal = {\apjl},
         year = 1971,
        month = jan,
       volume = {163},
        pages = {L35},
          doi = {10.1086/180662},
       adsurl = {https://ui.adsabs.harvard.edu/abs/1971ApJ...163L..35T}
}

@ARTICLE{2005ApJ...634.1126M,
       author = {{Milam}, S.~N. and {Savage}, C. and {Brewster}, M.~A. and {Ziurys}, L.~M. and {Wyckoff}, S.},
        title = "{The $^{12}$C/$^{13}$C Isotope Gradient Derived from Millimeter Transitions of CN: The Case for Galactic Chemical Evolution}",
      journal = {\apj},
         year = 2005,
        month = dec,
       volume = {634},
       number = {2},
        pages = {1126-1132},
          doi = {10.1086/497123},
       adsurl = {https://ui.adsabs.harvard.edu/abs/2005ApJ...634.1126M}
}

@article{Vastel2018,
  title = {Sulphur chemistry in the L1544 pre-stellar core},
  volume = {478},
  ISSN = {1365-2966},
  url = {http://dx.doi.org/10.1093/mnras/sty1336},
  DOI = {10.1093/mnras/sty1336},
  number = {4},
  journal = {Monthly Notices of the Royal Astronomical Society},
  publisher = {Oxford University Press (OUP)},
  author = {Vastel,  Charlotte and Quénard,  D and Le Gal,  R and Wakelam,  V and Andrianasolo,  A and Caselli,  P and Vidal,  T and Ceccarelli,  C and Lefloch,  B and Bachiller,  R},
  year = {2018},
  month = may,
  pages = {5514–5532}
}

@article{Virtanen2020,
  title = {SciPy 1.0: fundamental algorithms for scientific computing in Python},
  volume = {17},
  ISSN = {1548-7105},
  url = {http://dx.doi.org/10.1038/s41592-019-0686-2},
  DOI = {10.1038/s41592-019-0686-2},
  number = {3},
  journal = {Nature Methods},
  publisher = {Springer Science and Business Media LLC},
  author = {Virtanen,  Pauli and Gommers,  Ralf and Oliphant,  Travis E. and Haberland,  Matt and Reddy,  Tyler and Cournapeau,  David and Burovski,  Evgeni and Peterson,  Pearu and Weckesser,  Warren and Bright,  Jonathan and van der Walt,  Stéfan J. and Brett,  Matthew and Wilson,  Joshua and Millman,  K. Jarrod and Mayorov,  Nikolay and Nelson,  Andrew R. J. and Jones,  Eric and Kern,  Robert and Larson,  Eric and Carey,  C J and Polat,  İlhan and Feng,  Yu and Moore,  Eric W. and VanderPlas,  Jake and Laxalde,  Denis and Perktold,  Josef and Cimrman,  Robert and Henriksen,  Ian and Quintero,  E. A. and Harris,  Charles R. and Archibald,  Anne M. and Ribeiro,  Ant\^onio H. and Pedregosa,  Fabian and van Mulbregt,  Paul and Vijaykumar,  Aditya and Bardelli,  Alessandro Pietro and Rothberg,  Alex and Hilboll,  Andreas and Kloeckner,  Andreas and Scopatz,  Anthony and Lee,  Antony and Rokem,  Ariel and Woods,  C. Nathan and Fulton,  Chad and Masson,  Charles and H\"{a}ggstr\"{o}m,  Christian and Fitzgerald,  Clark and Nicholson,  David A. and Hagen,  David R. and Pasechnik,  Dmitrii V. and Olivetti,  Emanuele and Martin,  Eric and Wieser,  Eric and Silva,  Fabrice and Lenders,  Felix and Wilhelm,  Florian and Young,  G. and Price,  Gavin A. and Ingold,  Gert-Ludwig and Allen,  Gregory E. and Lee,  Gregory R. and Audren,  Hervé and Probst,  Irvin and Dietrich,  J\"{o}rg P. and Silterra,  Jacob and Webber,  James T and Slavič,  Janko and Nothman,  Joel and Buchner,  Johannes and Kulick,  Johannes and Sch\"{o}nberger,  Johannes L. and de Miranda Cardoso,  José Vinícius and Reimer,  Joscha and Harrington,  Joseph and Rodríguez,  Juan Luis Cano and Nunez-Iglesias,  Juan and Kuczynski,  Justin and Tritz,  Kevin and Thoma,  Martin and Newville,  Matthew and K\"{u}mmerer,  Matthias and Bolingbroke,  Maximilian and Tartre,  Michael and Pak,  Mikhail and Smith,  Nathaniel J. and Nowaczyk,  Nikolai and Shebanov,  Nikolay and Pavlyk,  Oleksandr and Brodtkorb,  Per A. and Lee,  Perry and McGibbon,  Robert T. and Feldbauer,  Roman and Lewis,  Sam and Tygier,  Sam and Sievert,  Scott and Vigna,  Sebastiano and Peterson,  Stefan and More,  Surhud and Pudlik,  Tadeusz and Oshima,  Takuya and Pingel,  Thomas J. and Robitaille,  Thomas P. and Spura,  Thomas and Jones,  Thouis R. and Cera,  Tim and Leslie,  Tim and Zito,  Tiziano and Krauss,  Tom and Upadhyay,  Utkarsh and Halchenko,  Yaroslav O. and Vázquez-Baeza,  Yoshiki},
  year = {2020},
  month = feb,
  pages = {261–272}
}

@article{Ohashi2015,
  title = {Chemical evolution of the HC3N and N2H+ molecules in dense cores of the Vela C giant molecular cloud complex},
  volume = {68},
  ISSN = {0004-6264},
  url = {http://dx.doi.org/10.1093/pasj/psv104},
  DOI = {10.1093/pasj/psv104},
  number = {1},
  journal = {Publications of the Astronomical Society of Japan},
  publisher = {Oxford University Press (OUP)},
  author = {Ohashi,  Satoshi and Tatematsu,  Ken’ichi and Fujii,  Kosuke and Sanhueza,  Patricio and Nguyen Luong,  Quang and Choi,  Minho and Hirota,  Tomoya and Mizuno,  Norikazu},
  year = {2015},
  month = nov 
}

@article{Lu2025,
  title = {Modeling Complex Organic Molecules’ Formation in Cold Cores: Multiphase Models with Nonthermal Mechanisms},
  volume = {277},
  ISSN = {1538-4365},
  url = {http://dx.doi.org/10.3847/1538-4365/ad9b88},
  DOI = {10.3847/1538-4365/ad9b88},
  number = {1},
  journal = {The Astrophysical Journal Supplement Series},
  publisher = {American Astronomical Society},
  author = {Lu,  Yang and Quan,  Donghui and Chang,  Qiang and Chen,  Long-Fei and Li,  Di},
  year = {2025},
  month = feb,
  pages = {8}
}

@ARTICLE{2016A&A...592L..11S,
       author = {{Spezzano}, S. and {Bizzocchi}, L. and {Caselli}, P. and {Harju}, J. and {Br{\"u}nken}, S.},
        title = "{Chemical differentiation in a prestellar core traces non-uniform illumination}",
      journal = {AAP},
         year = 2016,
        month = aug,
       volume = {592},
          eid = {L11},
        pages = {L11},
          doi = {10.1051/0004-6361/201628652},
archivePrefix = {arXiv},
       eprint = {1607.03242},
 primaryClass = {astro-ph.GA},
       adsurl = {https://ui.adsabs.harvard.edu/abs/2016A&A...592L..11S}
}

@article{Hirahara1992,
  title = {Mapping observations of sulfur-containing carbon-chain molecules in Taurus Molecular Cloud 1 (TMC-1)},
  volume = {394},
  ISSN = {1538-4357},
  url = {http://dx.doi.org/10.1086/171605},
  DOI = {10.1086/171605},
  journal = {The Astrophysical Journal},
  publisher = {American Astronomical Society},
  author = {Hirahara,  Yasuhiro and Suzuki,  Hiroko and Yamamoto,  Satoshi and Kawaguchi,  Kentarou and Kaifu,  Norio and Ohishi,  Masatoshi and Takano,  Shuro and Ishikawa,  Shin-Ichi and Masuda,  Akimasa},
  year = {1992},
  month = aug,
  pages = {539}
}

@ARTICLE{1990A&A...231..466M,
       author = {{Millar}, T.~J. and {Herbst}, E.},
        title = "{Organo-sulphur chemistry in dense interstellar clouds.}",
      journal = {\aap},
         year = 1990,
        month = may,
       volume = {231},
        pages = {466-472},
       adsurl = {https://ui.adsabs.harvard.edu/abs/1990A&A...231..466M}
}

@ARTICLE{1996ApJ...465..795W,
       author = {{Woon}, David E. and {Herbst}, Eric},
        title = "{On the Stability of Interstellar Carbon Clusters: The Rate of the Reaction between C 3 and O}",
      journal = {\apj},
         year = 1996,
        month = jul,
       volume = {465},
        pages = {795},
          doi = {10.1086/177463},
       adsurl = {https://ui.adsabs.harvard.edu/abs/1996ApJ...465..795W}
}

@article{Crapsi2007,
  title = {Observing the gas temperature drop in the high-density nucleus   of L 1544},
  volume = {470},
  ISSN = {1432-0746},
  url = {http://dx.doi.org/10.1051/0004-6361:20077613},
  DOI = {10.1051/0004-6361:20077613},
  number = {1},
  journal = {Astronomy &amp; Astrophysics},
  publisher = {EDP Sciences},
  author = {Crapsi,  A. and Caselli,  P. and Walmsley,  M. C. and Tafalla,  M.},
  year = {2007},
  month = may,
  pages = {221–230}
}

@ARTICLE{1997ApJ...477..204W,
       author = {{Woon}, David E. and {Herbst}, Eric},
        title = "{The Rate of the Reaction between CN and C$_{2}$H$_{2}$ at Interstellar Temperatures}",
      journal = {\apj},
         year = 1997,
        month = mar,
       volume = {477},
       number = {1},
        pages = {204-208},
          doi = {10.1086/303707},
       adsurl = {https://ui.adsabs.harvard.edu/abs/1997ApJ...477..204W}
}

@article{Cherchneff1993,
  title = {The formation of cyanopolyyne molecules in IRC + 10216},
  volume = {410},
  ISSN = {1538-4357},
  url = {http://dx.doi.org/10.1086/172737},
  DOI = {10.1086/172737},
  journal = {The Astrophysical Journal},
  publisher = {American Astronomical Society},
  author = {Cherchneff,  Isabelle and Glassgold,  Alfred E. and Mamon,  Gary A.},
  year = {1993},
  month = jun,
  pages = {188}
}

@article{Suzuki1988,
  title = {Sulfur-bearing carbon chain molecules in interstellar clouds},
  volume = {31},
  ISSN = {0083-6656},
  url = {http://dx.doi.org/10.1016/0083-6656(88)90247-4},
  DOI = {10.1016/0083-6656(88)90247-4},
  journal = {Vistas in Astronomy},
  publisher = {Elsevier BV},
  author = {Suzuki,  Hiroko and Ohishi,  Masatoshi and Kaifu,  Norio and Kasuga,  Takashi and Ishikawa,  Shin-ichi and Miyaji,  Takeshi},
  year = {1988},
  pages = {459–462}
}

@article{Fukuzawa1997,
  title = {Molecular Orbital Study of Neutral‐Neutral Reactions concerning HC3N Formation in Interstellar Space},
  volume = {489},
  ISSN = {1538-4357},
  url = {http://dx.doi.org/10.1086/304782},
  DOI = {10.1086/304782},
  number = {1},
  journal = {The Astrophysical Journal},
  publisher = {American Astronomical Society},
  author = {Fukuzawa,  Kaori and Osamura,  Yoshihiro},
  year = {1997},
  month = nov,
  pages = {113–121}
}

@ARTICLE{1988A&A...200..191S,
       author = {{Smith}, D. and {Adams}, N.~G. and {Giles}, K. and {Herbst}, E.},
        title = "{Organo-sulfur chemistry in dense interstellar clouds via S+ -hydrocarbon reactions.}",
      journal = {\aap},
         year = 1988,
        month = jul,
       volume = {200},
        pages = {191-194},
       adsurl = {https://ui.adsabs.harvard.edu/abs/1988A&A...200..191S}
}

@article{Clary1993,
  title = {Rate constants for chemical reactions of radicals at low temperatures},
  volume = {89},
  ISSN = {1364-5455},
  url = {http://dx.doi.org/10.1039/FT9938902185},
  DOI = {10.1039/ft9938902185},
  number = {13},
  journal = {Journal of the Chemical Society,  Faraday Transactions},
  publisher = {Royal Society of Chemistry (RSC)},
  author = {Clary,  David C. and Stoecklin,  Thierry S. and Wickham,  Andrew G.},
  year = {1993},
  pages = {2185}
}

@article{Smith1995_IJMSIP_149_231,
  author       = {Smith, I. W. M.},
  title        = {---}, 
  journal      = {International Journal of Mass Spectrometry and Ion Processes},
  volume       = {149--150},
  pages        = {231--?}, 
  year         = {1995},
  publisher    = {Elsevier},
  note         = {Issue 149/150, p.\,231}
}

@article{Rowe1993,
  title = {Rate coefficients for interstellar gas-phase chemistry},
  volume = {89},
  ISSN = {1364-5455},
  url = {http://dx.doi.org/10.1039/FT9938902193},
  DOI = {10.1039/ft9938902193},
  number = {13},
  journal = {Journal of the Chemical Society,  Faraday Transactions},
  publisher = {Royal Society of Chemistry (RSC)},
  author = {Rowe,  Bertrand R. and Canosa,  Andr� and Sims,  Ian R.},
  year = {1993},
  pages = {2193}
}

@ARTICLE{2008ApJ...672..371S,
       author = {{Sakai}, Nami and {Sakai}, Takeshi and {Hirota}, Tomoya and {Yamamoto}, Satoshi},
        title = "{Abundant Carbon-Chain Molecules toward the Low-Mass Protostar IRAS 04368+2557 in L1527}",
      journal = {\apj},
         year = 2008,
        month = jan,
       volume = {672},
       number = {1},
        pages = {371-381},
          doi = {10.1086/523635},
       adsurl = {https://ui.adsabs.harvard.edu/abs/2008ApJ...672..371S}
}

@article{deGregorioMonsalvo2006,
  title = {CCS and NH3Emission Associated with Low‐Mass Young Stellar Objects},
  volume = {642},
  ISSN = {1538-4357},
  url = {http://dx.doi.org/10.1086/500657},
  DOI = {10.1086/500657},
  number = {1},
  journal = {The Astrophysical Journal},
  publisher = {American Astronomical Society},
  author = {de Gregorio‐Monsalvo,  Itziar and Gomez,  Jose F. and Suarez,  Olga and Kuiper,  Thomas B. H. and Rodriguez,  Luis F. and Jimenez‐Bailon,  Elena},
  year = {2006},
  month = may,
  pages = {319–329}
}

@INPROCEEDINGS{2007prpl.conf...17D,
       author = {{di Francesco}, J. and {Evans}, N.~J., II and {Caselli}, P. and
         {Myers}, P.~C. and {Shirley}, Y. and {Aikawa}, Y. and {Tafalla}, M.},
        title = "{An Observational Perspective of Low-Mass Dense Cores I: Internal Physical and Chemical Properties}",
    booktitle = {Protostars and Planets V},
         year = 2007,
       editor = {{Reipurth}, Bo and {Jewitt}, David and {Keil}, Klaus},
        month = jan,
        pages = {17},
archivePrefix = {arXiv},
       eprint = {astro-ph/0602379},
 primaryClass = {astro-ph},
       adsurl = {https://ui.adsabs.harvard.edu/abs/2007prpl.conf...17D}
}

@INPROCEEDINGS{2014prpl.conf...27A,
       author = {{Andr{\'e}}, P. and {Di Francesco}, J. and {Ward-Thompson}, D. and
         {Inutsuka}, S. -I. and {Pudritz}, R.~E. and {Pineda}, J.~E.},
        title = "{From Filamentary Networks to Dense Cores in Molecular Clouds: Toward a New Paradigm for Star Formation}",
    booktitle = {Protostars and Planets VI},
         year = 2014,
       editor = {{Beuther}, Henrik and {Klessen}, Ralf S. and {Dullemond}, Cornelis P. and
         {Henning}, Thomas},
        month = jan,
        pages = {27},
          doi = {10.2458/azu_uapress_9780816531240-ch002},
archivePrefix = {arXiv},
       eprint = {1312.6232},
 primaryClass = {astro-ph.GA},
       adsurl = {https://ui.adsabs.harvard.edu/abs/2014prpl.conf...27A}
}

@ARTICLE{2016ApJ...830L...6J,
       author = {{Jim{\'e}nez-Serra}, Izaskun and {Vasyunin}, Anton I. and
         {Caselli}, Paola and {Marcelino}, Nuria and {Billot}, Nicolas and
         {Viti}, Serena and {Testi}, Leonardo and {Vastel}, Charlotte and
         {Lefloch}, Bertrand and {Bachiller}, Rafael},
        title = "{The Spatial Distribution of Complex Organic Molecules in the L1544 Pre-stellar Core}",
      journal = {ApJL},
         year = 2016,
        month = oct,
       volume = {830},
       number = {1},
          eid = {L6},
        pages = {L6},
          doi = {10.3847/2041-8205/830/1/L6},
archivePrefix = {arXiv},
       eprint = {1609.05045},
 primaryClass = {astro-ph.SR},
       adsurl = {https://ui.adsabs.harvard.edu/abs/2016ApJ...830L...6J}
}

@ARTICLE{2020ApJ...891...73S,
       author = {{Scibelli}, Samantha and {Shirley}, Yancy},
        title = "{Prevalence of Complex Organic Molecules in Starless and Prestellar Cores within the Taurus Molecular Cloud}",
      journal = {ApJ},
         year = 2020,
        month = mar,
       volume = {891},
       number = {1},
          eid = {73},
        pages = {73},
          doi = {10.3847/1538-4357/ab7375},
archivePrefix = {arXiv},
       eprint = {2002.02469},
 primaryClass = {astro-ph.GA},
       adsurl = {https://ui.adsabs.harvard.edu/abs/2020ApJ...891...73S}
}

@ARTICLE{2013ApJ...774...22P,
       author = {{Plunkett}, Adele L. and {Arce}, H{\'e}ctor G. and {Corder}, Stuartt A. and {Mardones}, Diego and {Sargent}, Anneila I. and {Schnee}, Scott L.},
        title = "{CARMA Observations of Protostellar Outflows in NGC 1333}",
      journal = {\apj},
         year = 2013,
        month = sep,
       volume = {774},
       number = {1},
          eid = {22},
        pages = {22},
          doi = {10.1088/0004-637X/774/1/22},
archivePrefix = {arXiv},
       eprint = {1307.3558},
 primaryClass = {astro-ph.SR},
       adsurl = {https://ui.adsabs.harvard.edu/abs/2013ApJ...774...22P}
}

@INPROCEEDINGS{2023ASPC..534..233P,
       author = {{Pineda}, J.~E. and {Arzoumanian}, D. and {Andre}, P. and {Friesen}, R.~K. and {Zavagno}, A. and {Clarke}, S.~D. and {Inoue}, T. and {Chen}, C. and {Lee}, Y. and {Soler}, J.~D. and {Kuffmeier}, M.},
        title = "{From Bubbles and Filaments to Cores and Disks: Gas Gathering and Growth of Structure Leading to the Formation of Stellar Systems}",
    booktitle = {Protostars and Planets VII},
         year = 2023,
       editor = {{Inutsuka}, S. and {Aikawa}, Y. and {Muto}, T. and {Tomida}, K. and {Tamura}, M.},
       series = {Astronomical Society of the Pacific Conference Series},
       volume = {534},
        month = jul,
        pages = {233},
          doi = {10.48550/arXiv.2205.03935},
archivePrefix = {arXiv},
       eprint = {2205.03935},
 primaryClass = {astro-ph.GA},
       adsurl = {https://ui.adsabs.harvard.edu/abs/2023ASPC..534..233P}
}

@INPROCEEDINGS{2000prpl.conf...59A,
       author = {{Andre}, P. and {Ward-Thompson}, D. and {Barsony}, M.},
        title = "{From Prestellar Cores to Protostars: the Initial Conditions of Star Formation}",
    booktitle = {Protostars and Planets IV},
         year = 2000,
       editor = {{Mannings}, V. and {Boss}, A.~P. and {Russell}, S.~S.},
        month = may,
        pages = {59},
          doi = {10.48550/arXiv.astro-ph/9903284},
archivePrefix = {arXiv},
       eprint = {astro-ph/9903284},
 primaryClass = {astro-ph},
       adsurl = {https://ui.adsabs.harvard.edu/abs/2000prpl.conf...59A}
}

@article{WardThompson2007,
  title = {The James Clerk Maxwell Telescope Legacy Survey of Nearby Star‐forming Regions in the Gould Belt},
  volume = {119},
  ISSN = {1538-3873},
  url = {http://dx.doi.org/10.1086/521277},
  DOI = {10.1086/521277},
  number = {858},
  journal = {Publications of the Astronomical Society of the Pacific},
  publisher = {IOP Publishing},
  author = {Ward‐Thompson,  D. and Di Francesco,  J. and Hatchell,  J. and Hogerheijde,  M. R. and Nutter,  D. and Bastien,  P. and Basu,  S. and Bonnell,  I. and Bowey,  J. and Brunt,  C. and Buckle,  J. and Butner,  H. and Cavanagh,  B. and Chrysostomou,  A. and Curtis,  E. and Davis,  C. J. and Dent,  W. R. F. and van Dishoeck,  E. and Edmunds,  M. G. and Fich,  M. and Fiege,  J. and Fissel,  L. and Friberg,  P. and Friesen,  R. and Frieswijk,  W. and Fuller,  G. A. and Gosling,  A. and Graves,  S. and Greaves,  J. S. and Helmich,  F. and Hills,  R. E. and Holland,  W. S. and Houde,  M. and Jayawardhana,  R. and Johnstone,  D. and Joncas,  G. and Kirk,  H. and Kirk,  J. M. and Knee,  L. B. G. and Matthews,  B. and Matthews,  H. and Matzner,  C. and Moriarty‐Schieven,  G. H. and Naylor,  D. and Padman,  R. and Plume,  R. and Rawlings,  J. M. C. and Redman,  R. O. and Reid,  M. and Richer,  J. S. and Shipman,  R. and Simpson,  R. J. and Spaans,  M. and Stamatellos,  D. and Tsamis,  Y. G. and Viti,  S. and Weferling,  B. and White,  G. J. and Whitworth,  A. P. and Wouterloot,  J. and Yates,  J. and Zhu,  M.},
  year = {2007},
  month = aug,
  pages = {855–870}
}

@ARTICLE{2005ApJ...632..982S,
       author = {{Shirley}, Yancy L. and {Nordhaus}, Miranda K. and {Grcevich}, Jana M. and
         {Evans}, Neal J., II and {Rawlings}, Jonathan M.~C. and
         {Tatematsu}, Ken'ichi},
        title = "{Modeling the Physical Structure of the Low-Density Pre-Protostellar Core Lynds 1498}",
      journal = {\apj},
         year = 2005,
        month = oct,
       volume = {632},
       number = {2},
        pages = {982-1000},
          doi = {10.1086/431963},
archivePrefix = {arXiv},
       eprint = {astro-ph/0505171},
 primaryClass = {astro-ph},
       adsurl = {https://ui.adsabs.harvard.edu/abs/2005ApJ...632..982S}
}

@ARTICLE{2015PASP..127..266M,
       author = {{Mangum}, Jeffrey G. and {Shirley}, Yancy L.},
        title = "{How to Calculate Molecular Column Density}",
      journal = {\pasp},
         year = 2015,
        month = mar,
       volume = {127},
       number = {949},
        pages = {266},
          doi = {10.1086/680323},
archivePrefix = {arXiv},
       eprint = {1501.01703},
 primaryClass = {astro-ph.IM},
       adsurl = {https://ui.adsabs.harvard.edu/abs/2015PASP..127..266M}
}

@ARTICLE{2020A&A...633A.118L,
       author = {{Lattanzi}, Valerio and {Bizzocchi}, Luca and {Vasyunin}, Anton I. and
         {Harju}, Jorma and {Giuliano}, Barbara M. and {Vastel}, Charlotte and
         {Caselli}, Paola},
        title = "{Molecular complexity in pre-stellar cores: a 3 mm-band study of L183 and L1544}",
      journal = {AAP},
         year = 2020,
        month = jan,
       volume = {633},
          eid = {A118},
        pages = {A118},
          doi = {10.1051/0004-6361/201936884},
       adsurl = {https://ui.adsabs.harvard.edu/abs/2020A&A...633A.118L}
}

@ARTICLE{2013A&A...558A..33A,
       author = {{Astropy Collaboration} and {Robitaille}, Thomas P. and
         {Tollerud}, Erik J. and {Greenfield}, Perry and {Droettboom}, Michael and
         {Bray}, Erik and {Aldcroft}, Tom and {Davis}, Matt and
         {Ginsburg}, Adam and {Price-Whelan}, Adrian M. and
         {Kerzendorf}, Wolfgang E. and {Conley}, Alexander and {Crighton}, Neil and
         {Barbary}, Kyle and {Muna}, Demitri and {Ferguson}, Henry and
         {Grollier}, Fr{\'e}d{\'e}ric and {Parikh}, Madhura M. and
         {Nair}, Prasanth H. and {Unther}, Hans M. and {Deil}, Christoph and
         {Woillez}, Julien and {Conseil}, Simon and {Kramer}, Roban and
         {Turner}, James E.~H. and {Singer}, Leo and {Fox}, Ryan and
         {Weaver}, Benjamin A. and {Zabalza}, Victor and {Edwards}, Zachary I. and
         {Azalee Bostroem}, K. and {Burke}, D.~J. and {Casey}, Andrew R. and
         {Crawford}, Steven M. and {Dencheva}, Nadia and {Ely}, Justin and
         {Jenness}, Tim and {Labrie}, Kathleen and {Lim}, Pey Lian and
         {Pierfederici}, Francesco and {Pontzen}, Andrew and {Ptak}, Andy and
         {Refsdal}, Brian and {Servillat}, Mathieu and {Streicher}, Ole},
        title = "{Astropy: A community Python package for astronomy}",
      journal = {AAP},
         year = 2013,
        month = oct,
       volume = {558},
          eid = {A33},
        pages = {A33},
          doi = {10.1051/0004-6361/201322068},
archivePrefix = {arXiv},
       eprint = {1307.6212},
 primaryClass = {astro-ph.IM},
       adsurl = {https://ui.adsabs.harvard.edu/abs/2013A&A...558A..33A}
}

@ARTICLE{2007CSE.....9...90H,
       author = {{Hunter}, John D.},
        title = "{Matplotlib: A 2D Graphics Environment}",
      journal = {Computing in Science and Engineering},
         year = 2007,
        month = may,
       volume = {9},
       number = {3},
        pages = {90-95},
          doi = {10.1109/MCSE.2007.55},
       adsurl = {https://ui.adsabs.harvard.edu/abs/2007CSE.....9...90H}
}

@ARTICLE{2020arXiv200610256H,
       author = {{Harris}, Charles R. and {Jarrod Millman}, K. and
         {van der Walt}, St{\'e}fan J. and {Gommers}, Ralf and
         {Virtanen}, Pauli and {Cournapeau}, David and {Wieser}, Eric and
         {Taylor}, Julian and {Berg}, Sebastian and {Smith}, Nathaniel J. and
         {Kern}, Robert and {Picus}, Matti and {Hoyer}, Stephan and
         {van Kerkwijk}, Marten H. and {Brett}, Matthew and {Haldane}, Allan and
         {Fern{\'a}ndez del R{\'\i}o}, Jaime and {Wiebe}, Mark and
         {Peterson}, Pearu and {G{\'e}rard-Marchant}, Pierre and
         {Sheppard}, Kevin and {Reddy}, Tyler and {Weckesser}, Warren and
         {Abbasi}, Hameer and {Gohlke}, Christoph and {Oliphant}, Travis E.},
        title = "{Array Programming with NumPy}",
      journal = {arXiv e-prints},
         year = 2020,
        month = jun,
          eid = {arXiv:2006.10256},
        pages = {arXiv:2006.10256},
archivePrefix = {arXiv},
       eprint = {2006.10256},
 primaryClass = {cs.MS},
       adsurl = {https://ui.adsabs.harvard.edu/abs/2020arXiv200610256H}
}

@ARTICLE{2022arXiv220613270C,
       author = {{Ceccarelli}, C. and {Codella}, C. and {Balucani}, N. and {Bockel{\'e}e-Morvan}, D. and {Herbst}, E. and {Vastel}, C. and {Caselli}, P. and {Favre}, C. and {Lefloch}, B. and {{\"O}berg}, K.},
        title = "{Organic chemistry in the first phases of Solar-type protostars}",
      journal = {arXiv e-prints},
         year = 2022,
        month = jun,
          eid = {arXiv:2206.13270},
        pages = {arXiv:2206.13270},
          doi = {10.48550/arXiv.2206.13270},
archivePrefix = {arXiv},
       eprint = {2206.13270},
 primaryClass = {astro-ph.SR},
       adsurl = {https://ui.adsabs.harvard.edu/abs/2022arXiv220613270C}
}

@ARTICLE{2025ApJS..281....9X,
       author = {{Xue}, Ci and {Byrne}, Alex N. and {Morgan}, Larry and {Wenzel}, Gabi and {Changala}, P. Bryan and {Fried}, Zachary T.~P. and {Loomis}, Ryan A. and {Remijan}, Anthony and {Bergin}, Edwin A. and {Cooke}, Ilsa R. and {Frayer}, David and {Burkhardt}, Andrew M. and {Charnley}, Steven B. and {Cordiner}, Martin A. and {Lipnicky}, Andrew and {McCarthy}, Michael C. and {McGuire}, Brett A.},
        title = "{The Molecular Inventory of TMC-1 with GOTHAM Observations}",
      journal = {\apjs},
         year = 2025,
        month = nov,
       volume = {281},
       number = {1},
          eid = {9},
        pages = {9},
          doi = {10.3847/1538-4365/ae04e5},
archivePrefix = {arXiv},
       eprint = {2509.06256},
 primaryClass = {astro-ph.GA},
       adsurl = {https://ui.adsabs.harvard.edu/abs/2025ApJS..281....9X}
}


\appendix
\raggedbottom
\onecolumn 

\section{Source Catalog}
Additional details about the 15 prestellar and starless cores in the Perseus Molecular Cloud are listed here in Table \ref{sources}. $^2$RA and DEC coordinates correspond with \textit{Herschel} source information. $^3$Source distance, $D$, is calculated as the mean distance for each region from Table 5 in \cite{2021A&A...645A..55P} and is listed in units of parsecs. $^4\mathrm{H_2}$ column and volume densities in this table are derived from the median values of \textit{Herschel} maps within the ARO 12m 62" beam. 
$^5$Kinetic temperatures, $T_{\rm k}$, were computed via RADEX calculations with NH$_{3}$ \citep{2008ApJS..175..509R}. The 15 sources are split into four separate star-forming regions of Perseus: NGC 1333, B1, IC 348, and B5.

\begin{table*}
\centering
\caption{$^1$Catalog of Perseus prestellar and starless cores from \textit{Herschel} \citep{2021A&A...645A..55P}. 
This table information is adopted from table 1 in \citealt{Scibelli2024}.}
\begin{tabular}{cccccccc}
Region & Core \#$^1$ & RA$^2$ (J200) & DEC$^2$ (J200) & $D^3$ (pc) & $N(\rm H_2)^4 \times 10^{22}\,\mathrm{cm}^{-2}$ & $n(\rm H_2)^4 \times 10^{5}\,\mathrm{cm}^{-3}$ & $T_{\rm k}$ (K) \\ \hline 
NGC 1333 & 264 &  03:28:47.14 & +31:15:11.4 & 294 & 2.06 & 0.75 & 10.8 \\
& 317 & 03:29:04.93 & +31:18:44.4 & 294 & 2.63 & 0.96  & 13.6 \\
& 321 & 03:29:07.17 &  +31:17:22.1 & 294 & 3.65 &  1.33 & 12.6 \\
& 326 & 03:29:08.97 & +31:15:17.2 & 294 & 7.88 & 2.87 & 12.3 \\
B1 & 413 & 03:30:46.74 & +30:52:44.8 & 297 & 1.17 & 0.42 & 10.5 \\
& 504 &  03:33:25.31 & +31:05:37.5 & 297 & 1.80 & 0.65 & 9.7 \\
IC 348 & 615 & 03:40:14.92 & +32:01:40.8 & 314 & 1.03 & 0.35 &  10.3\\
& 627 &  03:40:49.53 & +31:48:40.5 & 314 & 0.93 & 0.32 & 12.4 \\
& 709 & 03:43:38.06 &  +32:03:07.4 & 314 & 1.48 & 0.51 & 13.5 \\
& 715 & 03:43:46.34 & +32:01:43.5 & 314 & 1.45 & 0.49 & 14.8 \\
& 746 &  03:44:14.38 & +31:58:00.7 & 314 & 1.42 & 0.49 & 10.6 \\
& 752 & 03:44:23.10 & +32:10:01.0 & 314 & 0.63 & 0.22 & 14.7 \\
& 768 &  03:44:48.83 & +32:00:31.6 & 314 & 1.43 & 0.49 & 10.8 \\
B5 & 799 & 03:47:31.31 & +32:50:56.9 & 325 & 1.01 & 0.33 & 10.4 \\
 & 800 & 03:47:38.97 & +32:52:16.6 & 325 & 1.93 & 0.64 & 11.7
\end{tabular}
\label{sources}
\end{table*}

\section{Diagnostic Plots}
\begin{figure}
    \centering
    \includegraphics[width=18.0cm]{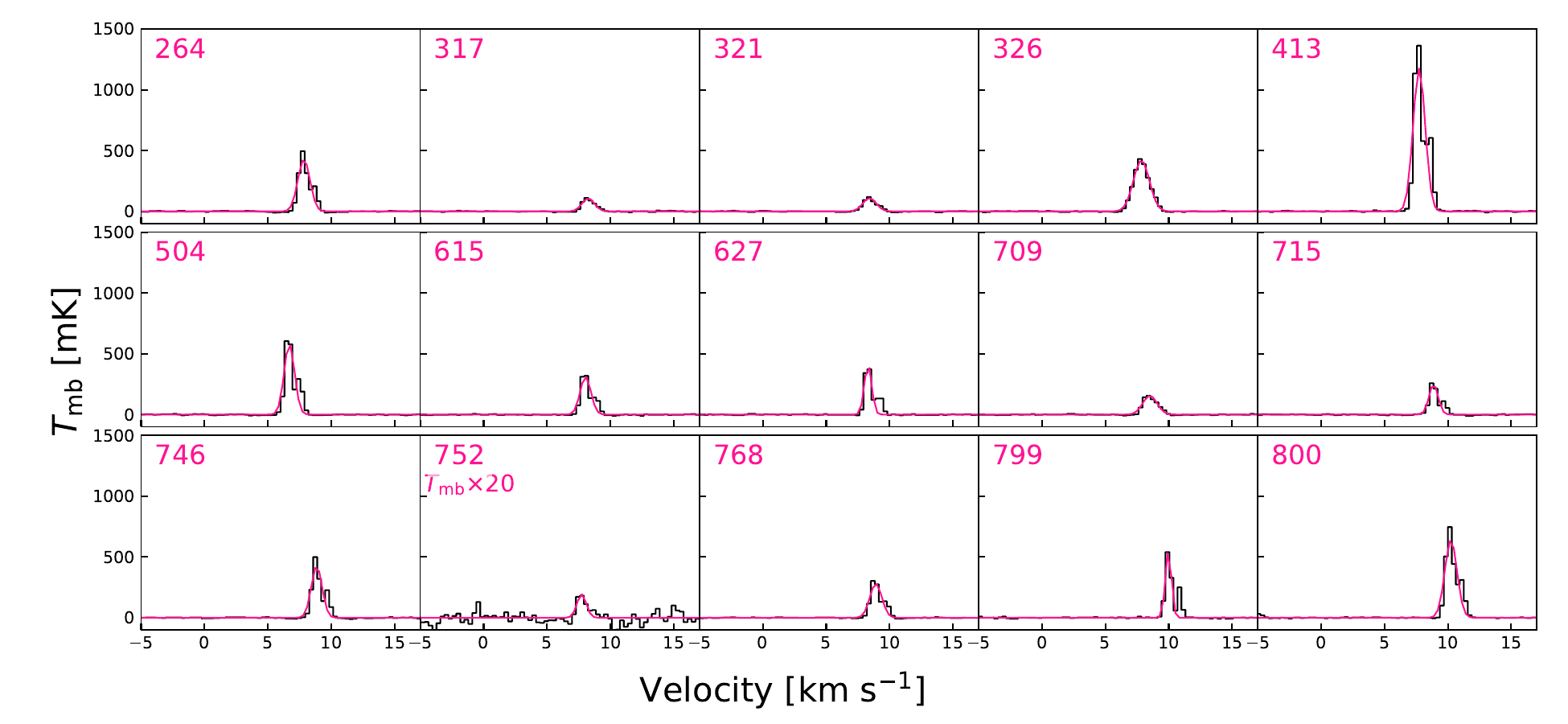}
    \caption{Diagnostic plots of cyanoacetylene, HC$_3$N (J=4-3, F=3-2). The Gaussian fit detected in all 15 cores produced by our data pipeline, using Pyspeckit \citep{2011ascl.soft09001G, Ginsburg2022}, is shown in magenta. Core 752 has a 3$\sigma$ preliminary detection, so the intensity has been magnified by 20$\times$ for visual and comparison purposes. We do not have high enough resolution to distinguish hyperfine components or the smaller peaks, so we only fit the main peak of HC$_3$N transitions.}
    \label{HC3Ndiagnostic}
\end{figure}

\begin{figure}
    \centering
    \includegraphics[width=\textwidth]{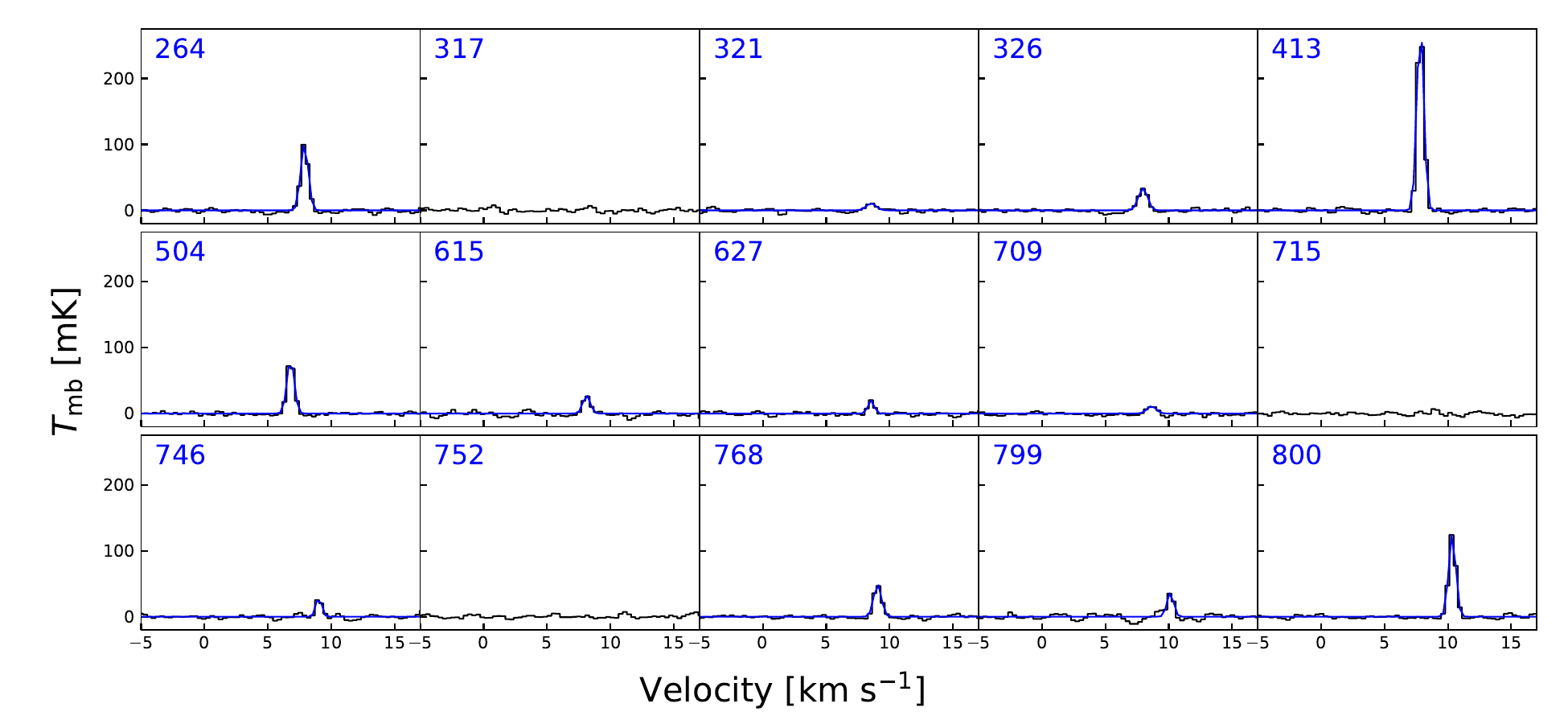}
    \caption{Diagnostic plots of HC$_5$N (J=13-12). Gaussian fits produced by our data pipeline, using Pyspeckit \citep{2011ascl.soft09001G, Ginsburg2022}, are shown in blue. The HC$_5$N J=13-12 line was not detected in cores 317, 715, and 752. } \label{HC5N_diagnostic}
\end{figure}

Here, we present diagnostic plots for detections of HC$_3$N and HC$_5$N in Figures \ref{HC3Ndiagnostic} and \ref{HC5N_diagnostic}, respectively. The HC$_3$N (J=4-3, F=3-2) lines contain some hyperfine structure; however, we would need higher resolution to be able to resolve the individual hyperfine components. Therefore, we only fit the main peak for HC$_3$N lines, shown in Figure \ref{HC3Ndiagnostic}. Core 752 exhibits weaker HC$_3$N emission compared to the other cores, so we magnify its spectrum by 20$\times$ for better visual comparison. For the HC$_5$N (J=13-12) transition, we find either confirmed or tentative detections in all cores except 317, 715, and 752 (Figure \ref{HC5N_diagnostic}). We chose to present this particular transition because it contained the most detections in the Perseus core sample out of all available HC$_5$N lines.

\newpage 

\section{CS Column Density and $^{12}$C/$^{13}$C Correction Methods} \label{appendix:CS}

As discussed in Section \ref{sec:3.2}, we explore multiple methods for calculating column densities and abundances due to the breakdown of the optically thin assumption for CS. Column densities are first derived via LTE under an optically thin assumption (via the FT method). However, after comparing their $^{12}\rm C/^{13}C$ ratios to the ISM value of 68, we determine that the optically thin assumption underestimates CS column densities for most cores. 

We employ \texttt{pythonradex}\footnote{\url{https://github.com/gica3618/pythonradex}} as the second method to optimize CS column density values. \texttt{pythonradex} is a Python re-implementation of the non-LTE radiative transfer code, RADEX. We note that in the default static spherical geometry, RADEX derives the Rayleigh-Jeans brightness temperature from the optical depth using the escape probabilities for a uniform slab, given by the standard expression: 
\begin{equation}
    I_\nu(\tau) = B_\nu (T_{\rm ex})(1-e^{-\tau}),
\end{equation}
where $I_\nu$ is the specific intensity, $B_\nu$ is the Planck function evaluated at the excitation temperature of the transition ($T_{\rm ex})$, and $\tau$ is the optical depth. For a static sphere, the correct intensity expression should also account for the background emission. In the optically thin limit, the slab expression overestimates the flux by a factor of 3/2 with respect to that of the static sphere, implying that the derived column densities are 2/3 lower than expected. However, the optical depth reported by RADEX for static sphere geometry corresponds to the maximum value along the line-of-sight through the center of the sphere. We would expect for RADEX to instead give an average column density since the sphere is contained within the telescope beam. For a sphere, however, the average column density and optical depth are 2/3 of the maximum value found at the center of the sphere. Geometrically, this follows from the fact that the average chord length through a sphere is 2/3 of its diameter, so the surface-averaged optical depth and column density are 2/3 of the central value. We find that these two effects compensate each other, and the use of the slab intensity formula reduces the derived column density by a factor of 2/3 relative to the correct spherical solution and converts the central to surface-averaged column density by the same factor. Consequently, in the optically thin regime, the geometric inconsistencies in RADEX become effectively negligible, as they yield similar numerical results.
 
Since \texttt{pythonradex} uses correct escape probabilities for all geometries, we use it in place of RADEX to reduce systematic uncertainty in our column density derivations. We adopt a static slab geometry to remain consistent with the slab geometry assumed by our previous LTE analysis. Collisional rate coefficients used in \texttt{pythonradex} were obtained from the Leiden Atomic and Molecular Database
(LAMBDA; \citealt{2005A&A...432..369S}). The CS collisional rates with ortho and para H$_2$ were adopted from \cite{DenisAlpizar2018}. We assume that H$_2$ is in the para state, consistent with the low ortho-to-para ratios expected in cold, dense prestellar cores. Under these conditions, para-H$_2$ dominates the collisional excitation of CS \citep{Sipil2013}. This is standard practice, but we confirm that the assumption of purely para-H$_2$ yields very similar results compared to using the thermal average. The background radiation temperature was set to 2.73 K, and dust emission was neglected, and we assume the line profiles to follow a Gaussian distribution. For each of the 15 sources, we compute a grid of 500 models log-spaced in $N$(CS) from 10$^{9}$ to 10$^{15}$ cm$^{-2}$, using the kinetic temperatures (T$_{\rm kin}$) and \textit{Herschel} \textit{n}(H$_2$) densities from Table 1, column 9 of \cite{Scibelli2024} and line widths derived from observed CS line profiles. The best-fit column density is determined by matching the modeled background-subtracted Rayleigh-Jeans brightness temperature to the observed main-beam temperature, $T_{\rm mb}$. Column density errors for \texttt{pythonradex} are estimated by accounting for $T_{\rm mb}$ errors and 20\% uncertainties in \textit{Herschel}-derived volume densities, \textit{n}(H$_2$), and kinetic temperatures, $T_{\rm kin}$ \citep{Scibelli2024}.

Given the additional assumptions introduced by non-LTE modeling, we explored an independent optical depth correction approach. Using Equations \ref{eq:5} and \ref{eq:6}, we derived an optical thick correction factor normalized to the ISM $^{12}\rm C/^{13}C$ value of 68 by first estimating the optical depth of CS based on the integrated intensities and column densities of $^{13}$CS and CS \citep{1999ApJ...517..209G,2015PASP..127..266M}. The column density derived under the optically thin limit ($N_{\rm FT}$) was then corrected based on the calculated optical depth, $\tau$, seen in Figure \ref{cs} and Table \ref{appendix:CS}. After comparing the three methods, we decided to use the optical depth correction factor for CS column densities and abundances throughout the paper, since \texttt{pythonradex} introduces additional uncertainties due to the collisional rate coefficients. Additionally, higher $^{12}\rm C/^{13}\rm C$ ratios by \texttt{pythonradex} reveal that it might be overestimating CS content. Core 752 is the only core in which we continue to use the column density with the optically thin assumption ($N_{\rm FT}$) because its derived $\tau$ suggested that it was indeed already optically thin. 

\begin{table*} 
\centering
\caption{CS column density values compared between optical thick (OT) correction factor, fixed temperature LTE (FT), and \texttt{pythonradex} (RADEX) methods. Column densities are in units of cm$^{-2}$. No value was calculated for core 752 using the OT method because CS is likely optically thin, so the correction factor is deemed unnecessary, and the FT column density is used throughout the study instead. We include more significant digits for the uncertainties of $N_{\rm FT}$ due to its smaller range of uncertainties compared to other methods.}
\begin{tabular}{ccccccc}
\multicolumn{1}{l}{Core \#} & $N_{\rm OT}$ ($\times$ 10$^{12}$) & \multicolumn{1}{l}{$N_{\rm FT}$ ($\times$ 10$^{12}$)} & \multicolumn{1}{l}{$N_{\rm RADEX}$ ($\times$ 10$^{12}$)} & \multicolumn{1}{l}{$^{12}$C/$^{13}$C (OT)} & \multicolumn{1}{l}{$^{12}$C/$^{13}$C (FT)} & \multicolumn{1}{l}{$^{12}$C/$^{13}$C (RADEX)} \\ \hline 
264 & 14.58 (0.09) & 10.549 (0.003) & 42.66 (2.36) & 63.01 (0.89) & 45.59 (0.58) & 184.85 (10.48) \\
317 & 16.47 (0.33) & 8.031 (0.003) & 36.06 (1.66) & 62.90 (1.45) & 30.86 (0.35) & 138.85 (6.57) \\
321 & 18.51 (0.27) & 9.542 (0.003) & 51.17 (2.95) & 62.51 (1.08) & 32.23 (0.31) & 172.97 (10.1) \\
326 & 66.02 (1.59) & 20.96 (0.003) & 155.96 (23.34) & 63.60 (1.55) & 20.19 (0.06) & 150.00 (22.41) \\
413 & 21.84 (5.44) & 4.477 (0.003) & 19.45 (1.12) & 64.27 (16.02) & 13.18 (0.09) & 57.35 (3.32) \\
504 & 17.05 (0.27) & 8.462 (0.003) & 41.40 (2.00) & 63.00 (1.15) & 31.27 (0.28) & 152.77 (7.51) \\
615 & 15.85 (0.01) & 5.785 (0.003) & 25.35 (1.87) & 66.50 (0.67) & 24.27 (0.25) & 106.72 (7.93)\\
627 & 9.60 (0.41) & 3.9343 (0.0002) & 18.16 (1.96) & 63.13 (2.83) & 25.88 (0.35) & 119.74 (13.00) \\
709 & 9.91 (0.01) & 9.514 (0.003) & 48.08 (4.21) & 62.03 (1.01) & 59.57 (0.97) & 300.62 (26.76) \\
715 & 10.77 (0.07) & 7.658 (0.003) & 38.02 (2.71) & 62.29 (0.92) & 44.29 (0.58) & 219.65 (15.93) \\
746 & 10.01 (0.09) & 7.010 (0.003) & 31.62 (1.97) & 62.61 (1.11) & 43.84 (0.68) & 197.5 (12.69) \\
752 & -- & 8.174 (0.003) & 34.04 (3.61) & -- & 63.55 (1.37) & 263.57 (28.55) \\
768 & 9.24 (0.04) & 7.658 (0.003) & 31.55 (2.11) & 62.79 (1.22) & 52.06 (0.98) & 214.97 (14.92) \\
799 & 10.31 (0.26) & 4.693 (0.003) & 19.32 (1.38) & 65.69 (1.85) & 30.99 (0.38) & 122.93 (8.92) \\
800 & 21.51 (0.65) & 8.208 (0.003) & 38.02 (3.06) & 63.35 (1.96) & 23.57 (0.16) & 111.76 (9.03)
\end{tabular}
\label{CSmethods}
\end{table*}

\section{Column Density, Abundances, and Excitation Temperature Tables}
In Tables \ref{tab:CS} - \ref{tab:HC7N}, we list column densities, abundances, and excitation temperatures for each detected molecule in the sample of 15 prestellar cores. We also list the method used to derive each column density, including fixed temperature LTE (FT), optically thick correction factor (OT), and rotation diagram (RD). The OT method technically applies the FT method and adds an optical depth correction factor to it. All three methods are LTE, but the FT and OT methods assume a previously fixed excitation temperature, and the RD method derives an independent excitation temperature. We include upper limits for non-detections, denoted by the $\textless$ symbol. When the FT method was utilized, we assume a $T_{\rm ex}$ from CCS or HC$_3$N, so the errors are not explicitly listed but can be found in Tables \ref{tab:CCS} and \ref{tab:HC3N}, respectively. While we find this assumption reasonable, we also recalculate the column densities derived from the FT method over a spread of excitation temperatures, ranging from 5--20 K, in order to evaluate the variation. For species CS, $^{13}$CS, C$^{34}$S, C$^{13}$CS, CC$^{34}$S, and CCCS, column densities vary on average by factors of 1.4$\times$, 1.4$\times$, 1.4$\times$, 5.4$\times$, 1.9$\times$, and 3.6$\times$, respectively. For species HC$_3$N, DC$_3$N, H$^{13}$CCCN, HC$^{13}$CCN, HCC$^{13}$CN, HC$_5$N, HC$^{13}$CCCCN, HCC$^{13}$CCCN, HCCC$^{13}$CCN, HCCCCC$^{15}$N, DCCCCCN, and HC$_7$N, column densities vary on average by factors of 1.8$\times$, 2.5$\times$, 2.6$\times$, 2.1$\times$, 2.5$\times$, 5.7$\times$, 4.1$\times$, 14$\times$, 9.6$\times$, 35$\times$, 6.5$\times$, and 53$\times$, respectively. For species HCC$^{13}$CCCN, HCCCCC$^{15}$N, and HC$_7$N, these column densities likely vary more because most of these cores were classified as tentative detections. In future analysis, more observations of additional transitions can help us confirm our estimates on excitation temperatures.

\begin{table*}
\centering
\caption{CS and isotopologues: column densities, abundances, and excitation temperatures with upper limit calculations. Isotopologue column densities calculated via LTE were derived using excitation temperatures obtained by CCS transitions.} \label{tab:CS}
\setlength{\tabcolsep}{12pt}
\renewcommand{\arraystretch}{1} 
 \setlength{\tabcolsep}{12pt}
	\begin{tabular}{rrrrrrr} 
     \tabcolsep=0.4cm
Core \# & $N(\rm X)\times$10$^{13}$ [cm$^{-2}$]& $N(\rm X)$/$N(\rm H_2)$$\times$10$^{-10}$ & $T_{\rm ex}$ [K] & Method\\
\hline 
\multicolumn{5}{c}{CS} \\ 
\hline 
264 & 1.458 (0.009) & 7.1 (1.5) & 4.05 & OT \\
317 & 1.64 (0.03) & 6.2 (1.4) & 5.52 & OT \\
321 & 1.85 (0.03) & 5.1 (1.1) & 7.11 & OT \\
326 & 6.60 (0.16) & 8.4 (1.9) & 5.36 & OT \\
413 & 2.2 (0.5) & 19 (8) & 5.98 & OT \\
504 & 1.71 (0.03) & 9.5 (2.0) & 5.15 & OT \\
615 & 1.5855 (0.0006) & 15 (3) & 4.35 & OT \\
627 & 1.0 (0.4) & 10.3 (2.5) & 5.55 & OT \\
709 & 0.9906 (0.0008) & 6.7 (1.3) & 6.26 & OT \\
715 & 1.077 (0.007) & 7.4 (1.5) & 6.30 & OT \\
746 & 1.001 (0.009) & 7.1 (1.5) & 5.08 & OT \\
752 & 0.8174 (0.0003) & 13 (3) & 4.27 & FT \\
768 & 0.924 (0.004) & 6.5 (1.3) & 4.27 & OT \\
799 & 1.03 (0.03) & 10.2 (2.3) & 4.41 & OT \\
800 & 2.15 (0.07) & 11.2 (2.6) & 5.01 & OT \\
\hline 
\hline 
Core \# & $N(\rm X)\times$10$^{11}$ [cm$^{-2}$]& $N(\rm X)$/$N(\rm H_2)$$\times$10$^{-11}$ & $T_{\rm ex}$ [K] & Method\\
\hline 
\multicolumn{5}{c}{$^{13}$CS} \\ 
\hline 
264 & 2.31 (0.03) & 1.12 (0.24) & 4.05 & FT\\
317 & 2.60 (0.03) & 0.99 (0.21) & 5.52 & FT\\
321 & 2.96 (0.03) & 0.81 (0.17) & 7.11 & FT\\
326 & 10.38 (0.03) & 1.32 (0.27) & 5.36 & FT\\
413 & 3.398 (0.025) & 2.9 (0.6) & 5.98 & FT\\
504 & 2.71 (0.03) & 1.5 (0.3) & 5.15 & FT\\
615 & 2.38 (0.03) & 2.4 (0.5) & 4.35 & FT\\
627 & 1.520 (0.022) & 1.6 (0.4) & 5.55 & FT\\
709 & 1.60 (0.03) & 1.08 (0.24) & 6.26 & FT\\
715 & 1.729 (0.024) & 1.19 (0.26) & 6.30 & FT\\
746 & 1.60 (0.03) & 1.13 (0.24) & 5.08 & FT\\
752 & 1.29 (0.03) & 2.0 (0.5) & 4.27 & FT\\
768 & 1.47 (0.03) & 1.03 (0.23) & 4.27 & FT\\
799 & 1.570 (0.021) & 1.56 (0.33) & 4.41 & FT\\
800 & 3.40 (0.03) & 1.8 (0.4) & 5.01 & FT\\
\hline 
\hline 
Core \# & $N(\rm X)\times$10$^{11}$ [cm$^{-2}$]& $N(\rm X)$/$N(\rm H_2)$$\times$10$^{-11}$ & $T_{\rm ex}$ [K]& Method\\
\hline 
\multicolumn{5}{c}{C$^{34}$S} \\ 
\hline 
264 & 10.60 (0.03) & 5.2 (1.1) & 4.05 & FT\\
317 & 7.76 (0.03) & 3.0 (0.6) & 5.52 & FT\\
321 & 10.00 (0.03) & 2.7 (0.6) & 7.11 & FT\\
326 & 30.59 (0.03) & 3.9 (0.8) & 5.36 & FT\\
413 & 9.129 (0.023) & 7.8 (1.6) & 5.98 & FT\\
504 & 13.39 (0.03) & 7.4 (1.5) & 5.15 & FT\\
615 & 13.02 (0.03) & 13 (3) & 4.35 & FT\\
627 & 7.572 (0.021) & 8.1 (1.7) & 5.55 & FT\\
709 & 6.74 (0.03) & 4.6 (0.9) & 6.26 & FT\\
715 & 8.819 (0.022) & 6.1 (1.2) & 6.30 & FT\\
746 & 7.05 (0.03) & 5.0 (1.0) & 5.08 & FT\\
752 & 6.13 (0.03) & 9.7 (2.0) & 4.27 & FT\\
768 & 6.75 (0.03) & 4.7 (1.0) & 4.27 & FT\\
799 & 7.365 (0.021) & 7.3 (1.5) & 4.41 & FT\\
800 & 14.1 (0.19) & 7.5 (1.6) & 5.01 & FT \\
\end{tabular}
\end{table*}

\begin{table*}
\centering
\caption{CCS and isotopologues: column densities, abundances, and excitation temperatures with upper limit calculations. Upper limits are denoted by the `\textless' symbol. Isotopologue column densities calculated via FT were derived using excitation temperatures obtained by CCS transitions.} \label{tab:CCS}
\setlength{\tabcolsep}{12pt}
\renewcommand{\arraystretch}{1} 
 \setlength{\tabcolsep}{12pt}
	\begin{tabular}{rrrrrrr} 
     \tabcolsep=0.4cm
Core \# & $N(\rm X)\times$10$^{11}$ [cm$^{-2}$] & $N(\rm X)$/$N(\rm H_2)$$\times$10$^{-11}$ & $T_\mathrm{ex}$ [K] & Method \\
\hline 
\multicolumn{5}{c}{CCS} \\ 
\hline 
264 & 7.9 (2.3)& 3.8 (1.9)& 4.05 (0.59) & RD\\
317 & 4.6 (0.7)& 1.8 (0.6)& 5.52 (0.50) & RD\\
321 & 4.5 (0.8)& 1.2 (0.5)& 7.11 (0.90) & RD\\
326 & 14 (3)& 1.8 (0.8)& 5.36 (0.68) & RD\\
413 & 20 (3)& 17 (6)  & 5.98 (0.48) & RD \\
504 & 9.4 (1.1)& 5.2 (1.7)& 5.15 (0.34) & RD\\
615 & 14 (2)& 13 (5)  & 4.35 (0.33) & RD\\
627 & 6.6 (1.8)& 7.1 (3.4)& 5.55 (0.90) & RD\\
709 & 2.8 (0.8)& 1.9 (0.9)& 6.26 (1.38) & RD\\
715 & 3.9 (0.6) & 2.7 (1.0)& 6.30 (0.68) & RD\\
746 & 6.8 (0.9)& 4.8 (1.6)& 5.02 (0.35) & RD\\
752 & 0.88 (0.18) & 1.4 (0.6)& 4.27 (0.79) & RD\\
768 & 7.0 (2.1)& 4.9 (2.4)& 4.27 (0.57) & RD\\
799 & 9.6 (1.6)& 9.5 (3.5)& 4.41 (0.34) & RD\\
800 & 15 (2)& 7.6 (2.7)& 5.01 (0.41) & RD\\
\hline 
\hline
Core \# & N(X) x 10$^{10}$ [cm$^{-2}$] & N(X)/N(H$_2$) x 10$^{-12}$ & $T_{\rm ex}$ [K] & Method\\
\hline
\multicolumn{5}{c}{C$^{13}$CS} \\ 
\hline 
264 & \textless 372 & \textless 181 & 4.05 & FT\\
317 & \textless 84 & \textless 32 & 5.52 & FT\\
321 & \textless 21 & \textless 5.9 & 7.11 & FT\\
326 & \textless 71 & \textless 9.0 & 5.36 & FT\\
413 & 6.1 (0.9) & 5.23 (1.83) & 5.98 & FT\\
504 & 54 (7) & 30 (10) & 5.15 & FT\\
615 & \textless 267 & \textless 259 & 4.35 & FT\\
627 & \textless 82 & \textless 88 & 5.55 & FT\\
709 & \textless 49 & \textless 33 & 6.26 & FT\\
715 & \textless 44 & \textless 30 & 6.30 & FT\\
746 & \textless 116 & \textless 82 & 5.02 & FT\\
752 & \textless 246 & \textless 390 & 4.27 & FT\\
768 & 60 (10) & 42 (15) & 4.27 & FT\\
799 & \textless 215 & \textless 213 & 4.41 & FT\\
800 & 20 (5) & 10 (5) & 5.01 & FT\\
\hline
\hline
Core \# & N(X) x 10$^{10}$ [cm$^{-2}$] & N(X)/N(H$_2$) x 10$^{-12}$ & $T_{\rm ex}$ [K] & Method\\
\hline
\multicolumn{5}{c}{CC$^{34}$S} \\
\hline 
264 & \textless 21 & \textless 0.19 & 4.05 & FT\\
317 & \textless 8.0 & \textless 3.0 & 5.52 & FT\\
321 & \textless 3.3 & \textless 0.9 & 7.11 & FT\\
326 & \textless 5.3 & \textless 0.7 & 5.36 & FT\\
413 & 11 (2) & 9.7 (3.4) & 4.43 (0.61) & RD\\
504 & 4.0 (1.5) & 2.23 (1.29) & 6.28 (3.01) & RD\\
615 & 9.5 (0.3) & 9.25 (2.16) & 4.35 & FT\\
627 & 4.2 (0.6) & 4.6 (1.5) & 5.55 & FT\\
709 & \textless 5.1 & \textless 3.5 & 6.26 & FT\\
715 & \textless 4.8 & \textless 3.3 & 6.30 & FT\\
746 & \textless 9.8 & \textless 6.9 & 5.02 & FT\\
752 & \textless 14 & \textless 23 & 4.27 & FT\\
768 & 4.7 (0.5) & 3.3 (1.0) & 4.27 & FT\\
799 & 2.9 (0.5) & 2.9 (1.0) & 4.41 & FT\\
800 & 7.9 (0.4) & 4.1 (1.0) & 5.01 & FT
\end{tabular}
\end{table*}

\begin{table*}
\centering
\caption{CCCS: Column Densities, Abundances, and Excitation Temperatures with Upper Limit Calculations. Upper limits are denoted by the `\textless' symbol. CCCS column densities calculated via FT were derived using excitation temperatures obtained by CCS transitions.} \label{tab:CCCS}
\setlength{\tabcolsep}{12pt}
\renewcommand{\arraystretch}{1} 
 \setlength{\tabcolsep}{12pt}
	\begin{tabular}{rrrrrrr} 
     \tabcolsep=0.4cm
Core \# & $N(\rm X)\times$10$^{10}$ [cm$^{-2}$] & $N(\rm X)$/$N(\rm H_2)$$\times$10$^{-12}$ & $T_{\rm ex}$ [K] & Method\\
\hline
\multicolumn{5}{c}{CCCS} \\
\hline 
264 & 5.2 (1.5) & 2.5 (1.2) & 8.06 (2.36) & RD \\
317 & 5.1 (0.3) & 2.0 (0.5) & 5.52 & FT \\
321 & 2.1 (0.3) & 0.59 (0.19) & 7.11 & FT \\
326 & 7.2 (1.1) & 0.9 (0.3) & 11.88 (2.59) & RD \\
413 & 20 (2) & 17 (5) & 13.06 (1.86) & RD \\
504 & 7.0 (0.5) & 3.9 (1.1) & 9.62 (0.83) & RD \\
615 & 8.1 (0.4) & 7.8 (1.9) & 10.76 (0.64) & RD \\
627 & 10.0 (0.23) & 11 (2) & 5.55 & FT \\
709 & 8.3 (3.9) & 5.6 (3.8) & 3.18 (0.69) & RD \\
715 & 4.93 (0.20) & 3.4 (0.8) & 6.30 & FT \\
746 & 4.6 (1.7) & 3.3 (1.8) & 7.52 (2.57) & RD \\
752 & \textless 5.3 & \textless 8.4 & \textless 4.27 & FT\\
768 & 2.7436 (0.0013) & 1.9 (0.4) & 28.13 (0.05) & RD \\
799 & 5.13 (0.05) & 5.1 (1.1) & 13.08 (0.22) & RD \\
800 & 14 (1) & 7.2 (2.0) & 10.61 (1.03) & RD
\end{tabular}
\end{table*}

\begin{table*}
\centering
\caption{HC$_{3}$N and isotopologues: column densities, abundances, and excitation temperatures with upper limit calculations. Upper limits are denoted by the `\textless' symbol. Isotopologue column densities calculated via FT were derived using excitation temperatures obtained by HC$_3$N transitions.} \label{tab:HC3N}
\setlength{\tabcolsep}{12pt}
\renewcommand{\arraystretch}{1} 
 \setlength{\tabcolsep}{12pt}
	\begin{tabular}{rrrrrrr} 
     \tabcolsep=0.4cm
Core \# & $N(\rm X)\times$10$^{12}$ [cm$^{-2}$] & $N(\rm X)$/$N(\rm H_2)$$\times$10$^{-11}$ & $T_{\rm{ex}}$ [K] & Method\\
\hline
\multicolumn{5}{c}{HC$_3$N} \\
\hline 
264 & 6.18 (0.05) & 3.0 (0.6) & 3.41 (0.02) & RD \\
317 & 0.90 (0.03) & 0.34 (0.08) & 6.12 (0.21) & RD \\
321 & 0.79 (0.03) & 0.22 (0.05) & 8.90 (0.40) & RD \\
326 & 5.19 (0.04) & 0.66 (0.14) & 5.16 (0.03) & RD \\
413 & 12.2 (0.04) & 11 (2) & 4.58 (0.01) & RD \\
504 & 4.93 (0.03) & 1.4 (0.6) & 4.86 (0.03) & RD \\
615 & 2.75 (0.04) & 2.7 (0.6) & 4.75 (0.05) & RD \\
627 & 1.907 (0.022) & 2.1 (0.4) & 6.65 (0.08) & RD \\
709 & 1.132 (0.024) & 0.77 (0.17) & 10.00 (0.34) & RD \\
715 & 1.197 (0.020) & 0.83 (0.18) & 9.34 (0.23) & RD \\
746 & 2.137 (0.019) & 1.5 (0.3) & 12.56 (0.23) & RD \\
752 & 0.042 (0.006) & 0.067 (0.022) & 12.56 & FT \\
768 & 1.511 (0.020) & 1.06 (0.23) & 13.84 (0.41) & RD \\
799 & 1.954 (0.018) & 1.9 (0.4) & 7.87 (0.10) & RD \\
800 & 3.914 (0.022) & 2.0 (0.4) & 10.46 (0.10) & RD \\
\hline
\hline
Core \# & $N(\rm X)\times$10$^{11}$ [cm$^{-2}$] & $N(\rm X)$/$N(\rm H_2)$$\times$10$^{-11}$ & $T_{\rm ex}$ [K] & Method\\
\hline
\multicolumn{5}{c}{DC$_3$N} \\
\hline 
264 & 1.22 (0.26) & 0.59 (0.24) & 9.74 (3.49) & RD \\
317 & 0.92 (0.06) & 0.35 (0.09) & 6.12 & FT \\
321 & 0.98 (0.05) & 0.23 (0.06) & 7.11 & FT \\
326 & 1.0 (0.4) & 0.12 (0.07) & 13.34 (4.29) & RD \\
413 & 7.24 (0.09) & 6.2 (1.3) & 4.58 & FT \\
504 & 4.53 (0.06) & 2.5 (0.5) & 4.86 & FT \\
615 & 1.15 (0.07) & 1.1 (0.3) & 4.75 & FT \\
627 & 1.49 (0.04) & 1.6 (0.4) & 6.65 & FT \\
709 & 0.49 (0.04) & 0.33 (0.09) & 10.00 & FT \\
715 & 0.58 (0.04) & 0.40 (0.11) & 9.34 & FT \\
746 & 1.71 (0.04) & 1.20 (0.27) & 12.56 & FT \\
752 & \textless 2.4 & \textless 3.9 & 12.56 & FT \\
768 & 0.72 (0.07) & 0.19 (0.06) & 4.27 & FT \\
799 & 1.36 (0.04) & 1.4 (0.3) & 7.87 & FT \\
800 & 2.89 (0.04) & 1.5 (0.3) & 10.46 & FT \\
\hline
\hline
Core \# & $N(\rm X)\times$10$^{10}$ [cm$^{-2}$] & $N(\rm X)$/$N(\rm H_2)$$\times$10$^{-12}$ & $T_{\rm ex}$ [K] & Method\\
\hline
\multicolumn{5}{c}{H$^{13}$CCCN} \\
\hline 
264 & \textless 111 & \textless 54 & 3.41 & FT \\
317 & \textless 11 & \textless 4.1 & 6.12 & FT \\
321 & \textless 3.5 & \textless 1.0 & 8.90 & FT \\
326 & \textless 11.5 & \textless 1.4 & 5.16 & FT \\
413 & 34 (1) & 29 (6) & 4.58 & FT \\
504 & 11 (1) & 6.1 (1.7) & 4.86 & FT \\
615 & \textless 23 & \textless 22 & 4.75 & FT \\
627 & \textless 8.1 & \textless 8.7 & 6.65 & FT \\
709 & \textless 3.1 & \textless 2.1 & 10.00 & FT \\
715 & \textless 3.4 & \textless 2.3 & 9.34 & FT \\
746 & \textless 2.1 & \textless 1.5 & 12.56 & FT \\
752 & \textless 1.6 & \textless 2.6 & 12.56 & FT \\
768 & \textless 1.0 & \textless 0.7 & 13.84 & FT \\
799 & \textless 4.9 & \textless 4.8 & 7.87 & FT \\
800 & 4.9 (0.3) & 2.6 (0.7) & 10.46 & FT
\end{tabular}
\end{table*}

\begin{table*}
\caption{HC$_{3}$N and isotopologues: column densities, abundances, and excitation temperatures with upper limit calculations. Upper limits are denoted by the `\textless' symbol. Isotopologue and upper limit column densities calculated via FT were derived using excitation temperatures obtained by HC$_3$N transitions.} \label{tab:HC3N2}
\setlength{\tabcolsep}{12pt}
\renewcommand{\arraystretch}{1} 
 \setlength{\tabcolsep}{12pt}
	\begin{tabular}{rrrrrrr} 
     \tabcolsep=0.4cm
Core \# & $N(\rm X)\times$10$^{10}$ [cm$^{-2}$] & $N(\rm X)$/$N(\rm H_2)$$\times$10$^{-12}$ & $T_{\rm ex}$ [K] & Method\\
\hline
\multicolumn{5}{c}{HC$^{13}$CCN} \\
\hline 
264 & \textless 92.42 & \textless 44.90 & 3.41 & FT \\
317 & \textless 8.84 & \textless 3.36 & 6.12 & FT \\
321 & \textless 2.82 & \textless 0.77 & 8.90 & FT \\
326 & 7.8 (0.7) & 0.99 (0.29) & 5.16 & FT \\
413 & 13 (0.8) & 11 (3) & 4.58 & FT \\
504 & 5.3 (1.0) & 3.0 (1.1) & 4.86 & FT \\
615 & \textless 18.8 & \textless 18.3 & 4.8 & FT \\
627 & \textless 6.6 & \textless 7.1 & 6.65 & FT \\
709 & \textless 2.5 & \textless 1.7 & 10.00 & FT \\
715 & \textless 2.7 & \textless 1.9 & 9.34 & FT \\
746 & \textless 1.7 & \textless 1.2 & 12.56 & FT \\
752 & \textless 1.3 & \textless 2.1 & 12.56 & FT \\
768 & \textless 0.8 & \textless 0.6 & 13.84 & FT \\
799 & \textless 4.0 & \textless 3.9 & 7.87 & FT \\
800 & 2.6 (0.4) & 1.3 (0.5) & 10.46 & FT \\
\hline
\hline
Core \# & $N(\rm X)\times$10$^{10}$ [cm$^{-2}$] & $N(\rm X)$/$N(\rm H_2)$$\times$10$^{-12}$ & $T_{\rm ex}$ [K] & Method\\
\hline
\multicolumn{5}{c}{HCC$^{13}$CN} \\
\hline 
264 & 5.0 (2.2) & 2.4 (1.6) & 8.49 (5.32) & RD \\
317 & \textless 9.7 & \textless 3.7 & 6.12 & FT \\
321 & \textless 3.1 & \textless 0.8 & 8.90 & FT \\
326 & 10 (1) & 1.3 (0.4) & 5.16 & FT \\
413 & 13 (2) & 11 (4) & 10.81 (2.65) & RD \\
504 & 7.1 (2.0) & 4.0 (2.0) & 6.57 (2.10) & RD \\
615 & \textless 21 & \textless 20 & 4.75 & FT \\
627 & \textless 7.2 & \textless 7.7 & 6.65 & FT \\
709 & \textless 2.7 & \textless 1.8 & 10.00 & FT \\
715 & \textless 3.0 & \textless 2.1 & 9.34 & FT \\
746 & \textless 1.8 & \textless 1.3 & 12.56 & FT \\
752 & \textless 1.4 & \textless 2.3 & 12.56 & FT \\
768 & 2.0 (0.26) & 1.4 (0.5) & 13.84 & FT \\
799 & \textless 4.3 & \textless 4.3 & 7.87 & FT \\
800 & 4.9 (0.3) & 2.5 (0.7) & 10.46 & FT 
\end{tabular}
\end{table*}

\begin{table*}
\caption{HC$_{5}$N and isotopologues: column densities, abundances, and excitation temperatures with upper limit calculations. Upper limits are denoted by the `\textless' symbol. Isotopologue and upper limit column densities calculated via FT were derived using excitation temperatures obtained by HC$_3$N transitions.} \label{tab:HC5N}
\begin{tabular}{lllll}
Core \# & $N(\rm X)\times$10$^{10}$ [cm$^{-2}$] & $N(\rm X)$/$N(\rm H_2)$$\times$10$^{-12}$ & $T_{\rm ex}$ [K] & Method\\
\hline
\multicolumn{5}{c}{HC$_5$N} \\
\hline
264 & 12 (0.5) & 5.6 (1.4) & 12.81 (0.45) & RD \\
317 & 3.1 (0.4) & 1.2 (0.4) & 6.12 & FT \\
321 & 1.4 (0.8) & 0.4 (0.3) & 13.09 (6.91) & RD \\
326 & 5.7 (1.5) & 0.7 (0.3) & 11.11 (2.10) & RD \\
413 & 30 (1) & 26 (6) & 13.57 (0.41) & RD \\
504 & 10 (1) & 5.8 (1.5) & 11.37 (0.40) & RD \\
615 & 3.7 (1.1) & 3.6 (1.8) & 10.64 (2.20) & RD \\
627 & 2.2 (0.5) & 2.4 (1.0) & 13.98 (2.82) & RD \\
709 & 2.1 (0.3) & 1.4 (0.5) & 10.00 & FT \\
715 & 1.2 (0.2) & 0.8 (0.3) & 9.34 & FT \\
746 & 2.2 (0.7) & 1.5 (0.8) & 16.12 (5.12) & RD \\
752 & \textless 17 & \textless 27 & 12.56 & FT \\
768 & 4.7 (0.4) & 3.3 (1.0) & 18.15 (1.91) & RD \\
799 & 5.6 (0.6) & 5.5 (1.7) & 9.26 (0.57) & RD \\
800 & 12 (1) & 6.2 (1.8) & 14.86 (1.18) & RD \\
\hline
\hline
Core \# & $N(\rm X)\times$10$^{10}$ [cm$^{-2}$] & $N(\rm X)$/$N(\rm H_2)$$\times$10$^{-13}$ & $T_{\rm ex}$ [K] & Method\\
\hline
\multicolumn{5}{c}{HC$^{13}$CCCCN} \\
\hline
264 & \textless 33 & \textless 161 & 3.41 & FT \\
317 & \textless 2.0 & \textless 7.5 & 6.12 & FT \\
321 & \textless 1.0 & \textless 2.8 & 8.90 & FT \\
326 & 2.0 (0.9) & 2.6 (1.7) & 5.16 & FT \\
413 & \textless 6.2 & \textless 53 & 4.58 & FT \\
504 & \textless 3.9 & \textless 22 & 4.86 & FT \\
615 & \textless 5.5 & \textless 53 & 4.75 & FT \\
627 & \textless 1.6 & \textless 17 & 6.65 & FT \\
709 & \textless 0.5 & \textless 3.4 & 10.00 & FT \\
715 & \textless 0.6 & \textless 3.8 & 9.34 & FT \\
746 & \textless 0.3 & \textless 2.0 & 12.56 & FT \\
752 & \textless 0.3 & \textless 4.0 & 12.56 & FT \\
768 & \textless 6.5 & \textless 46 & 13.84 & FT \\
799 & \textless 0.9 & \textless 8.6 & 7.87 & FT \\
800 & \textless 2.1 & \textless 11 & 10.46 & FT \\
\hline
\hline
Core \# & $N(\rm X)\times$10$^{10}$ [cm$^{-2}$] & $N(\rm X)$/$N(\rm H_2)$$\times$10$^{-13}$ & $T_{\rm ex}$ [K] & Method\\
\hline
\multicolumn{5}{c}{HCC$^{13}$CCCN} \\
\hline
264 & \textless 108 & \textless 526 & 3.41 & FT \\
317 & \textless 4.3 & \textless 16 & 6.12 & FT \\
321 & \textless 1.9 & \textless 5.2 & 8.90 & FT \\
326 & \textless 6.8 & \textless 8.6 & 5.16 & FT \\
413 & 4.4 (0.7) & 38 (14) & 4.58 & FT \\
504 & 39 (3) & 215 (60) & 4.86 & FT \\
615 & \textless 13 & \textless 124 & 4.75 & FT \\
627 & \textless 3.2 & \textless 34 & 6.65 & FT \\
709 & \textless 0.8 & \textless 5.4 & 10.00 & FT \\
715 & \textless 0.9 & \textless 6.4 & 9.34 & FT \\
746 & \textless 0.5 & \textless 3.3 & 12.56 & FT \\
752 & \textless 0.4 & \textless 5.9 & 12.56 & FT \\
768 & 0.79 (0.16) & 5.5 (2.2) & 13.84 & FT \\
799 & \textless 1.78 & \textless 17.70 & 7.87 & FT \\
800 & \textless 0.68 & \textless 3.53 & 10.46 & FT
\end{tabular}
\end{table*}

\begin{table*}
\caption{HC$_{5}$N and isotopologues: column densities, abundances, and excitation temperatures with upper limit calculations. Upper limits are denoted by the `\textless' symbol. Isotopologue and upper limit column densities calculated via FT were derived using excitation temperatures obtained by HC$_3$N transitions.} \label{tab:HC5N2}
\begin{tabular}{lllll}
Core \# & $N(\rm X)\times$10$^{10}$ [cm$^{-2}$] & $N(\rm X)$/$N(\rm H_2)$$\times$10$^{-13}$ & $T_{\rm ex}$ [K] & Method\\
\hline
\multicolumn{5}{c}{HCCC$^{13}$CCN} \\
\hline
264 & \textless 90 & \textless 435 & 3.41 & FT \\
317 & \textless 2.7 & \textless 10 & 6.12 & FT \\
321 & \textless 1.4 & \textless 3.8 & 8.90 & FT \\
326 & \textless 5.2 & \textless 6.6 & 5.16 & FT \\
413 & 5.7 (2.7) & 49 (33) & 4.58 & FT \\
504 & \textless 6.9 & \textless 38 & 4.86 & FT \\
615 & \textless 9.4 & \textless 92 & 4.75 & FT \\
627 & \textless 1.9 & \textless 20 & 6.65 & FT \\
709 & \textless 0.5 & \textless 3.5 & 10.00 & FT \\
715 & \textless 0.6 & \textless 4.1 & 9.34 & FT \\
746 & \textless 0.3 & \textless 1.9 & 12.56 & FT \\
752 & \textless 0.2 & \textless 3.4 & 12.56 & FT \\
768 & \textless 15 & \textless 107 & 13.84 & FT \\
799 & \textless 1.0 & \textless 9.8 & 7.87 & FT \\
800 & \textless 0.4 & \textless 2.2 & 10.46 & FT \\
\hline
\hline
Core \# & $N(\rm X)\times$10$^{10}$ [cm$^{-2}$] & $N(\rm X)$/$N(\rm H_2)$$\times$10$^{-12}$ & $T_{\rm ex}$ [K] & Method\\
\hline
\multicolumn{5}{c}{HCCCCC$^{15}$N} \\
\hline

264 & \textless 3810 & \textless 1850 & 3.41 & FT \\
317 & \textless 36 & \textless 14 & 6.12 & FT \\
321 & \textless 12 & \textless 3.3 & 8.90 & FT \\
326 & \textless 75 & \textless 9.5 & 5.16 & FT \\
413 & \textless 194 & \textless 166 & 4.58 & FT \\
504 & \textless 100 & \textless 56 & 4.86 & FT \\
615 & 105 (4) & 102 (22) & 4.75 & FT \\
627 & \textless 28 & \textless 30 & 6.65 & FT \\
709 & \textless 3.6 & \textless 2.4 & 10.00 & FT \\
715 & \textless 4.7 & \textless 3.3 & 9.34 & FT \\
746 & \textless 2.1 & \textless 1.5 & 12.56 & FT \\
752 & \textless 1.5 & \textless 2.4 & 12.56 & FT \\
768 & \textless 289 & \textless 202 & 13.84 & FT \\
799 & \textless 9.2 & \textless 9.1 & 7.87 & FT \\
800 & \textless 2.9 & \textless 1.5 & 10.46 & FT \\
\hline
\hline
Core \# & $N(\rm X)\times$10$^{10}$ [cm$^{-2}$] & $N(\rm X)$/$N(\rm H_2)$$\times$10$^{-13}$ & $T_{\rm ex}$ [K] & Method\\
\hline
\multicolumn{5}{c}{DCCCCCN} \\
\hline
264 & \textless 70 & \textless 340 & 3.41 & FT \\
317 & \textless 3.0 & \textless 11 & 6.12 & FT \\
321 & \textless 1.4 & \textless 3.7 & 8.90 & FT \\
326 & \textless 4.6 & \textless 5.9 & 5.16 & FT \\
413 & 4.8 (0.8) & 41 (15) & 4.58 & FT \\
504 & \textless 6.1 & \textless 34 & 4.86 & FT \\
615 & \textless 11 & \textless 102 & 4.75 & FT \\
627 & \textless 2.3 & \textless 24 & 6.65 & FT \\
709 & \textless 0.6 & \textless 3.9 & 10.00 & FT \\
715 & \textless 0.7 & \textless 4.6 & 9.34 & FT \\
746 & \textless 0.3 & \textless 2.4 & 12.56 & FT \\
752 & \textless 0.3 & \textless 4.3 & 12.56 & FT \\
768 & \textless 12 & \textless 83 & 13.84 & FT \\
799 & \textless 1.3 & \textless 13 & 7.87 & FT \\
800 & \textless 0.5 & \textless 2.6 & 10.46 & FT
\end{tabular}
\end{table*}

\begin{table*}
\caption{HC$_{7}$N: column densities, abundances, and excitation temperatures with upper limit calculations. Upper limits are denoted by the `\textless' symbol. HC$_7$N column densities calculated via FT were derived using excitation temperatures obtained by HC$_3$N transitions.} \label{tab:HC7N}
\begin{tabular}{lllll}
Core \# & $N(\rm X)\times$10$^{10}$ [cm$^{-2}$] & $N(\rm X)$/$N(\rm H_2)$$\times$10$^{-12}$ & $T_{\rm ex}$ [K] & Method\\
\hline
\multicolumn{5}{c}{HC$_7$N} \\
\hline
264 & 29 (20) & 14 (12) & 7.97 (1.66) & RD \\
317 & \textless 36 & \textless 14 & 6.12 & FT \\
321 & \textless 1.5 & \textless 0.4 & 8.90 & FT \\
326 & \textless 2.3 & \textless 0.3 & 5.16 & FT \\
413 & 16 (11) & 14 (12) & 14 (5) & RD \\
504 & \textless 2.2 & \textless 1.2 & 4.86 & FT \\
615 & \textless 1.9 & \textless 1.8 & 4.75 & FT \\
627 & \textless 1.1 & \textless 1.2 & 6.65 & FT \\
709 & \textless 4.3 & \textless 2.9 & 10.00 & FT \\
715 & \textless 5.1 & \textless 3.5 & 9.34 & FT \\
746 & \textless 1.8 & \textless 1.2 & 12.56 & FT \\
752 & \textless 1.7 & \textless 2.6 & 12.56 & FT \\
768 & 4.6 (1.9) & 3.2 (2.0) & 13.84 & FT \\
799 & \textless 4.3 & \textless 4.2 & 7.87 & FT \\
800 & 15 (1) & 7.8 (2.3) & 10.46 & FT
\end{tabular}
\end{table*}

\section{Line Fitting Parameters} \label{ap:E}
In this section, we detail the molecular line parameters derived by Gaussian fits from the pipeline using \texttt{Pyspeckit} \citep{Ginsburg2022}. We include the main beam temperature ($T_{\rm mb}$), local standard of rest velocity ($v_{\rm lsr}$), line width (FWHM), and RMS values in Tables \ref{info1} - \ref{info2}. Absorption features from the frequency-switched data were not removed (see Figure \ref{spectrum}) because we do not observe any contamination from overlapping emission in our `line poor' spectra. Both confirmed and tentative detections are noted in these tables. Uncertainties are rounded to two decimal places, so 0.00 uncertainties imply that the uncertainty is $<$ 0.005. Laboratory spectroscopy information at the frequencies cited in this study is detailed in the following studies. CS: \citealt{1955PhRv...98.1837M,1963JChPh..39.2856K,1968JMoSp..28..266W,1982JMoSp..95...35B}. CCS: \citealt{1990ApJ...361..318Y}. CCCS: \citealt{1987ApJ...317L.119Y}. HC$_3$N: \citealt{Lafferty1978, 1995ZNatA..50.1179Y}. HC$_5$N: \citealt{1976JMoSp..62..175A, 2013A&A...559A..47B}. HC$_7$N: \citealt{1978ApJ...219L.133K, 1980MNRAS.192..651R, 1981A&A....95..143T, 1986A&A...160..181C, 1988A&A...196..194O, 1988A&A...199..291T}. 

\begin{table*}
\caption{Gaussian fit line parameters for C$_n$S family.}
\label{info1}
\begin{tabular}{ccccccccc}
 & \multicolumn{4}{c}{$\mathbf{^{13}CS}$ \textbf{J=1-0 (46247.567 MHz)}} & \multicolumn{4}{c}{$\mathbf{C^{34}S}$\textbf{ J=1-0 (48206.915 MHz)}} \\ \hline
Core \# & \begin{tabular}[c]{@{}c@{}}$T_{\rm mb}$\\ (mK)\end{tabular} & \begin{tabular}[c]{@{}c@{}}$v_{\rm lsr}$\\ (km s$^{-1}$)\end{tabular} & \begin{tabular}[c]{@{}c@{}}FWHM\\ (km s$^{-1}$)\end{tabular} & \begin{tabular}[c]{@{}c@{}}RMS\\ (mK)\end{tabular} & \begin{tabular}[c]{@{}c@{}}$T_{\rm mb}$\\ (mK)\end{tabular} & \begin{tabular}[c]{@{}c@{}}$v_{\rm lsr}$\\ (km s$^{-1}$)\end{tabular} & \begin{tabular}[c]{@{}c@{}}FWHM\\ (km s$^{-1}$)\end{tabular} & \begin{tabular}[c]{@{}c@{}}RMS\\ (mK)\end{tabular} \\ \hline
264 & 85.63 (0.78) & 7.90 (0.00) & 0.80 (0.01) & 6.6 & 312.52 (0.67) & 7.67 (0.00) & 1.07 (0.00) & 8.8 \\
317 & 87.42 (0.69) & 8.19 (0.00) & 1.02 (0.01) & 7.1 & 292.79 (0.70) & 8.06 (0.00) & 0.97 (0.00) & 9.3 \\
321 & 101.42 (0.67) & 8.55 (0.00) & 1.10 (0.01) & 5.3 & 367.94 (0.66) & 8.33 (0.00) & 1.10 (0.00) & 6.8 \\
326 & 367.94 (0.65) & 7.83 (0.00) & 1.16 (0.00) & 4.5 & 923.64 (0.63) & 7.64 (0.00) & 1.20 (0.01) & 6.7 \\
413 & 179.31 (0.86) & 7.83 (0.00) & 0.67 (0.00) & 6.1 & 499.21 (0.83) & 7.58 (0.00) & 0.69 (0.00) & 7.3 \\
504 & 131.37 (0.85) & 6.77 (0.00) & 0.69 (0.01) & 4.7 & 628.19 (0.79) & 6.54 (0.00) & 0.76 (0.00) & 5.9 \\
615 & 122.76 (0.89) & 8.18 (0.00) & 0.63 (0.01) & 6.9 & 633.05 (0.84) & 7.96 (0.00) & 0.67 (0.00) & 9.1 \\
627 & 103.85 (1.03) & 8.48 (0.00) & 0.51 (0.01) & 7.1 & 592.19 (1.03) & 8.28 (0.00) & 0.47 (0.00) & 11.2 \\
709 & 66.85 (0.76) & 8.69 (0.01) & 0.86 (0.01) & 6.8 & 275.70 (0.71) & 8.49 (0.00) & 0.94 (0.00) & 9.0 \\
715 & 93.92 (0.86) & 8.97 (0.00) & 0.67 (0.01) & 6.6 & 523.27 (0.85) & 8.73 (0.01) & 0.65 (0.00) & 9.1 \\
746 & 77.12 (0.85) & 8.90 (0.00) & 0.69 (0.01) & 6.9 & 338.14 (0.80) & 8.69 (0.00) & 0.74 (0.00) & 11.4 \\
752 & 53.27 (0.82) & 7.59 (0.01) & 0.74 (0.01) & 5.6 & 244.72 (0.76) & 7.45 (0.00) & 0.81 (0.00) & 8.2 \\
768 & 60.01 (0.81) & 9.08 (0.01) & 0.75 (0.01) & 4.4 & 287.29 (0.79) & 8.89 (0.00) & 0.76 (0.00) & 6.2 \\
799 & 111.28 (1.00) & 10.15 (0.00) & 0.44 (0.00) & 6.6 & 508.10 (0.91) & 10.00 (0.05) & 0.48 (0.00) & 8.5 \\
800 & 176.03 (0.87) & 10.27 (0.16) & 0.75 (0.00) & 6.8 & 704.35 (0.08) & 10.09 (0.38) & 0.69 (0.01) & 8.6 \\ \hline \hline
 & \multicolumn{4}{c}{\textbf{CS J=1-0 (48990.978 MHz)}} & \multicolumn{4}{c}{$\mathbf{C^{13}CS}$ \textbf{N=2-1, J=3-2, F=7/2-5/2 (33613.339 MHz)}} \\ \hline
Core \# & \begin{tabular}[c]{@{}c@{}}$T_{\rm mb}$\\ (mK)\end{tabular} & \begin{tabular}[c]{@{}c@{}}$v_{\rm lsr}$\\ (km s$^{-1}$)\end{tabular} & \begin{tabular}[c]{@{}c@{}}FWHM\\ (km s$^{-1}$)\end{tabular} & \begin{tabular}[c]{@{}c@{}}RMS\\ (mK)\end{tabular} & \begin{tabular}[c]{@{}c@{}}$T_{\rm mb}$\\ (mK)\end{tabular} & \begin{tabular}[c]{@{}c@{}}$v_{\rm lsr}$\\ (km s$^{-1}$)\end{tabular} & \begin{tabular}[c]{@{}c@{}}FWHM\\ (km s$^{-1}$)\end{tabular} & \begin{tabular}[c]{@{}c@{}}RMS\\ (mK)\end{tabular} \\ \hline
264 & 2929.57 (0.63) & 7.99 (0.00) & 1.16 (0.00) & 8.8 & -- & -- & -- & 2.5 \\
317 & 3001.73 (0.68) & 8.30 (0.00) & 1.01 (0.00) & 9.3 & -- & -- & -- & 2.7 \\
321 & 3235.05 (0.62) & 8.61 (0.00) & 1.22 (0.00) & 6.8 & -- & -- & -- & 2.3 \\
326 & 5691.94 (0.58) & 7.90 (0.00) & 1.37 (0.00) & 6.7 & -- & -- & -- & 2.2 \\
413 & 1971.39 (0.73) & 7.82 (0.00) & 0.88 (0.00) & 7.3 & 9.34 (0.91) & 6.12 (0.04) & 0.82 (0.09) & 2.7 \\
504 & 3032.29 (0.68) & 6.79 (0.00) & 1.02 (0.00) & 5.9 & -- & -- & -- & 2.2 \\
615 & 2596.89 (0.79) & 8.19 (0.00) & 0.75 (0.00) & 9.1 & -- & -- & -- & 2.8 \\
627 & 2699.87 (0.92) & 8.57 (0.00) & 0.55 (0.00) & 11.2 & -- & -- & -- & 2.8 \\
709 & 3566.24 (0.66) & 8.82 (0.00) & 1.06 (0.00) & 9.0 & -- & -- & -- & 2.8 \\
715 & 3640.54 (0.75) & 8.95 (0.00) & 0.83 (0.00) & 9.1 & -- & -- & -- & 2.6 \\
746 & 2702.76 (0.70) & 8.96 (0.00) & 0.94 (0.00) & 11.4 & -- & -- & -- & 2.6 \\
752 & 2788.75 (0.69) & 7.79 (0.00) & 0.97 (0.00) & 8.2 & -- & -- & -- & 2.3 \\
768 & 2749.84 (0.71) & 9.19 (0.00) & 0.93 (0.00) & 6.2 & -- & -- & -- & 2.0 \\
799 & 2175.80 (0.80) & 10.27 (0.00) & 0.73 (0.00) & 8.5 & -- & -- & -- & 2.6 \\
800 & 3282.32 (0.72) & 10.36 (0.00) & 0.90 (0.00) & 8.6 & -- & -- & -- & 2.7
\end{tabular}
\end{table*}

\begin{table*}
\begin{tabular}{ccccccccc}
 & \multicolumn{4}{c}{\textbf{\begin{tabular}[c]{@{}c@{}}$\mathbf{C^{13}CS}$ N=3-2, J=3-2, F=7/2-5/2 (38682.482 MHz)\end{tabular}}} & \multicolumn{4}{c}{\textbf{$\mathbf{C^{13}CS}$ N=4-3, J=3-2, F=5/2-3/2 (43746.043 MHz)}} \\ \hline
Core \# & \begin{tabular}[c]{@{}c@{}}$T_{\rm mb}$\\ (mK)\end{tabular} & \begin{tabular}[c]{@{}c@{}}$v_{\rm lsr}$\\ (km s$^{-1}$)\end{tabular} & \begin{tabular}[c]{@{}c@{}}FWHM\\ (km s$^{-1}$)\end{tabular} & \begin{tabular}[c]{@{}c@{}}RMS\\ (mK)\end{tabular} & \begin{tabular}[c]{@{}c@{}}$T_{\rm mb}$\\ (mK)\end{tabular} & \begin{tabular}[c]{@{}c@{}}$v_{\rm lsr}$\\ (km s$^{-1}$)\end{tabular} & \begin{tabular}[c]{@{}c@{}}FWHM\\ (km s$^{-1}$)\end{tabular} & \begin{tabular}[c]{@{}c@{}}RMS\\ (mK)\end{tabular} \\ \hline
264 & -- & -- & -- & 2.8 & -- & -- & -- & 4.2 \\
317 & -- & -- & -- & 3.2 & -- & -- & -- & 4.7 \\
321 & -- & -- & -- & 2.5 & -- & -- & -- & 3.2 \\
326 & -- & -- & -- & 2.4 & -- & -- & -- & 3.5 \\
413 & -- & -- & -- & 3.0 & -- & -- & -- & 3.3 \\
504 & -- & -- & -- & 2.4 & 10.42 (0.84) & 6.93 (0.29) & 0.74 (0.07) & 2.7 \\
615 & -- & -- & -- & 3.0 & -- & -- & -- & 4.6 \\
627 & -- & -- & -- & 3.3 & -- & -- & -- & 4.7 \\
709 & -- & -- & -- & 3.0 & -- & -- & -- & 4.6 \\
715 & -- & -- & -- & 2.9 & -- & -- & -- & 4.2 \\
746 & -- & -- & -- & 3.1 & -- & -- & -- & 4.5 \\
752 & -- & -- & -- & 2.5 & -- & -- & -- & 3.8 \\
768 & -- & -- & -- & 2.3 & 9.20 (0.99) & 8.73 (0.03) & 0.55 (0.07) & 3.0 \\
799 & -- & -- & -- & 3.3 & -- & -- & -- & 4.0 \\
800 & 9.22 (1.40) & 8.33 (0.03) & 0.41 (0.08) & 2.9 & -- & -- & -- & 4.3 \\ \hline \hline
 & \multicolumn{4}{c}{\textbf{$\mathbf{CC^{34}S}$ N=2-1, J=3-2 (33111.839 MHz)}} & \multicolumn{4}{c}{\textbf{$\mathbf{CC^{34}S}$ N=3-2, J=4-3 (44497.599 MHz)}} \\ \hline
Core \# & \begin{tabular}[c]{@{}c@{}}$T_{\rm mb}$\\ (mK)\end{tabular} & \begin{tabular}[c]{@{}c@{}}$v_{\rm lsr}$\\ (km s$^{-1}$)\end{tabular} & \begin{tabular}[c]{@{}c@{}}FWHM\\ (km s$^{-1}$)\end{tabular} & \begin{tabular}[c]{@{}c@{}}RMS\\ (mK)\end{tabular} & \begin{tabular}[c]{@{}c@{}}$T_{\rm mb}$\\ (mK)\end{tabular} & \begin{tabular}[c]{@{}c@{}}$v_{\rm lsr}$\\ (km s$^{-1}$)\end{tabular} & \begin{tabular}[c]{@{}c@{}}FWHM\\ (km s$^{-1}$)\end{tabular} & \begin{tabular}[c]{@{}c@{}}RMS\\ (mK)\end{tabular} \\ \hline
264 & -- & -- & -- & 2.5 & -- & -- & -- & 3.2 \\
317 & -- & -- & -- & 2.7 & -- & -- & -- & 3.6 \\
321 & -- & -- & -- & 2.3 & -- & -- & -- & 3.4 \\
326 & -- & -- & -- & 2.2 & -- & -- & -- & 3.1 \\
413 & 30.67 (1.07) & 7.83 (0.01) & 0.65 (0.03) & 2.7 & 39.09 (0.94) & 7.83 (0.01) & 0.57 (0.02) & 3.5 \\
504 & 8.65 (0.8) & 6.72 (0.05) & 1.01 (0.11) & 2.2 & 20.22 (0.97) & 6.69 (0.01) & 0.56 (0.03) & 3.0 \\
615 & -- & -- & -- & 2.8 & 25.54 (0.84) & 8.26 (0.00) & 0.72 (0.00) & 3.6 \\
627 & 10.04 (0.90) & 8.47 (0.04) & 0.86 (0.09) & 2.8 & -- & -- & -- & 3.5 \\
709 & -- & -- & -- & 2.8 & -- & -- & -- & 3.9 \\
715 & -- & -- & -- & 2.6 & -- & -- & -- & 3.7 \\
746 & -- & -- & -- & 2.6 & -- & -- & -- & 3.8 \\
752 & -- & -- & -- & 2.3 & -- & -- & -- & 3.0 \\
768 & -- & -- & -- & 2.0 & 14.75 (0.92) & 9.24 (0.02) & 0.61 (0.04) & 3.1 \\
799 & 9.50 (1.01) & 10.24 (0.04) & 0.54 (0.06) & 2.6 & -- & -- & -- & 3.5 \\
800 & -- & -- & -- & 2.7 & 32.85 (1.0) & 10.36 (0.01) & 0.55 (0.02) & 3.6\\ \hline \hline
 & \multicolumn{4}{c}{\textbf{CCS N=2-1, J=3-2 (33751.3737 MHz)}} & \multicolumn{4}{c}{\textbf{CCS N=3-2, J=3-2 (38866.4232 MHz)}} \\ \hline
Core \# & \begin{tabular}[c]{@{}c@{}}$T_{\rm mb}$\\ (mK)\end{tabular} & \begin{tabular}[c]{@{}c@{}}$v_{\rm lsr}$\\ (km s$^{-1}$)\end{tabular} & \begin{tabular}[c]{@{}c@{}}FWHM\\ (km s$^{-1}$)\end{tabular} & \begin{tabular}[c]{@{}c@{}}RMS\\ (mK)\end{tabular} & \begin{tabular}[c]{@{}c@{}}$T_{\rm mb}$\\ (mK)\end{tabular} & \begin{tabular}[c]{@{}c@{}}$v_{\rm lsr}$\\ (km s$^{-1}$)\end{tabular} & \begin{tabular}[c]{@{}c@{}}FWHM\\ (km s$^{-1}$)\end{tabular} & \begin{tabular}[c]{@{}c@{}}RMS\\ (mK)\end{tabular} \\ \hline
264 & 127.37 (0.87) & 7.96 (0.00) & 0.91 (0.01) & 2.5 & 28.53 (1.09) & 7.95 (0.01) & 0.48 (0.02) & 2.8\\
317 & 91.55 (0.85) & 8.28 (0.00) & 0.94 (0.01) & 2.7 & 26.65 (0.95) & 8.26 (0.01) & 0.66 (0.03) & 3.2\\
321 & 97.83 (0.83) & 8.54 (0.00) & 0.98 (0.01) & 2.3 & 33.85 (0.83) & 8.45 (0.01) & 0.85 (0.02) & 2.5\\
326 & 197.40 (0.72) & 7.90 (0.00) & 1.31 (0.01) & 2.2 & 43.74 (0.73) & 7.91 (0.01) & 1.09 (0.02) & 2.4\\
413 & 624.98 (1.02) & 7.84 (0.00) & 0.64 (0.00) & 2.7 & 194.75 (1.01) & 7.84 (0.00) & 0.48 (0.00) & 3.0\\
504 & 259.60 (0.99) & 6.82 (0.00) & 0.66 (0.00) & 2.2 & 65.57 (1.12) & 6.85 (0.00) & 0.50 (0.01) & 2.4\\
615 & 322.08 (0.97) & 8.18 (0.00) & 0.68 (0.00) & 2.8 & 57.33 (1.02) & 8.16 (0.01) & 0.50 (0.01) & 3.0\\
627 & 203.55 (1.02) & 8.49 (0.00) & 0.57 (0.00) & 2.8 & 65.13 (1.00) & 8.48 (0.01) & 0.37 (0.01) & 3.3\\
709 & 52.46 (0.81) & 8.71 (0.01) & 1.03 (0.02) & 2.8 & -- & -- & -- & 3.0\\
715 & 113.84 (1.02) & 9.03 (0.00) & 0.69 (0.01) & 2.6 & 51.06 (1.23) & 9.03 (0.01) & 0.38 (0.01) & 2.9\\
746 & 160.30 (0.95) & 8.96 (0.00) & 0.76 (0.01) & 2.6 & 27.10 (0.90) & 8.95 (0.01) & 0.80 (0.01) & 3.1\\
752 & 16.38 (0.85) & 7.76 (0.02) & 0.95 (0.06) & 2.3 & -- & -- & -- & 2.5\\
768 & 128.71 (0.92) & 9.07 (0.00) & 0.81 (0.01) & 2.0 & 25.26 (1.15) & 8.99 (0.01) & 0.48 (0.03) & 2.3\\
799 & 265.37 (1.17) & 10.22 (0.00) & 0.60 (0.00) & 2.6 & 60.83 (3.29) & 10.17 (0.01) & 0.32 (0.02) & 3.3\\
800 & 369.15 (0.97) & 10.31 (0.00) & 0.69 (0.00) & 2.7 & 86.91 (1.08) & 10.31 (0.00) & 0.54 (0.01) & 2.9
\end{tabular}
\end{table*}

\begin{table*}
\label{info3}
\begin{tabular}{ccccccccc}
\multicolumn{5}{c}{\textbf{CCS N=4-3, J=3-2 (43981.027 MHz)}} & \multicolumn{4}{c}{\textbf{CCS N=3-2, J=4-3 (45379.033 MHz)}} \\ \hline
Core \# & \begin{tabular}[c]{@{}c@{}}$T_{\rm mb}$\\ (mK)\end{tabular} & \begin{tabular}[c]{@{}c@{}}$v_{\rm lsr}$\\ (km s$^{-1}$)\end{tabular} & \begin{tabular}[c]{@{}c@{}}FWHM\\ (km s$^{-1}$)\end{tabular} & \begin{tabular}[c]{@{}c@{}}RMS\\ (mK)\end{tabular} & \begin{tabular}[c]{@{}c@{}}$T_{\rm mb}$\\ (mK)\end{tabular} & \begin{tabular}[c]{@{}c@{}}$v_{\rm lsr}$\\ (km s$^{-1}$)\end{tabular} & \begin{tabular}[c]{@{}c@{}}FWHM\\ (km s$^{-1}$)\end{tabular} & \begin{tabular}[c]{@{}c@{}}RMS\\ (mK)\end{tabular} \\ \hline
264 & -- & -- & -- & 4.2 & 238.30 (0.84) & 7.97 (0.00) & 0.72 (0.00) & 5.7\\
317 & 33.96 (0.96) & 8.23 (0.01) & 0.56 (0.18) & 4.7 & 182.71 (0.82) & 8.26 (0.00) & 0.75 (0.00) & 6.5\\
321 & 43.38 (0.82) & 8.50 (0.01) & 0.77 (0.02) & 3.2 & 183.73 (0.74) & 8.55 (0.00) & 0.92 (0.00) & 5.3\\
326 & 58.27 (0.74) & 7.90 (0.01) & 1.07 (0.02) & 3.5 & 344.12 (0.64) & 7.88 (0.00) & 1.247 (0.00) & 4.0\\
413 & 173.62 (0.93) & 7.83 (0.00) & 0.59 (0.00) & 3.3 & 1151.34 (0.95) & 7.85 (0.00) & 0.54 (0.00) & 5.7 \\
504 & 56.74 (0.96) & 6.84 (0.01) & 0.57 (0.01) & 2.7 & 458.09 (0.96) & 6.82 (0.00) & 0.55 (0.00) & 4.2\\
615 & 61.06 (1.00) & 8.21 (0.00) & 0.52 (0.01) & 4.6 & 559.83 (0.94) & 8.1831 (0.0004) & 0.5665 (0.0011) & 6.0\\
627 & 83.84 (1.07) & 8.50 (0.00) & 0.37 (0.01) & 4.7 & 504.39 (1.23) & 8.52 (0.00) & 0.43 (0.00) & 6.1\\
709 & 22.27 (0.88) & 8.62 (0.01) & 0.68 (0.03) & 4.6 & 100.67 (0.72) & 8.61 (0.00) & 0.98 (0.01) & 6.0\\
715 & 54.79 (1.00) & 9.06 (0.01) & 0.42 (0.01) & 4.2 & 360.73 (1.40) & 9.07 (0.00) & 0.36 (0.00) & 5.7\\
746 & 49.45 (1.00) & 9.03 (0.01) & 0.46 (0.01) & 4.5 & 323.43 (0.96) & 8.96 (0.00) & 0.55 (0.00) & 6.2\\
752 & -- & -- & -- & 3.8 & 25.7 (0.9) & 7.772 (0.011) & 0.65 (0.03) & 4.9\\
768 & 30.75 (0.97) & 9.02 (0.01) & 0.55 (0.02) & 3.7 & 241.80 (0.85) & 9.05 (0.00) & 0.70 (0.00) & 3.7\\
799 & 55.00 (0.02) & 10.21 (0.00) & 0.45 (0.01) & 4.0 & 550.09 (0.02) & 10.23 (0.00) & 0.39 (0.00) & 5.6\\
800 & 93.19 (1.03) & 10.35 (0.00) & 0.52 (0.01) & 4.3 & 667.44 (0.92) & 10.34 (0.40) & 0.59 (0.01) & 5.6 \\ \hline \hline
\multicolumn{5}{c}{\textbf{CCCS J=6-5 (34684.368 MHz)}} & \multicolumn{4}{c}{\textbf{CCCS J=7-6 (40465.014 MHz)}} \\ \hline
Core \# & \begin{tabular}[c]{@{}c@{}}$T_{\rm mb}$\\ (mK)\end{tabular} & \begin{tabular}[c]{@{}c@{}}$v_{\rm lsr}$\\ (km s$^{-1}$)\end{tabular} & \begin{tabular}[c]{@{}c@{}}FWHM\\ (km s$^{-1}$)\end{tabular} & \begin{tabular}[c]{@{}c@{}}RMS\\ (mK)\end{tabular} & \begin{tabular}[c]{@{}c@{}}$T_{\rm mb}$\\ (mK)\end{tabular} & \begin{tabular}[c]{@{}c@{}}$v_{\rm lsr}$\\ (km s$^{-1}$)\end{tabular} & \begin{tabular}[c]{@{}c@{}}FWHM\\ (km s$^{-1}$)\end{tabular} & \begin{tabular}[c]{@{}c@{}}RMS\\ (mK)\end{tabular} \\ \hline
264 & 26.73 (0.86) & 7.92 (0.01) & 0.90 (0.03) & 2.7 & 43.92 (0.91) & 7.91 (0.01) & 0.67 (0.02) & 3.2 \\
317 & -- & -- & -- & 3.2 & -- & -- & -- & 3.6 \\
321 & 11.08 (0.88) & 8.53 (0.03) & 0.85 (0.08) & 2.5 & 12.72 (0.80) & 8.66 (0.03) & 0.89 (0.06) & 3.4 \\
326 & 40.87 (0.79) & 7.93 (0.01) & 1.05 (0.02) & 2.4 & 54.65 (0.76) & 7.95 (0.01) & 0.98 (0.02) & 3.1 \\
413 & 206.24 (0.98) & 7.82 (0.00) & 0.64 (0.00) & 3.0 & 287.93 (1.11) & 7.81 (0.00) & 0.52 (0.00) & 3.5 \\
504 & 58.20 (0.98) & 6.80 (0.01) & 0.65 (0.01) & 2.4 & 91.57 (1.03) & 6.81 (0.00) & 0.48 (0.01) & 3.0 \\
615 & 72.44 (1.05) & 8.11 (0.00) & 0.64 (0.01) & 3.0 & 95.04 (0.97) & 8.15 (0.00) & 0.57 (0.01) & 3.6 \\
627 & 28.05 (1.06) & 8.48 (0.01) & 0.63 (0.03) & 3.2 & 70.26 (1.62) & 8.51 (0.01) & 0.37 (0.01) & 3.5 \\
709 & 11.94 (0.77) & 8.66 (0.04) & 1.11 (0.08) & 3.0 & 17.47 (1.01) & 8.92 (0.02) & 0.56 (0.04) & 3.9 \\
715 & 19.48 (1.13) & 9.03 (0.02) & 0.57 (0.04) & 2.9 & 35.46 (1.00) & 8.97 (0.01) & 0.51 (0.02) & 3.7 \\
746 & 30.28 (0.97) & 8.90 (0.01) & 0.74 (0.03) & 3.1 & 32.52 (0.98) & 8.96 (0.01) & 0.62 (0.02) & 3.8 \\
752 & -- & -- & -- & 2.5 & -- & -- & -- & 3.0 \\
768 & 31.28 (0.97) & 9.01 (0.01) & 0.71 (0.03) & 2.3 & 53.01 (0.98) & 9.007 (0.005) & 0.535 (0.011) & 3.1 \\
799 & 61.55 (1.13) & 10.21 (0.01) & 0.53 (0.01) & 3.3 & 93.73 (1.05) & 10.18 (0.00) & 0.41 (0.01) & 3.5 \\
800 & 111.23 (0.95) & 10.29 (0.00) & 0.73 (0.01) & 2.9 & 168.25 (1.09) & 10.30 (0.00) & 0.53 (0.00) & 3.6 \\ \hline \hline
 & \multicolumn{4}{c}{\textbf{CCCS J=8-7 (46245.623 MHz)}} & \multicolumn{4}{c}{\textbf{}} \\ \hline
Core \# & \begin{tabular}[c]{@{}c@{}}$T_{\rm mb}$\\ (mK)\end{tabular} & \begin{tabular}[c]{@{}c@{}}$v_{\rm lsr}$\\ (km s$^{-1}$)\end{tabular} & \begin{tabular}[c]{@{}c@{}}FWHM\\ (km s$^{-1}$)\end{tabular} & \begin{tabular}[c]{@{}c@{}}RMS\\ (mK)\end{tabular} & \textbf{} & \textbf{} & \textbf{} & \textbf{} \\ \hline
264 & 42.77 (0.91) & 7.95 (0.01) & 0.60 (0.02) & 6.6 &  &  &  &  \\
317 & 23.04 (0.87) & 8.04 (0.01) & 0.65 (0.03) & 7.1 &  &  &  &  \\
321 & 21.67 (0.70) & 8.67 (0.02) & 1.00 (0.04) & 5.3 &  &  &  &  \\
326 & 57.81 (0.72) & 7.89 (0.01) & 0.94 (0.01) & 4.5 &  &  &  &  \\
413 & 337.8 (1.0) & 7.83 (0.00) & 0.51 (0.00) & 6.1 &  &  &  &  \\
504 & 94.36 (1.06) & 6.80 (0.00) & 0.47 (0.00) & 4.7 &  &  &  &  \\
615 & 106.90 (0.96) & 8.13 (0.00) & 0.53 (0.01) & 6.9 &  &  &  &  \\
627 & 80.13 (1.17) & 8.50 (0.00) & 0.37 (0.01) & 7.1 &  &  &  &  \\
709 & -- & -- & -- & 6.8 &  &  &  &  \\
715 & 57.12 (1.46) & 9.01 (0.01) & 0.32 (0.01) & 6.6 &  &  &  &  \\
746 & 34.78 (0.86) & 8.93 (0.01) & 0.66 (0.02) & 6.9 &  &  &  &  \\
752 & -- & -- & -- & 5.6 &  &  &  & \\
768 & 54.59 (0.89) & 9.00 (0.01) & 0.63 (0.01) & 4.4 &  &  &  & \\
799 & 98.58 (1.01) & 10.19 (0.00) & 0.44 (0.01) & 6.6 &  &  &  &  \\
800 & 188.30 (0.97) & 10.33 (0.00) & 0.52 (0.00) & 6.8 &  &  &  &
\end{tabular}
\end{table*}

\begin{table*}
\caption{Gaussian fit line parameters for the HC$_n$N family.}
\label{info2}
\begin{tabular}{ccccccccc}
\multicolumn{5}{c}{\textbf{$\mathbf{HC_3N}$ J=4-3, F=3-2 (36392.326 MHz)}} & \multicolumn{4}{c}{\textbf{$\mathbf{HC_3N}$ J=5-4, F=5-4 (45490.310 MHz)}} \\ \hline
Core \# & \begin{tabular}[c]{@{}c@{}}$T_{\rm mb}$\\ (mK)\end{tabular} & \begin{tabular}[c]{@{}c@{}}$v_{\rm lsr}$\\ (km s$^{-1}$)\end{tabular} & \begin{tabular}[c]{@{}c@{}}FWHM\\ (km s$^{-1}$)\end{tabular} & \begin{tabular}[c]{@{}c@{}}RMS\\ (mK)\end{tabular} & \begin{tabular}[c]{@{}c@{}}$T_{\rm mb}$\\ (mK)\end{tabular} & \begin{tabular}[c]{@{}c@{}}$v_{\rm lsr}$\\ (km s$^{-1}$)\end{tabular} & \begin{tabular}[c]{@{}c@{}}FWHM\\ (km s$^{-1}$)\end{tabular} & \begin{tabular}[c]{@{}c@{}}RMS\\ (mK)\end{tabular} \\ \hline
264 & 428.82 (0.75) & 7.83 (0.00) & 1.12 (0.00) & 2.7 & 895.13 (0.83) & 8.07 (0.00) & 0.74 (0.00) & 5.7 \\
317 & 107.19 (0.74) & 8.24 (0.00) & 1.15 (0.01) & 3.2 & 204.28 (0.75) & 8.25 (0.00) & 0.89 (0.00) & 6.5 \\
321 & 109.27 (0.71) & 8.42 (0.00) & 1.24 (0.01) & 2.5 & 253.12 (0.75) & 8.51 (0.00) & 0.89 (0.00) & 5.3 \\
326 & 420.01 (0.65) & 7.84 (0.00) & 1.48 (0.00) & 2.4 & 717.96 (0.65) & 7.86 (0.00) & 1.20 (0.00) & 4.0 \\
413 & 1177.73 (0.75) & 7.74 (0.00) & 1.12 (0.00) & 3.0 & 2222.67 (0.80) & 7.84 (0.00) & 0.78 (0.00) &  5.7 \\
504 & 582.00 (0.81) & 6.68 (0.00) & 0.96 (0.00) & 2.4 & 997.88 (0.81) & 6.80 (0.00) & 0.76 (0.00) & 4.2 \\
615 & 313.14 (0.80) & 8.03 (0.00) & 0.98 (0.00) & 3.0 & 562.94 (0.83) & 8.15 (0.00) & 0.73 (0.00) & 6.0 \\
627 & 411.40 (0.97) & 8.30 (0.00) & 0.67 (0.00) & 3.2 & 764.45 (0.86) & 8.63 (0.00) & 0.68 (0.00) & 6.1\\
709 & 154.89 (0.69) & 8.52 (0.00) & 1.32 (0.01) & 3.0 & 374.52 (0.74) & 8.49 (0.00) & 0.93 (0.00) & 6.0 \\
715 & 248.96 (0.86) & 8.87 (0.00) & 0.84 (0.00) & 2.9 & 543.01 (0.88) & 8.99 (0.00) & 0.65 (0.00) &  5.7\\
746 & 425.08 (0.79) & 8.84 (0.00) & 0.99 (0.00) & 3.1 & 867.7 (0.84) & 8.58 (0.00) & 0.70 (0.00) & 6.2\\
752 & 9.66 (0.86) & 7.72 (0.04) & 0.86 (0.09) & 2.5 & -- & -- & -- & 4.9\\
768 & 281.86 (0.76) & 8.91 (0.00) & 1.10 (0.00) & 2.3 & 566.85 (0.79) & 8.64 (0.00) & 0.80 (0.00) & 3.7\\
799 & 559.69 (1.1) & 9.98 (0.00) & 0.56 (0.00) & 3.3 & 773.88 (0.88) & 10.18 (0.00) & 0.65 (0.00) & 5.6\\
800 & 640.54 (0.75) & 10.24 (0.00) & 1.13 (0.00) & 2.9 & 1287.17 (0.80) & 10.00 (0.00) &  0.78 (0.00) &  5.6\\ \hline \hline
\multicolumn{5}{c}{\textbf{$\mathbf{DC_3N}$ J=4-3, F=3-2 (33772.541 MHz)}} & \multicolumn{4}{c}{\textbf{$\mathbf{DC_3N}$ J=5-4, F=4-3 (42215.595 MHz)}} \\ \hline
Core \# & \begin{tabular}[c]{@{}c@{}}$T_{\rm mb}$\\ (mK)\end{tabular} & \begin{tabular}[c]{@{}c@{}}$v_{\rm lsr}$\\ (km s$^{-1}$)\end{tabular} & \begin{tabular}[c]{@{}c@{}}FWHM\\ (km s$^{-1}$)\end{tabular} & \begin{tabular}[c]{@{}c@{}}RMS\\ (mK)\end{tabular} & \begin{tabular}[c]{@{}c@{}}$T_{\rm mb}$\\ (mK)\end{tabular} & \begin{tabular}[c]{@{}c@{}}$v_{\rm lsr}$\\ (km s$^{-1}$)\end{tabular} & \begin{tabular}[c]{@{}c@{}}FWHM\\ (km s$^{-1}$)\end{tabular} & \begin{tabular}[c]{@{}c@{}}RMS\\ (mK)\end{tabular} \\ \hline
264 & 18.78 (0.81) & 7.88 (0.02) & 1.03 (0.05) & 2.5 & 38.46 (0.88) & 7.97 (0.01) & 0.70 (0.02) & 4.2 \\
317 & -- & -- & -- & 2.7 & 18.30 (0.80) & 8.46 (0.02) & 0.77 (0.04) & 4.7 \\
321 & -- & -- & -- & 2.3 & 17.47 (0.74) & 8.68 (0.02) & 0.98 (0.05) &  3.2\\
326 & 12.38 (0.81) & 8.04 (0.03) & 1.02 (0.08) & 2.2 & 34.24 (0.74) & 8.11 (0.01) & 1.00 (0.03) & 3.5 \\
413 & 52.35 (0.98) & 7.74 (0.01) & 0.72 (0.02) & 2.7 & 104.13 (0.84) & 7.91 (0.00) & 0.76 (0.01) & 3.3 \\
504 & 38.78 (0.96) & 6.75 (0.01) & 0.73 (0.02) & 2.2 & 73.75 (0.86) & 6.88 (0.00) & 0.73 (0.01) & 2.7\\
615 & 10.58 (0.98) & 8.10 (0.03) & 0.67 (0.07) & 2.8 & 18.94 (0.88) & 8.22 (0.02) & 0.70 (0.04) &  4.6\\
627 & 19.03 (0.97) & 8.41 (0.02) & 0.72 (0.04) & 2.8 & 45.30 (1.00) & 8.52 (0.01) & 0.54 (0.01) &  4.7\\
709 & 10.51 (1.00) & 8.56 (0.03) & 0.67 (0.07) & 2.8 & 17.70 (0.93) & 8.50 (0.02) & 0.63 (0.04) &  4.6 \\
715 & 111.33 (1.26) & 8.97 (0.03) & 0.58 (0.08) & 2.6 & 17.82 (0.88) & 8.99 (0.02) & 0.70 (0.04) &  4.2\\
746 & 25.18 (0.97) & 8.93 (0.02) & 0.89 (0.04) & 2.6 & 57.27 (0.84) & 9.03 (0.01) & 0.76 (0.01) &  4.5\\
752 & -- & -- & -- & 2.3 & -- & -- & -- & 3.8\\
768 &  & &  & 2.0 & 13.30 (0.98) & 8.90 (0.02) & 0.54 (0.04) & 3.0\\
799 & 21.30 (1.04) & 10.10 (0.02) & 0.63 (0.04) & 2.6 & 41.33 (0.93) & 10.30 (0.01) & 0.63 (0.02) & 4.0 \\
800 & 41.64 (0.92) & 10.27 (0.01) & 0.79 (0.02) & 2.7 & 89.53 (0.85) & 10.43 (0.00) & 0.74 (0.01) &  4.3 \\ \hline \hline
\multicolumn{5}{c}{\textbf{$\mathbf{H^{13}CCCN}$ J=4-3, F=3-2 (35267.403 MHz)}} & \multicolumn{4}{c}{\textbf{$\mathbf{H^{13}CCCN}$ J=5-4, F=4-3 (44084.171 MHz)}} \\ \hline
Core \# & \begin{tabular}[c]{@{}c@{}}$T_{\rm mb}$\\ (mK)\end{tabular} & \begin{tabular}[c]{@{}c@{}}$v_{\rm lsr}$\\ (km s$^{-1}$)\end{tabular} & \begin{tabular}[c]{@{}c@{}}FWHM\\ (km s$^{-1}$)\end{tabular} & \begin{tabular}[c]{@{}c@{}}RMS\\ (mK)\end{tabular} & \begin{tabular}[c]{@{}c@{}}$T_{\rm mb}$\\ (mK)\end{tabular} & \begin{tabular}[c]{@{}c@{}}$v_{\rm lsr}$\\ (km s$^{-1}$)\end{tabular} & \begin{tabular}[c]{@{}c@{}}FWHM\\ (km s$^{-1}$)\end{tabular} & \begin{tabular}[c]{@{}c@{}}RMS\\ (mK)\end{tabular} \\ \hline
264 & -- & -- & -- & 2.7 & -- & -- & -- & 5.7 \\
317 &  --& -- & -- &  3.2 & -- & -- & -- &  6.5 \\
321 & -- &  --& -- & 2.5 & -- & -- & -- & 5.3  \\
326 &  --& -- &--  &  2.4& -- & -- & -- &  4.0 \\
413 & 13.02 (0.95) & 7.66 (0.03) & 0.71 (0.06) & 3.0 & 45.84 (0.79) & 7.90 (0.01) & 0.82 (0.02) & 5.7 \\
504 & 8.70 (0.83) & 6.73 (0.04) & 0.94 (0.10) & 2.4 & 18.19 (0.84) & 6.77 (0.02) & 0.74 (0.04) & 4.2\\
615 & -- & -- & -- & 3.0 & -- &  & -- &  6.0 \\
627 & -- & -- & -- & 3.2 & -- & -- & -- &  6.1 \\
709 & -- & -- & -- & 3.0 & -- & -- & -- &  6.0 \\
715 & -- & -- & -- & 2.9 & -- & -- & -- & 5.7\\
746 & -- & -- & -- & 3.1 & -- &  -- & -- & 6.2\\
752 & -- & -- & -- & 2.5 & -- & -- & -- &  4.9\\
768 & -- & -- & -- & 2.3 & -- & -- & -- & 3.7\\
799 & -- & -- & -- & 3.3 & -- & -- & -- & 5.6\\
800 & -- & -- & -- & 2.9 & 23.12 (0.98) & 10.39 (0.01) & 0.52 (0.03) & 5.6
\end{tabular}
\end{table*}


\begin{table*}
\begin{tabular}{ccccccccc}
\multicolumn{5}{c}{\textbf{$\mathbf{HC^{13}CCN}$ J=4-3, F=3-2 (36237.945 MHz)}} & \multicolumn{4}{c}{\textbf{$\mathbf{HC^{13}CCN}$ J=5-4, F=4-3 (45297.347 MHz)}} \\ \hline
Core \# & \begin{tabular}[c]{@{}c@{}}$T_{\rm mb}$\\ (mK)\end{tabular} & \begin{tabular}[c]{@{}c@{}}$v_{\rm lsr}$\\ (km s$^{-1}$)\end{tabular} & \begin{tabular}[c]{@{}c@{}}FWHM\\ (km s$^{-1}$)\end{tabular} & \begin{tabular}[c]{@{}c@{}}RMS\\ (mK)\end{tabular} & \begin{tabular}[c]{@{}c@{}}$T_{\rm mb}$\\ (mK)\end{tabular} & \begin{tabular}[c]{@{}c@{}}$v_{\rm lsr}$\\ (km s$^{-1}$)\end{tabular} & \begin{tabular}[c]{@{}c@{}}FWHM\\ (km s$^{-1}$)\end{tabular} & \begin{tabular}[c]{@{}c@{}}RMS\\ (mK)\end{tabular} \\ \hline
264 & -- & -- & -- & 2.7 & -- & -- & -- & 5.7 \\
317 & -- & -- & -- & 3.2 & -- & -- & -- &  6.5 \\
321 & -- & -- & -- & 2.5 & -- & -- & -- & 5.3  \\
326 & -- & -- & -- & 2.4 & 14.7 (0.8) & 7.901 (0.020) & 0.71 (0.05) &  4.0 \\
413 & 22.80 (1.01) & 7.74 (0.01) & 0.59 (0.03) & 3.0 & 34.06 (0.84) & 7.92 (0.01) & 0.72 (0.02) & 5.7 \\
504 & 8.70 (0.83) & 6.73 (0.04) & 0.94 (0.10) & 2.4 & 18.19 (0.84) & 6.77 (0.02) & 0.74 (0.04) & 4.2\\
615 & -- & -- & -- & 3.0 & -- & -- & -- &  6.0 \\
627 & -- & -- & -- & 3.2 & -- & -- & -- &   6.1\\
709 & -- & -- & -- & 3.0 & -- & -- & -- &  6.0 \\
715 & -- & -- & -- & 2.9 & -- & -- & -- & 5.7 \\
746 & -- & -- & -- & 3.1 & -- & -- & -- & 6.2 \\
752 & -- & -- & -- & 2.5 & -- & -- & -- & 4.9 \\
768 & -- & -- & -- & 2.3 & -- & -- & -- & 3.7 \\
799 & -- & -- & -- & 3.3 & -- & -- & -- & 5.6 \\
800 & 10.29 (1.05)  &  10.07 (0.03)& 0.45 (0.05) & 2.9 & 18.82 (0.84) & 10.52 (0.16) & 0.72 (0.04) & 5.6\\ \hline \hline
\multicolumn{5}{c}{\textbf{$\mathbf{HCC^{13}CN}$ J=4-3, F=3-2 (36241.435 MHz)}} & \multicolumn{4}{c}{\textbf{$\mathbf{HCC^{13}CN}$ J=5-4, F=4-3 (45301.711 MHz)}} \\ \hline
Core \# & \begin{tabular}[c]{@{}c@{}}$T_{\rm mb}$\\ (mK)\end{tabular} & \begin{tabular}[c]{@{}c@{}}$v_{\rm lsr}$\\ (km s$^{-1}$)\end{tabular} & \begin{tabular}[c]{@{}c@{}}FWHM\\ (km s$^{-1}$)\end{tabular} & \begin{tabular}[c]{@{}c@{}}RMS\\ (mK)\end{tabular} & \begin{tabular}[c]{@{}c@{}}$T_{\rm mb}$\\ (mK)\end{tabular} & \begin{tabular}[c]{@{}c@{}}$v_{\rm lsr}$\\ (km s$^{-1}$)\end{tabular} & \begin{tabular}[c]{@{}c@{}}FWHM\\ (km s$^{-1}$)\end{tabular} & \begin{tabular}[c]{@{}c@{}}RMS\\ (mK)\end{tabular} \\ \hline
264 & 9.39 (0.84) & 7.64 (0.04) & 0.89 (0.09) & 2.7 & 15.45 (0.84) & 7.96 (0.02) & 0.71 (0.05) & 5.7 \\
317 & -- & -- & -- & 3.2 & -- & -- & -- & 6.5\\
321 & -- & -- & -- & 2.5 & -- & -- & -- & 5.3 \\
326 & -- & -- & -- & 2.4 & 15.14 (0.75) & 7.74 (0.02) & 0.89 (0.05) & 4.0\\
413 & 32.57 (0.92) & 7.61 (0.01) & 0.75 (0.02) & 3.0 & 58.54 (0.93) & 7.89 (0.01) & 0.58 (0.01) & 5.7\\
504 & 18.32 (1.00) & 6.59 (0.02) & 0.56 (0.03) & 2.4 & 20.64 (0.91) & 6.88 (0.01) & 0.61 (0.03) & 4.2\\
615 & -- & -- & -- & 3.0 & -- & -- & -- & 6.0 \\
627 & -- & -- & -- & 3.2 & -- & -- & -- & 6.1\\
709 & -- & -- & -- & 3.0  & -- & -- & -- & 6.0\\
715 & -- & -- & -- & 2.9 & -- & -- & -- & 5.7\\
746 & -- & -- & -- & 3.1& -- & -- & -- & 6.2\\
752 & -- & -- & -- & 2.5 & -- & -- & -- & 4.9\\
768 & -- & -- & -- & 2.3 & 11.55 (0.99) & 9.00 (0.02) & 0.49 (0.05) & 3.7\\
799 & -- & -- & -- & 3.3 & -- & -- & -- & 5.6\\
800 & 10.14 (0.94) & 10.00 (0.03) & 0.72 (0.08) & 2.9 & 24.06 (1.01) & 10.30 (0.01) & 0.52 (0.03) & 5.6\\ \hline \hline
\multicolumn{5}{c}{\textbf{$\mathbf{HC^5N}$ J=12-11 (31951.772 MHz)}} & \multicolumn{4}{c}{\textbf{$\mathbf{HC^5N}$ J=13-12 (34614.387 MHz)}} \\ \hline
Core \# & \begin{tabular}[c]{@{}c@{}}$T_{\rm mb}$\\ (mK)\end{tabular} & \begin{tabular}[c]{@{}c@{}}$v_{\rm lsr}$\\ (km s$^{-1}$)\end{tabular} & \begin{tabular}[c]{@{}c@{}}FWHM\\ (km s$^{-1}$)\end{tabular} & \begin{tabular}[c]{@{}c@{}}RMS\\ (mK)\end{tabular} & \begin{tabular}[c]{@{}c@{}}$T_{\rm mb}$\\ (mK)\end{tabular} & \begin{tabular}[c]{@{}c@{}}$v_{\rm lsr}$\\ (km s$^{-1}$)\end{tabular} & \begin{tabular}[c]{@{}c@{}}FWHM\\ (km s$^{-1}$)\end{tabular} & \begin{tabular}[c]{@{}c@{}}RMS\\ (mK)\end{tabular} \\ \hline
264 & 104.41 (1.03) & 7.85 (0.00) & 0.62 (0.01) & 2.5 & 102.26 (0.97) & 7.87 (0.00) & 0.69 (0.01) & 2.7 \\
317 & -- & -- & -- & 2.7 & -- & -- & -- &  3.2 \\
321 & -- & -- & -- & 2.3 & 10.59 (0.85) & 8.55 (0.04) & 0.91 (0.08) & 2.5\\
326 & 28.28 (0.91) & 7.96 (0.01) & 0.86 (0.03) & 2.2 & 33.60 (0.88) & 7.97 (0.01) & 0.85 (0.03) &  2.4 \\
413 & 329.92 (1.08) & 7.80 (0.00) & 0.54 (0.00) & 2.7 & 284.37 (1.05) & 7.83 (0.00) & 0.64 (0.00) & 3.0 \\
504 & 104.65 (1.08) & 6.75 (0.00) & 0.49 (0.01) & 2.2 & 82.82 (1.02) & 6.78 (0.00) & 0.67 (0.01) & 2.4\\
615 & 33.50 (1.06) & 8.06 (0.01) & 0.54 (0.02) & 2.8 & 27.53 (0.99) & 8.11 (0.01) & 0.67 (0.03) &  3.0\\
627 & 28.70 (1.34) & 8.48 (0.01) & 0.50 (0.03) & 2.8 & 20.45 (1.00) & 8.54 (0.02) &  0.55 (0.03) & 3.2 \\
709 & -- & -- & -- & 2.8 & 11.26 (0.89) & 8.64 (0.03) & 0.84 (0.08) & 3.0 \\
715 & 10.00 (1.17) & 9.03 (0.04) & 0.50 (0.07) & 2.6 & -- & -- & -- & 2.9\\
746 & 27.82 (3.86) & 8.97 (0.01) & 0.45 (0.07) & 2.6 & 29.96 (1.41) & 9.00 (0.01)  & 0.53 (0.03) & 3.1\\
752 & -- & -- & -- & 2.3 & -- & -- & -- & 2.5 \\
768 & 51.62 (0.99) & 8.98 (0.01) & 0.64 (0.01) & 2.0 & 49.13 (0.96) & 9.05 (0.01) & 0.71 (0.02) & 2.3\\
799 & 67.37 (48.72) & 10.17 (0.02) & 0.35 (0.18) & 2.6 & 36.76 (1.03) & 10.12 (0.01) & 0.61 (0.02) & 3.3\\
800 & 121.45 (1.27) & 10.30 (0.00) & 0.58 (0.01) & 2.7 & 126.34 (0.99) & 10.35 (0.00) & 0.65 (0.01) & 2.9 
\end{tabular}
\end{table*}


\begin{table*}
\begin{tabular}{ccccccccc}
\multicolumn{5}{c}{\textbf{$\mathbf{HC^5N}$ J=14-13 (37276.994 MHz)}} & \multicolumn{4}{c}{\textbf{$\mathbf{HC^5N}$ J=15-14 (39939.994 MHz)}} \\ \hline
Core \# & \begin{tabular}[c]{@{}c@{}}$T_{\rm mb}$\\ (mK)\end{tabular} & \begin{tabular}[c]{@{}c@{}}$v_{\rm lsr}$\\ (km s$^{-1}$)\end{tabular} & \begin{tabular}[c]{@{}c@{}}FWHM\\ (km s$^{-1}$)\end{tabular} & \begin{tabular}[c]{@{}c@{}}RMS\\ (mK)\end{tabular} & \begin{tabular}[c]{@{}c@{}}$T_{\rm mb}$\\ (mK)\end{tabular} & \begin{tabular}[c]{@{}c@{}}$v_{\rm lsr}$\\ (km s$^{-1}$)\end{tabular} & \begin{tabular}[c]{@{}c@{}}FWHM\\ (km s$^{-1}$)\end{tabular} & \begin{tabular}[c]{@{}c@{}}RMS\\ (mK)\end{tabular} \\ \hline
264 & 125.89 (1.00) & 7.94 (0.00) & 0.53 (0.01) & 2.8 & 148.51 (1.01) & 8.00 (0.00) & 0.45 (0.00) &  3.2\\
317 & 10.42 (0.99) & 8.35 (0.03) & 0.56 (0.06) & 3.2 & 10.71 (0.90) & 8.01 (0.03) & 0.70 (0.07) & 3.6\\
321 & 9.23 (0.90) & 8.53 (0.04) & 0.76 (0.09) & 2.5 & 10.41 (0.82)  & 8.71 (0.03) & 0.86 (0.08) & 3.4\\
326 & 36.21 (0.85) & 8.07 (0.01) & 0.84 (0.02) & 2.4 & 31.00 (0.81) & 8.06 (0.01) & 0.88 (0.03) &  3.1 \\
413 & 340.52 (1.07) & 7.90 (0.00) & 0.56 (0.00) & 3.0 & 433.09 (1.01) & 7.96 (0.00) & 0.43 (0.00) & 3.5 \\
504 & 101.77 (1.13) & 6.85 (0.00) & 0.53 (0.01) & 2.4 & 119.46 (1.30) & 6.90 (0.00) & 0.46 (0.01) & 3.0\\
615 & 34.81 (1.34) & 8.14 (0.01) & 0.50 (0.03) & 3.0 & 40.80 (1.00) & 8.26 (0.01) & 0.45 (0.01) & 3.6\\
627 & 30.76 (1.00) & 8.58 (0.01) & 0.45 (0.02) & 3.3 & 41.12 (1.28) & 8.62 (0.01) & 0.41 (0.02) & 3.5\\
709 & -- & -- & -- & 3.0 & -- & -- & -- & 3.9 \\
715 & -- & -- & -- & 2.9 & -- & -- & -- & 3.7\\
746 & 32.22 (1.04) & 9.01 (0.01) & 0.60 (0.02) & 3.1 & 33.79 (1.00) & 9.13 (0.01) & 0.39 (0.01) & 3.8\\
752 & -- & -- & -- &  2.5& -- & -- & -- &  3.0\\
768 & 64.86 (1.03) & 9.07 (0.01) & 0.59 (0.01) & 2.3 & 89.54 (1.00) & 9.14 (0.00) & 0.46 (0.01) & 3.1\\
799 & 44.92 (1.61) & 10.25 (0.01) & 0.46 (0.02) &  3.3 & 54.81 (1.09) & 10.32 (0.01) & 0.40 (0.01) & 3.5\\
800 & 152.15 (1.03) & 10.40 (0.00) & 0.57 (0.00) & 2.9 & 164.26 (1.21) & 10.45 (0.00) & 0.50 (0.01) & 3.6\\ \hline \hline
\multicolumn{5}{c}{\textbf{$\mathbf{HC^5N}$ J=16-15 (42602.171 MHz)}} & \multicolumn{4}{c}{\textbf{$\mathbf{HC^5N}$ J=17-16 (45264.745 MHz)}} \\ \hline
Core \# & \begin{tabular}[c]{@{}c@{}}$T_{\rm mb}$\\ (mK)\end{tabular} & \begin{tabular}[c]{@{}c@{}}$v_{\rm lsr}$\\ (km s$^{-1}$)\end{tabular} & \begin{tabular}[c]{@{}c@{}}FWHM\\ (km s$^{-1}$)\end{tabular} & \begin{tabular}[c]{@{}c@{}}RMS\\ (mK)\end{tabular} & \begin{tabular}[c]{@{}c@{}}$T_{\rm mb}$\\ (mK)\end{tabular} & \begin{tabular}[c]{@{}c@{}}$v_{\rm lsr}$\\ (km s$^{-1}$)\end{tabular} & \begin{tabular}[c]{@{}c@{}}FWHM\\ (km s$^{-1}$)\end{tabular} & \begin{tabular}[c]{@{}c@{}}RMS\\ (mK)\end{tabular} \\ \hline
264 & 140.90 (1.20) & 8.02 (0.00) & 0.47 (0.01) & 4.2 & 144.85 (1.00) & 8.06 (0.00) & 0.43 (0.00) &  5.7\\
317 & -- & -- & -- & 4.7 & -- & -- & -- &  6.5\\
321 & 12.90 (0.90) & 8.48 (0.02) & 0.67 (0.05)  & 3.2 & -- & -- & -- &  5.3\\
326 & 37.91 (0.79) & 8.06 (0.01) & 0.85 (0.02) & 3.5 & 37.81 (0.83) & 8.10 (0.01) & 0.73 (0.02) & 4.0\\
413 & 315.63 (0.98) & 7.98 (0.00) & 0.56 (0.00) & 3.3 & 325.79 (0.98) & 8.01 (0.00) & 0.53 (0.00) &  5.7\\
504 & 98.71 (1.01) & 6.94 (0.00) & 0.50 (0.01) & 2.7 & 84.53 (0.98) & 7.00 (0.00) & 0.54 (0.01) & 4.2 \\
615 & 52.59 (1.78) & 8.26 (0.03) & 0.24 (0.01) & 4.6 & 28.43 (0.91) & 8.22 (0.01) & 0.61 (0.02) &  6.0\\
627 & 37.40 (1.00) & 8.68 (0.01) & 0.42 (0.01) & 4.7 & 41.89 (0.45) & 8.67 (0.01) & 0.28 (0.03) &  6.1\\
709 & 10.51 (1.00) & 8.56 (0.03) & 0.67 (0.07) & 4.6 & 17.70 (0.93) & 8.50 (0.02) & 0.63 (0.04) & 6.0 \\
715 & 111.33 (1.26) & 8.97 (0.03) & 0.58 (0.08) & 4.2 & 17.82 (0.88) & 8.99 (0.02) & 0.70 (0.04) &  5.7\\
746 & 26.82 (1.49) & 9.08 (0.01) & 0.78 (0.07) & 4.5 & 31.42 (1.34) & 9.13 (0.01) & 0.42 (0.02) &  6.2\\
752 & -- & -- & -- & 3.8 & -- & -- & -- & 4.9\\
768 & 78.15 (1.11) & 9.11 (0.00) & 0.50 (0.01) & 3.0 & 67.65 (0.99) & 9.16 (0.00) & 0.53 (0.01) & 3.7\\
799 & 47.7 (1.1) & 10.381 (0.006) & 0.379 (0.010) & 4.0 & 48.3 (1.0) & 10.361 (0.005) & 0.382 (0.008) &  5.6\\
800 & 157.08 (1.04) & 10.46 (0.00) & 0.49 (0.00) & 4.3 & 147.23 (0.99) & 10.52 (0.00) & 0.52 (0.00) &  5.6\\ \hline \hline
\multicolumn{5}{c}{\textbf{$\mathbf{HC^5N}$ J=18-17 (47927.306 MHz)}} & \multicolumn{4}{c}{\textbf{$\mathbf{HC^{13}CCCCN}$ 12-11 (31624.340 MHz)}} \\ \hline
Core \# & \begin{tabular}[c]{@{}c@{}}$T_{\rm mb}$\\ (mK)\end{tabular} & \begin{tabular}[c]{@{}c@{}}$v_{\rm lsr}$\\ (km s$^{-1}$)\end{tabular} & \begin{tabular}[c]{@{}c@{}}FWHM\\ (km s$^{-1}$)\end{tabular} & \begin{tabular}[c]{@{}c@{}}RMS\\ (mK)\end{tabular} & \begin{tabular}[c]{@{}c@{}}$T_{\rm mb}$\\ (mK)\end{tabular} & \begin{tabular}[c]{@{}c@{}}$v_{\rm lsr}$\\ (km s$^{-1}$)\end{tabular} & \begin{tabular}[c]{@{}c@{}}FWHM\\ (km s$^{-1}$)\end{tabular} & \begin{tabular}[c]{@{}c@{}}RMS\\ (mK)\end{tabular} \\ \hline
264 & 136.80 (0.99) & 8.11 (0.00) & 0.44 (0.00) & 6.6 & -- & -- & -- &  2.5 \\
317 & -- & -- & -- & 7.1 & -- & -- & -- &  2.7 \\
321 & -- & -- & -- &  5.3& -- & -- & -- &  2.3 \\
326 & 23.36 (0.81) & 8.309 (0.012) & 0.73 (0.03) & 4.5 & 9.39 (2.75) & 7.88 (0.08)  & 0.39 (0.14)  & 2.2  \\
413 & 300.75 (0.91) & 8.05 (0.00) & 0.57 (0.00) & 6.1 & -- & -- & -- & 2.7 \\
504 & 80.76 (0.95) & 7.01 (0.00) & 0.53 (0.01) & 4.7 & -- & -- & -- & 2.2\\
615 & -- & -- & -- & 6.9 & -- & -- & -- & 2.8 \\
627 & 34.49 (1.02) & 8.70 (0.01) & 0.37 (0.01) & 7.1 & -- & -- & -- &  2.8 \\
709 & -- & -- & -- & 6.8 & -- & -- & -- & 2.8 \\
715 & -- & -- & -- &  6.6 & -- & -- & -- & 2.6\\
746 & 34.69 (1.23) & 9.20 (0.13) & 0.46 (0.07) & 6.9 & -- & -- & -- & 2.6\\
752 & -- & -- & -- & 5.6 & -- & -- & -- & 2.3 \\
768 & 74.43 (0.92) & 9.21 (0.00) & 0.56 (0.01) & 4.4 & -- & -- & -- & 2.0 \\
799 & 58.45 (1.44) & 10.44 (0.02) & 0.24 (0.00) & 6.6 & -- & -- & -- & 2.6\\
800 & 143.03 (0.92) & 10.56 (0.00) & 0.56 (0.00) & 6.8 & -- & -- & -- & 2.7
\end{tabular}
\end{table*}


\begin{table*}
\label{info5}
\begin{tabular}{ccccccccc}
\multicolumn{5}{c}{\textbf{$\mathbf{HCC^{13}CCCN}$ 14-13 (37238.390 MHz)}} & \multicolumn{4}{c}{\textbf{$\mathbf{HCC^{13}CCCN}$ 18-17 (47877.653 MHz)}} \\ \hline
Core \# & \begin{tabular}[c]{@{}c@{}}$T_{\rm mb}$\\ (mK)\end{tabular} & \begin{tabular}[c]{@{}c@{}}$v_{\rm lsr}$\\ (km s$^{-1}$)\end{tabular} & \begin{tabular}[c]{@{}c@{}}FWHM\\ (km s$^{-1}$)\end{tabular} & \begin{tabular}[c]{@{}c@{}}RMS\\ (mK)\end{tabular} & \begin{tabular}[c]{@{}c@{}}$T_{\rm mb}$\\ (mK)\end{tabular} & \begin{tabular}[c]{@{}c@{}}$v_{\rm lsr}$\\ (km s$^{-1}$)\end{tabular} & \begin{tabular}[c]{@{}c@{}}FWHM\\ (km s$^{-1}$)\end{tabular} & \begin{tabular}[c]{@{}c@{}}RMS\\ (mK)\end{tabular} \\ \hline
264 & -- & -- & -- & 2.8 & -- & -- & -- &  6.6 \\
317 & -- & -- & -- &  3.2& -- & -- & -- &  7.1 \\
321 & -- & -- & -- & 2.5 & -- & -- & -- &  5.3 \\
326 & -- & -- & -- & 2.4 & -- & -- & -- &  4.5 \\
413 & 9.42 (1.03) & 7.96 (0.03) & 0.42 (0.05) & 3.0 & -- & -- & -- &  6.1 \\
504 & -- & -- & -- & 2.4  & 14.18 (0.75) & 7.21 (0.02) & 0.85 (0.05) &  4.7 \\
615 & -- & -- & -- & 3.0 & -- & -- & -- &  6.9\\
627 & -- & -- & -- & 3.3 & -- & -- & -- &  7.1 \\
709 & -- & -- & -- & 3.0 & -- & -- & -- & 6.8 \\
715 & -- & -- & -- & 2.9 & -- & -- & -- & 6.6\\
746 & -- & -- & -- & 3.1 & -- & -- & -- &  6.9 \\
752 & -- & -- & -- & 2.5 & -- & -- & -- &  5.6\\
768 & 7.03 (0.93) & 9.35 (0.05) & 0.71 (0.11) & 2.3 & -- & -- & -- &  5.6 \\
799 & -- & -- & -- & 3.3 & -- & -- & -- & 6.6  \\
800 & -- & -- & -- & 2.9 & -- & -- & -- &  6.8 \\ \hline \hline
\multicolumn{5}{c}{\textbf{$\mathbf{HCCC^{13}CCN}$ 15-14 (39903.082 MHz)}} & \multicolumn{4}{c}{\textbf{$\mathbf{HCCCCC^{15}N}$ 19-18 (49347.584 MHz)}} \\ \hline
Core \# & \begin{tabular}[c]{@{}c@{}}$T_{\rm mb}$\\ (mK)\end{tabular} & \begin{tabular}[c]{@{}c@{}}$v_{\rm lsr}$\\ (km s$^{-1}$)\end{tabular} & \begin{tabular}[c]{@{}c@{}}FWHM\\ (km s$^{-1}$)\end{tabular} & \begin{tabular}[c]{@{}c@{}}RMS\\ (mK)\end{tabular} & \begin{tabular}[c]{@{}c@{}}$T_{\rm mb}$\\ (mK)\end{tabular} & \begin{tabular}[c]{@{}c@{}}$v_{\rm lsr}$\\ (km s$^{-1}$)\end{tabular} & \begin{tabular}[c]{@{}c@{}}FWHM\\ (km s$^{-1}$)\end{tabular} & \begin{tabular}[c]{@{}c@{}}RMS\\ (mK)\end{tabular} \\ \hline
264& -- & -- & -- & 3.2 & -- & -- & -- & 8.8  \\
317 & -- & -- & -- & 3.6 & -- & -- & -- &   9.3\\
321& -- & -- & -- & 3.4 & -- & -- & -- & 6.8  \\
326 & -- & -- & -- &  3.1 & -- & -- & -- &  6.7 \\
413 & 12.36 (3.95) & 5.98 (0.05) & 0.31 (0.11) & 3.5 & -- & -- & -- & 7.3 \\
504 & -- & -- & -- & 3.0 & -- & -- & -- &  5.9 \\
615& -- & -- & -- & 3.6 & 33.92 (0.86) & 8.04 (0.01) & 0.62 (0.02) & 9.1 \\
627 & -- & -- & -- & 3.5 & -- & -- & -- &  11.2 \\
709 & -- & -- & -- & 3.9 & -- & -- & -- & 9.0 \\
715 & -- & -- & -- & 3.7 & -- & -- & -- & 9.1\\
746 & -- & -- & -- & 3.8 & -- & -- & -- & 11.4  \\
752 & -- & -- & -- & 3.0 & -- & -- & -- & 8.2 \\
768 & -- & -- & -- & 3.1& -- & -- & -- & 6.2  \\
799 & -- & -- & -- & 3.5  & -- & -- & -- &  8.5\\
800 & -- & -- & -- & 3.6 & -- & -- & -- &  8.6 \\ \hline \hline
\multicolumn{5}{c}{\textbf{$\mathbf{DCCCCCN}$ 14-13 (35589.324 MHz)}} & \multicolumn{4}{c}{\textbf{$\mathbf{HC_{7}N}$ J=28-27 (31583.704 MHz)}} \\ \hline
Core \# & \begin{tabular}[c]{@{}c@{}}$T_{\rm mb}$\\ (mK)\end{tabular} & \begin{tabular}[c]{@{}c@{}}$v_{\rm lsr}$\\ (km s$^{-1}$)\end{tabular} & \begin{tabular}[c]{@{}c@{}}FWHM\\ (km s$^{-1}$)\end{tabular} & \begin{tabular}[c]{@{}c@{}}RMS\\ (mK)\end{tabular} & \begin{tabular}[c]{@{}c@{}}$T_{\rm mb}$\\ (mK)\end{tabular} & \begin{tabular}[c]{@{}c@{}}$v_{\rm lsr}$\\ (km s$^{-1}$)\end{tabular} & \begin{tabular}[c]{@{}c@{}}FWHM\\ (km s$^{-1}$)\end{tabular} & \begin{tabular}[c]{@{}c@{}}RMS\\ (mK)\end{tabular} \\ \hline
264 & -- & -- & -- & 2.7 & 14.68 (1.00) & 7.79 (0.02) & 0.65 (0.05) & 2.5  \\
317 & -- & -- & -- & 3.2& -- & -- & -- & 2.7 \\
321 & -- & -- & -- & 2.5 & -- & -- & -- & 2.3  \\
326 & -- & -- & -- & 2.4 & -- & -- & -- &  2.2\\
413 & 9.61 (1.02) & 7.71 (0.03) & 0.47 (0.05) & 3.0 & 35.83 (1.22) & 7.78 (0.01) & 0.47 (0.02) &  2.7 \\
504 & -- & -- & -- &  2.4& -- & -- & -- & 2.2  \\
615 & -- & -- & -- & 3.0 & -- & -- & -- &  2.8\\
627& -- & -- & -- & 3.2 & -- & -- & -- & 2.8 \\
709 & -- & -- & -- &  3.0& -- & -- & -- & 2.8 \\
715& -- & -- & -- &2.9 & -- & -- & -- & 2.6 \\
746& -- & -- & -- & 3.1  & -- & -- & -- & 2.6 \\
752 & -- & -- & -- & 2.5 & -- & -- & -- &  2.3\\
768& -- & -- & -- & 2.3  & -- & -- & -- & 2.0  \\
799 & -- & -- & -- & 3.3 & -- & -- & -- & 2.6 \\
800& -- & -- & -- & 2.9 & 16.06 (2.29) & 10.34 (0.03) & 0.45 (0.08) & 2.7  \\
\end{tabular}
\end{table*}

\begin{table*}
\label{info6}
\begin{tabular}{ccccccccc}
\multicolumn{5}{c}{\textbf{$\mathbf{HC_{7}N}$ J=29-28 (32711.666 MHz)}} & \multicolumn{4}{c}{\textbf{$\mathbf{HC_{7}N}$ J=30-29 (33839.628 MHz)}} \\ \hline
Core \# & \begin{tabular}[c]{@{}c@{}}$T_{\rm mb}$\\ (mK)\end{tabular} & \begin{tabular}[c]{@{}c@{}}$v_{\rm lsr}$\\ (km s$^{-1}$)\end{tabular} & \begin{tabular}[c]{@{}c@{}}FWHM\\ (km s$^{-1}$)\end{tabular} & \begin{tabular}[c]{@{}c@{}}RMS\\ (mK)\end{tabular} & \begin{tabular}[c]{@{}c@{}}$T_{\rm mb}$\\ (mK)\end{tabular} & \begin{tabular}[c]{@{}c@{}}$v_{\rm lsr}$\\ (km s$^{-1}$)\end{tabular} & \begin{tabular}[c]{@{}c@{}}FWHM\\ (km s$^{-1}$)\end{tabular} & \begin{tabular}[c]{@{}c@{}}RMS\\ (mK)\end{tabular} \\ \hline
264 & -- & -- & -- & 2.5 & 11.2 (0.9) & 7.63 (0.03) & 0.79 (0.08) &  2.5 \\
317 & -- &--  & -- & 2.7 & -- & -- & -- &  2.7 \\
321 & -- & -- & -- & 2.3 & -- & -- & -- & 2.3  \\
326 & -- & -- & -- & 2.2 & -- & -- & -- & 2.2  \\
413 & 37.14 (1.09) & 7.85 (0.01) & 0.46 (0.02) & 2.7 & 27.11 (1.10) & 7.84 (0.01) & 0.54 (0.03) &  2.7 \\
504 & -- & -- & -- &  2.2& -- & -- & -- &  2.2 \\
615 & -- & -- & -- & 2.8 & -- & -- & -- & 2.8 \\
627 & -- & -- & -- & 2.8 & -- & -- & -- &  2.8 \\
709 & -- & -- & -- & 2.8 & -- & -- & -- & 2.8 \\
715 & -- & -- & -- & 2.6 & -- & -- & -- & 2.6\\
746 & -- & -- & -- & 2.6 & --& -- & -- &  2.6 \\
752 & -- & -- & -- & 2.3 & -- & -- & -- & 2.3 \\
768 & -- & -- & -- & 2.0 & -- & -- & -- &  2.0 \\
799 & -- & -- & -- & 2.6 & -- &  --& -- &  2.6 \\
800 & 16.91 (1.10) & 10.18 (0.02) & 0.54 (0.04) & 2.7 & -- & -- & -- &  2.7 \\\hline \hline
\multicolumn{5}{c}{\textbf{$\mathbf{HC_{7}N}$ J=31-30 (34967.585 MHz)}} & \multicolumn{4}{c}{\textbf{$\mathbf{HC_{7}N}$ J=32-31 (36095.541 MHz)}} \\ \hline
Core \# & \begin{tabular}[c]{@{}c@{}}$T_{\rm mb}$\\ (mK)\end{tabular} & \begin{tabular}[c]{@{}c@{}}$v_{\rm lsr}$\\ (km s$^{-1}$)\end{tabular} & \begin{tabular}[c]{@{}c@{}}FWHM\\ (km s$^{-1}$)\end{tabular} & \begin{tabular}[c]{@{}c@{}}RMS\\ (mK)\end{tabular} & \begin{tabular}[c]{@{}c@{}}$T_{\rm mb}$\\ (mK)\end{tabular} & \begin{tabular}[c]{@{}c@{}}$v_{\rm lsr}$\\ (km s$^{-1}$)\end{tabular} & \begin{tabular}[c]{@{}c@{}}FWHM\\ (km s$^{-1}$)\end{tabular} & \begin{tabular}[c]{@{}c@{}}RMS\\ (mK)\end{tabular} \\ \hline
264 & 13.67 (2.07) & 7.83 (0.02) & 0.47 (0.09) & 2.7  & 11.52 (1.92) & 7.96 (0.02) &  0.46 (0.10) &  2.7 \\
317 & -- & -- & -- & 3.2 & -- & -- &-- &  3.2\\
321 & -- & -- & -- & 2.5 & -- & -- & -- &  2.5 \\
326 & -- &  -- & -- & 2.4 & -- & -- & -- &   2.4\\
413 & 23.64 (1.20) & 7.80 (0.01) & 0.50 (0.03) & 3.0 & 26.08 (1.12) & 7.80 (0.01) & 0.58 (0.03) & 3.0\\
504 & -- & -- & -- & 2.4 & -- & -- & -- &  2.4 \\
615 & -- & -- & -- & 3.0 & -- & -- & -- & 3.0  \\
627 & --  & -- & -- & 3.2 & -- & -- & -- & 3.2  \\
709 & -- & -- &  -- & 3.0 & -- & -- & -- & 3.0 \\
715 & -- & -- & -- & 2.9 & -- & -- & -- &  2.9\\
746 & -- & -- & -- & 3.1 & -- & -- & -- &  3.1 \\
752 & -- & -- & -- &  2.5 & -- & -- & -- &  2.5\\
768 & -- & -- & -- & 2.3 & 9.55 (2.53) & 8.92 (0.03) & 0.42 (0.14) & 2.3  \\
799 & -- & -- & -- & 3.3 & -- & -- & -- &  3.3 \\
800 & -- & -- & -- & 2.9 & -- & -- & -- &  2.9\\
\end{tabular}
\end{table*}

\bsp	
\label{lastpage}
\end{document}